\documentclass[11pt]{article}
\usepackage[dvips]{graphicx}
\usepackage{color}
\usepackage[x11names,rgb]{xcolor}
\usepackage[margin=1cm]{geometry}
\usepackage{fancyvrb,xcolor}
\usepackage{amsfonts}
\usepackage{amsmath, bm}
\usepackage{amssymb}
\usepackage{amsthm}
\usepackage{enumerate}
\usepackage{venndiagram}
\usepackage{booktabs}
\usepackage{multirow}
\usepackage{authblk}
\usepackage[pdftex]{hyperref}
\usepackage{natbib}
\usepackage{fancybox}
\usepackage{hyperref}
\usepackage{authblk}
\usepackage[normalem]{ulem}
\usepackage[font={small},labelfont=bf]{caption}
\usepackage[draft]{changes}

\hypersetup{colorlinks,            citecolor=blue,            filecolor=blue,            linkcolor=blue,            urlcolor=blue,            }
\renewcommand{\vec}[1]{\mathbf{#1}}
\newtheorem{proposition}{{\bf \sc Proposition}}
\newtheorem{theorem}{{\bf \sc Theorem}}
\newtheorem{lemma}{{\bf \sc Lemma}}

\newtheorem{corollary}{{\bf \sc Corollary}}
\newtheorem{definition}{{\bf \sc Definition}}
\newtheorem{remark}{{\bf \sc Remark}}

\newtheorem{assumption}{{\bf \sc Assumption}}

\theoremstyle{definition}
\DeclareMathOperator*{\argmax}{argmax}
\newtheorem{examplex}{Example}
\newenvironment{example}
  {\pushQED{\qed}\examplex}
  {\popQED \endexamplex}
\input{tcilatex}

\begin{document}

\title{Learning and Selfconfirming Equilibria in Network Games\thanks{%
We thank Federico Bobbio, Davide Bordoli, Yann Bramoull\'{e}, Ben Golub,
Sebastiano Della Lena, Nicol\`{o} Generoso, Julien Manili, Paola
Moscariello, Alessandro Pavan, Giulio Principi, Sergey Savchenko, Yves Zenou, and seminar
participants at 24$^{th}$ CTN Workshop in Aix--en--Provence, Bergamo,
Bocconi, CISEI in Capri, Milano Bicocca, Milano Cattolica, 7$^{th}$ European
Meeting on Networks in Cambridge, Marseille, Nazarbayev University, NSE at
Indiana University, Pompeu Fabra, Siena, UTS Sydney, Venice. Pierpaolo
Battigalli, Fabrizio Panebianco, and Paolo Pin gratefully acknowledge
funding from, respectively, the European Research Council (ERC) grant
324219, the Spanish Ministry of Economia y Competitividad project
ECO2017-87245-R, and the Italian Ministry of Education Progetti di Rilevante
Interesse Nazionale (PRIN) grant 2017ELHNNJ.}}

\author[a,d,e]{Pierpaolo Battigalli}
\author[b,f]{Fabrizio Panebianco}
\author[c,e]{Paolo Pin}

\affil[a]{Department of Decision Sciences, Universit\`a  Bocconi, Milan,
Italy} 
\affil[b]{Department of Economics and Finance, Universit\`a
Cattolica, Milan, Italy} 
\affil[c]{Department of Economics and Statistics,
Universit\`a di Siena, Italy} 
\affil[d]{IGIER,  Universit\`a  Bocconi,
Milan, Italy} \affil[e]{BIDSA,  Universit\`a  Bocconi, Milan, Italy} %
\affil[f]{CLE,  Universit\`a Cattolica, Milan, Italy}

\date{{\bf Extended version} \\ July 2022}
\maketitle

\begin{abstract}
Consider a set of agents who play a network game repeatedly. Agents may not
know the network. They may even be unaware that they are interacting with
other agents in a network. Possibly, they just understand that their optimal
action depends on an unknown state that is, actually, an aggregate of the
actions of their neighbors. Each time, every agent chooses an action that
maximizes her instantaneous subjective expected payoff and then updates her
beliefs according to what she observes. In particular, we assume that each
agent only observes her realized payoff. A steady state of the resulting
dynamic is a \textbf{selfconfirming equilibrium} given the assumed feedback.
We characterize the structure of the set of selfconfirming equilibria in the
given class of network games, we relate selfconfirming and Nash equilibria,
and we analyze simple conjectural best-reply paths whose limit points are
selfconfirming equilibria. 
\end{abstract}


\affil[a]{Department of Decision Sciences, Universit\`a  Bocconi, Milan,
Italy} 
\affil[b]{Department of Economics and Finance, Universit\`a
Cattolica, Milan, Italy} 
\affil[c]{Department of Economics and Statistics,
Universit\`a di Siena, Italy} 
\affil[d]{IGIER,  Universit\`a  Bocconi,
Milan, Italy} \affil[e]{BIDSA,  Universit\`a  Bocconi, Milan, Italy} %
\affil[f]{CLE,  Universit\`a Cattolica, Milan, Italy}





\textbf{D83}, 
\textbf{D85}. 



\section{Introduction}

Social networks can be quite complex. Think about friendship networks,
networks of people interacting online (such as Twitter, Facebook, Instagram,
and so on){, or networks of firms (input-output or R\&D networks)}%
. These networks often consist of thousands (or millions) of agents or firms
interacting, and agents rarely know how the network is shaped.\footnote{%
For example, \cite{breza2018seeing} provide evidence from Indian rural
villages on the fact that people have actually limited knowledge about the
social networks of personal relations in which they are embedded, at odds
with many of the existing theoretical models of strategic interactions in
networks.} In this paper, we provide a novel approach to analyze how
incomplete information about the network affects behavior and learning
processes. We propose a framework in which agents may ignore how the network
affects their payoffs, how the network is shaped, or even that they are
interacting in a network.

The standard solution concept used to study the behavior of agents in
network games is Nash equilibrium, with the motivation that learning and
adaptation converge to a profile of actions in which every player best
responds to the actions of the other players. Nash equilibrium action
profiles are limit outcomes of learning paths when agents have perfect
feedback about the payoff relevant aspects of others' behavior. Yet, as we
shall argue, such perfect feedback hypothesis may be too strong for some
social networks applications and, if learning is based on imperfect
feedback, non-Nash action profiles may result as the steady-state limits of
learning paths. Indeed, such limits under (possibly) imperfect feedback are
characterized by the selfconfirming equilibrium concept. With this, we
analyze the effects of milder conditions on information feedback. To
illustrate, we consider examples where many agents interact and it is
plausible to assume that they cannot perfectly observe, whatever action they
take, the payoff-relevant aspects of the actions of the others. 

In our analysis we assume that the only feedback players receive is their
realized payoff. This implies that they do not always observe the
payoff-relevant aspects of the actions of others, represented by a payoff
state. Yet, each one of them understands how the payoff state and her action
determine her payoff and the feedback she is going to receive \emph{ex post}%
. We analyze how agents use the feedback they receive to update their
conjectures about the payoff state and best respond to them, and we
characterize limit behavior under different settings of local and global
externalities.

\subsection{Introductory example}

To be more specific about our modelling approach, let us introduce an
example that will guide us through the whole discussion. Consider an online
social network with many users, like Twitter, and a simultaneous-moves game
in which each user $i$ decides her level of activity $a_{i}\geq 0$ in the
social network.\footnote{%
Even if online social networks are now ubiquitous and relevant, there is a
very scarce literature based on game theory that models the incentives of
people to be active and interact on these platforms. We are aware of some
attempts by computer scientists, stemming from the early era of this form of
interaction, such as \cite{fu2007social}. In the economic literature, the
only paper we are aware of on this topic is \cite{tarbush2017social}, which
does not focus on the activity of users, but rather on the endogenous
formation of contacts.} The payoff that agents get from their activity
depends on the social interaction. We start considering the case in which
only local externalities are at play, eventually extending the model to the
case in which there are also global externalities. In particular, active
user $i$ receives idiosyncratic externalities---that can be positive or
negative---from the other users with whom she is in contact in the social
network. The externality from user $j$ to user $i$ is proportional to the
time that they both spend on the social network, $a_{i}$ and $a_{j}$.
Sticking to a quadratic specification, that yields linear best replies, we
assume that the payoff function of $i$ is\footnote{%
This is the class of \emph{linear--quadratic network} games originally
analyzed by \cite{ballester2006s}, as we discuss in the next section. We use 
\textbf{boldface }symbols to denote vectors (in this case, action profiles)
and matrices.} 
\begin{equation}
u_{i}(a_{i},\mathbf{a}_{-i})=\alpha _{i}a_{i}-\frac{1}{2}a_{i}^{2}+%
\displaystyle \sum_{j\in I\backslash \left \{ i\right \} }z_{ij}a_{i}a_{j}%
\text{.}  \label{eq:local_intro}
\end{equation}%
In equation (\ref{eq:local_intro}), $I$ is the set of agents, or
individuals, in the social network, $a_{i}$ is the level of activity of $%
i\in I$, $\mathbf{a}_{-i}$ is the profile of activities of all the other
users in $I$, and $\alpha _{i}>0$ represents the individual pleasure of $i$
from being active on the social network in isolation, which results in the 
\emph{bliss point} of activity in autarchy. 
For each $j\in I\backslash \left \{ i\right \} $, parameter $z_{ij}$
represents the intensity (absolute value) and type (sign) of the externality
from $j$ to $i$. We say that $j$ affects $i$, or that $j$ is a \textbf{peer}
(or a \textbf{neighbor}) of $i$, if $z_{ij}\neq 0$.

The network described by the matrix $\mathbf{Z}$ of all the $z_{ij}$'s is
assumed to be \emph{exogenous}. As a first approximation, this fits a \emph{%
directed} online social network like Twitter or Instagram, where users do
not have full control on who follows them.

Consider payoff as expressed in equation \eqref{eq:local_intro} (and in %
\eqref{eq:global_intro} below in the introduction). An endogenous directed
network in which player $i$ decides who to follow (the $z_{ji}$ entries of
matrix $\mathbf{Z}$) but not who is following her (the $z_{ij}$ entries of
matrix $\mathbf{Z}$) seems to us in line with our assumption of exogenous
network. That is because, in this modification of our model, a player
affects the payoff of those that she follows but her payoff is not affected
by their choices, including, if the network were endogenous, who they
follow. So, endogenizing $\mathbf{Z}$ would mean to endogenize choices (link
choices) that are payoff irrelevant for the players. 

Under this interpretation, $i$ receives positive or negative externalities
from those who follow her proportional to her activity. We do not assume
that player $i$ knows all the $z_{ij}$'s. She may not know them either
because she cannot observe who is following her,\footnote{%
There are online social networks, like \href{https://www.reddit.com/}{Reddit}%
, which actually do not provide this information at all to their users.
Reddit, in particular, provides a measure to each user, called \emph{karma},
which is apparently based on how many other people follow, and how much they
like, what that user posts. However, the algorithm on which this measure is
based is not public.} or because she knows her followers but she does not
know the sign or intensity of their externality. 
The payoff of $i$ represents 
both the pleasure that $i$ gets from participating in the platform and what $%
i$ can indirectly observe about her own popularity. 
We consider that $i$ cannot choose the style of what she writes, since she
just follows her exogenous nature. In this interpretation, $a_{i}$
represents both the amount of time that $i$ passes on the platform and the
amount of posts that $i$ writes, and this can make her more or less {%
appreciated}, according to how her style combines with the (typically
unobserved) tastes of each of her followers. 
{\ In our setting, player $i$ may also set $a_{i}=0$. Indeed, we interpret $%
\mathbf{Z}$ as a network of \emph{opportunities} of interaction, with
players deciding endogenously whether they want to be active or inactive.
When they are inactive, not only the network becomes irrelevant for them,
but they also become irrelevant for the payoffs of other players.}

{The feedback received by agents who have payoff given by %
\eqref{eq:local_intro} is such that, if a player $i$ decides to be inactive (%
$a_{i}=0$), then she cannot learn anything about the game and about what the
others are doing, whereas if she is active 
($a_{i}>0$) it is \emph{as if} she had perfect feedback: 
indeed, knowing }$a_{i}$ and the shape of $u_{i}$, {she can infer from her
realized payoff the aggregate activity of her peers }$\sum_{j\in I\backslash
\left \{ i\right \} }z_{ij}a_{j}$ (the payoff-relevant aspect of the
behavior of others) and understand whether $a_{i}$ was a best reply to it{.
Inactive players, instead, cannot observe whether inactivity was a best
reply to peers' activity. This simplified framework mimics the fact that,
for example, in an online social network active users are surrounded by
enough information to have a quasi-perfect feedback about what happens,
whereas inactive agents, because they opt out from the network, are likely
to ignore relevant information.}

We specify what agents observe after their choices because this affects how
they update their beliefs, and we are going to analyze learning dynamics and
their steady states. To fix ides, we shall refer to the social network
Twitter. Twitter user $i$, typically, does not observe the sign of the
externalities and the activity of others. However, she gets indirect
measures of her level of 
appreciation that come, for instance, from her conversations and experiences
in the real world, where her {activity on} Twitter affects her social and
professional real life. If the players are small firms using Twitter for
advertising, they will observe their actual profits. Players of this game
may have wrong beliefs about the details of the game they are playing (e.g.,
the structure of the network, or the value of the parameters) and about the
actions of other players. Consequently, they update their beliefs in
response to the feedback they receive, which is assumed to be their payoff, 
and maximize their instantaneous expected payoff given such updated beliefs.
This updating process yields \emph{learning paths} that do not necessarily
converge to a Nash equilibrium of the game.

Next we also consider an extra \textbf{global} term in the payoff function: 
\begin{equation}
u_{i}(a_{i},\mathbf{a}_{-i})=\alpha _{i}a_{i}-\frac{1}{2}a_{i}^{2}+%
\displaystyle \sum_{j\in I\backslash \left \{ i\right \}
}z_{ij}a_{i}a_{j}+\gamma \displaystyle \sum_{k\in I\backslash \left \{
i\right \} }a_{k}\text{.}  \label{eq:global_intro}
\end{equation}%
%
%
%
%
%
%
We can interpret this extra term $\gamma \displaystyle \sum_{k\in
I\backslash \left \{ i\right \} }a_{k}$ as an additional utility that $i$
gets, regardless of being active or inactive. In this case, what agents can
learn from being active or inactive radically changes with respect to the
previous case without global externalities, because even an active player
may not be able to disentangle what is the contribution of the global term.%
\newline
\bigskip

{We propose an online social network as our leading example, but
there are other possible cases in which incomplete information about the
network is key. For example, a network of firms, where the feedback is
observed profit and actions are levels of production, posted prices, or R\&D
activities.\footnote{%
These applications have been considered in the literature, each with
specific assumptions and different approaches from ours. For example, \cite%
{bimpikis2019cournot} consider Cournot competition, while \cite%
{nermuth2013informational}, \cite{lach2017asymmetric}, and \cite%
{heijnen2018price} consider Bertrand competition on multiple markets,
modelling the environment as a network with local externalities. This is the
same approach that \cite{westbrock2010natural} and \cite{konig2019r} use to
model R\&D local interactions between firms.} Many firms are competitors,
experiencing local substitutabilities in their choices, some are
complementors, and for some of them it may not be clear what kind of
strategic interaction is at play. Sometimes, the firm does not know of the
set of all its competitors or complementors. Moreover, firms often tend to
hide their investment plans and R\&D choices to some of their partners,
while each firm observes its own profits. In this case, firms ignore
important aspects of the network {and do not observe ex post the actions of
other firms. So,} incomplete information plays a critical role and, as we
are going to argue, objectively suboptimal choices may be implemented even
in a long-run steady state.\footnote{%
Anyway, incomplete information is not the only reason for non-Nash steady
states. As we formally argue in Appendix \ref{ration}, complete information
(i.e., common knowledge of the game) and strategic sophistication imply Nash
behavior in games with strategic complementarities and a unique Nash
equilibrium, but not otherwise.}} 

\subsection{Preview of the model and results}

Although we let agents largely ignore the nature and extent of network
externalities, we rely on the following minimal \emph{maintained assumption}%
: each agent knows how her payoff (utility) and information feedback depend
on her action and on a \textbf{payoff state}, which in turn depends on
neighbors' actions in the given network (but the agent may ignore the latter
dependence). With this, each agent best responds to her conjecture about the
payoff state, observes her realized payoff, and---in equilibrium---her
conjecture must be consistent with the feedback received, that is, \textbf{%
confirmed}. Note that conjectures may be confirmed without being correct. A
profile of actions and conjectures satisfying these requirements forms a 
\textbf{selfconfirming equilibrium}, whereby agents best respond to
conjectures that can be wrong, but are nonetheless believed to be true, as
they are consistent with the available evidence. 

In our analysis, we assume that \emph{agents observe only their realized
payoff}. Given the assumed properties of the payoff functions, it follows
that there exists a discontinuity in what agents learn from their feedback
about their neighborhood depending on whether they are active (choosing a
strictly positive action) or inactive (choosing a null action). In
particular, if externalities are only local (i.e.,~positive or negative peer
effects), active players are always able to exactly infer from the feedback
the realized payoff state, even if they may have a wrong conjecture about
how many neighbors they have or what their neighbors choose. Indeed, we say
that in a selfconfirming equilibrium \emph{active} agents have correct 
\textbf{shallow} conjectures about the payoff state, but possibly wrong%
\textbf{\ deep} conjectures about the parameters and the actions of others.
Actually, agents may even be unaware that the payoff state is determined by
others within an interactive network structure; in this case, they do not
hold deep conjectures. Given that network games without global externalities
are easier to analyze and relevent in their own right, we first study this
special case and then extend the analysis to games with both local and
global externalities.

Absent global externalities, an \emph{inactive} agent receives uninformative
feedback. If---given her conjecture---she finds it subjectively optimal to
be inactive, such lack of information about the payoff state creates an
\textquotedblleft inactivity trap\textquotedblright , allowing her possibly
wrong conjecture to persist. This has important consequences for
selfconfirming equilibrium action profiles. If being inactive is
dominated---e.g., because local externalities are positive and this is
known---, then Nash and selfconfirming equilibrium action profiles coincide.
However, if there are agents for whom being inactive is not
dominated---e.g., due to some negative local externalities---, then any
subset of this set of agents may be inactive in some selfconfirming
equilibrium. In this case inactivity is a best reply to confirmed, but
possibly false conjectures. Specifically, under the assumption that
externalities are only local, we characterize selfconfirming equilibrium
action profiles as Nash equilibrium profiles of fictitious reduced games
where inactive players are absent, augmented by the null actions of the
inactive players. We also discuss how the structure of the network adjacency
matrix (which may be unknown to the players) determines the existence of
such equilibria.\newline
%

We then study \textquotedblleft conjectural best-reply
paths\textquotedblright \ whereby each agent best responds to a shallow
conjecture that coincides with the payoff state of the previous period, if
it was revealed, or with the confirmed conjecture of the previous period, if
the payoff state was not revealed. It follows that the set of inactive
agents can only increase, because once an agent becomes inactive she gets
uninformative feedback and the conjecture to which she is best replying
persists. If such a process converges, the limit must be a selfconfirming
equilibrium. Conversely, every selfconfirming equilibrium
is---trivially---the limit of a constant conjectural best-reply path. More
interestingly, we provide conditions on the adjacency matrix for convergence
and stability of such paths. Again, what we find is the possibility of
\textquotedblleft inactivity traps.\textquotedblright \ Consider the case of
online social networks. If an agent experiences a negative payoff because
some of her followers whose externalities toward her are negative played 
\emph{high} actions (hence, giving negative feedback online), then she may
choose to abstain from interacting. Later, platform conditions may improve,
making it objectively profitable to be active, but the now inactive agent
cannot observe it.\footnote{%
Actually, for the application to online social networks, such inactivity
trap\ seems to be perceived by the platforms, to the point that many of
them, after some period of inactivity of agents, start sending emails about
what is happening on the online social network to provide a positive signal
and make agents more prone to be active again.} \newline

Models of games on networks have mainly focused on the impact of local
externalities, since global ones just change welfare without affecting the
best-reply functions. However, when agents observe only their realized
payoff, the presence of global externalities may impact the way in which
conjectures are confirmed or revised. Recall that in our setting \textit{a
game is not solely characterized by the best-reply functions, but also by
the structure of the payoff/feedback functions}. This implies that
additional selfconfirming equilibrium (SCE) action profiles are possible
compared to the case with only local externalities. Indeed, we show that the
SCE action profiles studied for the latter special case correspond to the
equilibria of games with local \textit{and} global externalities, in which
agents have correct conjectures about the global aggregate. But there are
other SCEs in which conjectures about global aggregates are wrong. For the
sake of simplicity, we focus on the case of \textit{positive} local and
global externalities, in which being inactive is dominated. Even in this
simple case, agents may have a continuum of confirmed conjectures about the
relative size of the two externalities. Indeed, there are multiple SCEs
because, even if they are active, players may have false but confirmed
conjectures making them choose actions that are not objective best replies.
In detail, we find that active agents are not able to perfectly infer the
size of the local externality due to the confound induced by the global
externality: the realized payoff, a one-dimensional feedback, does not allow
to retrieve a two-dimensional (local-global) externality. In particular,
since we assume positive externalities, we show that agents' perception of
their role in the network determines whether in a selfconfirming equilibrium
they are more or less active than predicted by a Nash equilibrium. Thus,
overall activity and (possibly) welfare are higher if agents think that
(externalities are positive and that) they are more linked than in reality.
If we consider the example of online social networks, this may help explain
why firms always try to send to their users messages to make them believe
that they are very connected, so as to increase their level of activity. 
{When considering a network of investing
firms, we may have under(over)--investment with respect to what would be the
optimal, as firms may under(over)--estimate what their neighbors do, without
being corrected by the feedback they receive.} 
{Even though
this equilibria multiplicity can shed light on some interesting phenomena of
games with global externalities, we also find an interesting relationship
linking the equilibrium action profiles of games with only local
externalities and also global ones. In details, the SCE action profiles of a
game with only local externalities selects the SCE action profiles of the
corresponding game with also global externalities in which conjectures about
global externalities are correct.}

\bigskip

The paper is structured as follows. In Section \ref{sec:literature} we
discuss the related literature. Section \ref{sec:basic} presents the basic
framework and equilibrium concept. In Section \ref{sec:local} we analyze
network games with only local externalities, whereas in 
Section \ref{sec:local_global} we analyze a more general model that accounts
for global externalities. Section \ref{sec:conclusion} concludes. 

We devote appendices to proofs and technical results. Appendix \ref{app:SCE}
analyzes properties of feedback and selfconfirming equilibria in a class of
games including as a special case the linear--quadratic network games that
we consider in the main text. 
{In Appendix \ref{ration} we study how
equilibria are affected when when the network (or some aspects of it) is
commonly known and players are strategically sophisticated.} Appendix \ref%
{app:interior} reports existing and novel results in linear algebra, that we
use to find sufficient conditions for unique and interior Nash equilibria in
network games. 
Appendix \ref{app:proofs} contains the proofs of the results presented in
the main text.

\section{Related literature}

\label{sec:literature}

We model interactions through \emph{linear--quadratic network games}. We
focus on this class of games because it has well-known properties, and it
has been used for modelling a variety of environments where strategic
interaction is local and can be described by a network structure, as
surveyed by \cite{zenou2016handbook} and \cite{bramoullegames}. Moreover,
these games belong to the larger class of \emph{nice games} %
\citep{moulin1984dominance}, for which we provide in Appendix \ref{app:SCE}
some general results. \cite{bramoulle2014strategic} show that other payoff
functions lead to the same best-reply functions, hence, to the same Nash
equilibria of linear--quadratic network games. However, we focus on \emph{%
selfconfirming equilibria} (SCE), and, since realized payoffs affect
feedback, the entire payoff function is relevant, not just the corresponding
best-reply function. Thus, we rely in our analysis on the specific original
payoff function of network games, as introduced in the economic literature
by \cite{ballester2006s}.

\bigskip

We call \textquotedblleft selfconfirming equilibria\textquotedblright \ the
steady states of learning processes when static or dynamic games are played
recurrently, independently of the specific assumptions about feedback
(monitoring) at the end of each one-period play (see also \citealp{BGM92rie}%
). This concept encompasses what used to be called \textquotedblleft
conjectural equilibrium\textquotedblright \ as well as the original
\textquotedblleft selfconfirming equilibrium\textquotedblright \ of \cite%
{fudenberg1993self}. In an SCE, agents best respond to confirmed conjectures
that may be inconsistent with sophisticated strategic reasoning. The latter
has been added to SCE relating it to rationalizability. See Section IV of 
\cite{battigalli2015self} and the relevant references therein for a more
detailed discussion of different versions of these concepts. Here we focus
on SCE, while we analyze SCE with rationalizable conjectures in Appendix 
\ref{ration}. 
\cite{lipnowski2017peer} apply a concept akin to
rationalizable SCE of games where feedback about the behavior of others is
described by a network topology: agents have correct conjectures about the
strategies of their peers (neighbors), but their payoff may depend on the
whole strategy profile and it is not observed ex post.\footnote{%
We interpret the recent model of \cite{bochet2020perceived} as another
interesting application of the SCE concept to a network game where agents
observe, besides their realized payoff, the behavior of their neighbors. In
this game agents play a Tullock contest with incomplete information about
the structure of externalities. We note that the equilibrium is, actually, a
refinement of SCE whereby agents wrongly believe that they compete for a
local rather than a global resource.} 
We instead assume that agents observe (only) their realized payoff and that
the network describes how the payoff of each agent is affected by the
actions of other her neighbors (with global externalities, there is also an
influence of other players on own payoffs not mediated by the network
structure).

\cite{mcbride2006imperfect} applies SCE 
to games of network formation with asymmetric information. In his model,
agents observe (only) the private information of other agents they link to,
and possibly of agents to whom they are indirectly linked. We instead assume
that the network is exogenous and actions are activity levels. We allow for
information incompleteness, but---with the partial exception of Section \ref%
{sec:local_global}---we do not assume that agents are necessarily aware of
the states of nature (e.g., the possible network structures), hence we do
not assume that agents necessarily reason about them.\footnote{\cite%
{de2015network} consider network formation games where players do not know
the externalities in the network, which are random, but their analysis
concerns Bayesian-Nash equilibria, and players have correct ex--ante beliefs.%
} \cite{frick2018dispersed} apply a refinement of rationalizable SCE to
analyze a model with asymmetric information and assortative matching. The
refinement is obtained by assuming that agents neglect the assortativity of
matching when they make inferences from feedback. 
\cite{foerster2018shadow} share elements of \cite{lipnowski2017peer} and of 
\cite{mcbride2006imperfect}. As in the former, agents observe the behavior
of those with whom they are linked; furthermore, they also observe public
links. As in the latter, theirs is a model of network formation. They assume
that beliefs satisfy a kind of rationalizable SCE condition. Unlike those
papers, however, \cite{foerster2018shadow} do not explicitly analyze the
equilibria of a non-cooperative game, but rather adopt a reduced-form notion
of stability akin to \cite{jackson1996strategic}. 

\section{Framework}

\label{sec:basic}

\subsection{Network games}

Consider a finite set of agents (or players) $I$, with cardinality $n=|I|$
and generic element $i$. Agents are located in a network $\mathbf{Z}\in 
\mathcal{Z}\subset \mathbb{R}^{I\times I}$, where $\mathcal{Z}$ is the \emph{%
compact} set of all possible weighted networks, here expressed as adjacency
matrices. Each agent $i\in I$ chooses an action $a_{i}$ from a \emph{compact}
\emph{interval} $A_{i}=[0,\bar{a}_{i}]$.\footnote{%
Note that in the network literature it is common to assume $A_{i}=\mathbb{R}%
_{+}$. For the case of local externalities with complementarities, we
consider constraints on the parameters so that assuming an upper bound on
actions is without loss of generality for the analysis of Nash equilibria
and of selfconfirming equilibria without global externalities. When
externalities are global the upper bound may become binding, and we discuss
this issue below in the paper.} For each $i\in I$, $\mathbf{A}_{-i}:=\times
_{j\neq i}A_{j}$ denotes the set of feasible action profiles $\mathbf{a}%
_{-i}=\left( a_{j}\right) _{j\in I\backslash \{i\}}$ for players different
from $i$. For each $i\in I$, we posit two \emph{compact} \emph{intervals} $%
X_{i}:=[\underline{x}_{i},\bar{x}_{i}]\subset \mathbb{R}$ and $Y_{i}:=[0,%
\bar{y}_{i}]\subset \mathbb{R}_{+}${\small \ }of \textbf{payoff states for }$%
i$, with the interpretation that $i$'s payoff is determined by her action $%
a_{i}$, the interaction between $a_{i}$ and state $x_{i}$, and the additive
term $y_{i}$ according to the quadratic utility function%
\begin{equation}
\begin{tabular}{llll}
$v_{i}:$ & $A_{i}\times X_{i}\times Y_{i}$ & $\rightarrow $ & $\mathbb{R}$,
\\ 
& $\left( a_{i},x_{i},y_i\right) $ & $\mapsto $ & $\alpha _{i}a_{i}-\frac{1}{%
2}a_{i}^{2}+a_{i}x_{i}+y_{i}$.%
\end{tabular}
\label{eq:LQ}
\end{equation}%
%
%
%
%
%

Payoff state $x_{i}$ is determined by the actions of $i$'s neighbors---the
agents with non-zero weight in adjacency matrix $\mathbf{Z}$---according to
the parameterized linear \textbf{aggregator}\footnote{%
In principle we can allow for non--linear aggregators, as in \cite%
{feri2018effect}. However, in this paper, we focus on the linear case. In
Appendix \ref{app:SCE} we provide results for the non-linear case.} 
\begin{equation}
\begin{tabular}{llll}
$\ell _{i}:$ & $\mathbf{A}_{-i}\times \mathcal{Z}$ & $\rightarrow $ & $X_{i}$%
, \\ 
& $\  \left( \mathbf{a}_{-i},\mathbf{Z}\right) $ & $\mapsto $ & $\sum_{j\neq
i}z_{ij}a_{j}$.%
\end{tabular}
\label{eq:L aggr}
\end{equation}%
Since the codomain of $\ell _{i}$ is $[\underline{x}_{i},\bar{x}_{i}]$, we
are effectively assuming that%
\begin{equation*}
\underline{x}_{i}\leq \sum_{j\in N_{i}^{-}}z_{ij}\bar{a}_{j}\text{, }\bar{x}%
_{i}\geq \sum_{j\in N_{i}^{+}}z_{ij}\bar{a}_{j}
\end{equation*}%
for every $\mathbf{Z}\in \mathcal{Z}$, where $N_{i}^{-}:=\left \{ j\in
I:z_{ij}<0\right \} $ denotes the set of neighbors\ of player $i$ that have
a negative effect on the payoff state of $i$, and $N_{i}^{+}:=\left \{ j\in
I:z_{ij}>0\right \} $ denotes the set of neighbors\ of player $i$ that have
a positive effect on the payoff state of $i$.

We also posit a \emph{compact }set $\mathcal{G}\subset \mathbb{R}_{+}$%
{\small \ }of nonnegative global externality parameter values. Payoff state $%
y_{i}$ is a non-strategic global externality determined by all the
co-players' actions according to the proportional aggregator:%
\begin{equation}
\begin{tabular}{llll}
$g_{i}:$ & $\mathbf{A}_{-i}\times \mathcal{G}$ & $\rightarrow $ & $Y_{i}$ \\ 
& $\  \  \ (\mathbf{a}_{-i},\gamma )$ & $\mapsto $ & $\gamma \displaystyle%
\sum_{j\neq i}a_{j}$%
\end{tabular}%
.  \label{eq:G aggr}
\end{equation}%
Since the codomain of $g_{i}$ is $[0,\bar{y}_{i}]$, we are assuming that%
\begin{equation*}
\max \mathcal{G}\displaystyle \sum_{j\neq i}\bar{a}_{j}\leq \bar{y}_{i}
\end{equation*}%
The special case of \textbf{no global externalities} obtains if $\bar{y}%
_{i}=0$.

With this, we derive the \textbf{parameterized payoff function }%
\begin{equation*}
\begin{tabular}{llll}
$u_{i}:$ & $A_{i}\times \mathbf{A}_{-i}\times \mathcal{Z\times B}$ & $%
\rightarrow $ & $\mathbb{R}$, \\ 
& $\  \  \  \left( a_{i},\mathbf{a}_{-i},\mathbf{Z},\gamma \right) $ & $\mapsto 
$ & $v_{i}\left( a_{i},\ell _{i}\left( \mathbf{a}_{-i},\mathbf{Z}\right)
,g_{i}\left( \mathbf{a}_{-i},\gamma \right) \right) $.%
\end{tabular}%
\end{equation*}%
Since $y_{i}$ does not interact with $a_{i}$, $x_{i}=\ell _{i}\left( \mathbf{%
a}_{-i},\mathbf{Z}\right) $ is the payoff-relevant state that $i$ has to
guess in order to choose a subjectively optimal action. We let 
\begin{equation}
r_{i}\left( x_{i}\right) =\left \{ 
\begin{tabular}{ll}
$0$, & if $x_{i}\leq -\alpha _{i}$, \\ 
$\alpha _{i}+x_{i}$, & if $-\alpha _{i}<x_{i}<\bar{a}_{i}-\alpha _{i}$, \\ 
$\bar{a}_{i}$, & if $x_{i}\geq \bar{a}_{i}-\alpha _{i}$.%
\end{tabular}%
\right.  \label{eq:BR to x if LQ}
\end{equation}%
denote the continuous and piecewise linear \textbf{best-reply function} of
player $i\in I$. Note that, since $\alpha _{i}>0$, we may have $r_{i}\left(
x_{i}\right) =0$ only if $\underline{x}_{i}<0$.

We assume that \emph{the game is repeatedly played by agents maximizing
their instantaneous payoff}. Each agent $i$ knows her utility function $%
v_{i}:A_{i}\times X_{i}\times Y_{i}\rightarrow \mathbb{R}$ as specified in
eq. (\ref{eq:LQ}), hence also its domain $A_{i}\times X_{i}\times Y_{i}=[0,%
\bar{a}_{i}]\times \lbrack \underline{x}_{i},\bar{x}_{i}]\times \left[ 0,%
\bar{y}_{i}\right] $ and the \textquotedblleft stand-alone\textquotedblright
\ parameter $\alpha _{i}$, but we do not assume that the aggregators
parameters $\left( \mathbf{Z},\gamma \right) $ are known. Actually, for most
of our analysis it does not even matter that agents understand that payoff
states aggregate the actions of others according to eq.s (\ref{eq:L aggr})
and (\ref{eq:G aggr}). After each play, agents get an imperfect feedback
about the payoff states. Specifically, we assume that \emph{each agent
observes only her realized utility/payoff}. What agent $i$ learns in a given
period after choosing action $a_{i}$ and observing her realized payoff $\hat{%
v}_{i}$ is that $\left( x_{i},y_{i}\right) \in \left \{ \left( x_{i}^{\prime
},y_{i}^{\prime }\right) :v_{i}\left( a_{i},x_{i}^{\prime },y_{i}^{\prime
}\right) =\hat{v}_{i}\right \} $, that is,%
\begin{equation*}
\left( x_{i},y_{i}\right) \in \left \{ 
\begin{tabular}{ll}
$\left \{ \left( x_{i}^{\prime },y_{i}^{\prime }\right) :y_{i}^{\prime }=%
\hat{v}_{i}\right \} $, & if $a_{i}=0$, \\ 
$\left \{ \left( x_{i}^{\prime },y_{i}^{\prime }\right) :\alpha _{i}a_{i}-%
\frac{1}{2}a_{i}^{2}+a_{i}x_{i}^{\prime }+y_{i}^{\prime }=\hat{v}_{i}\right
\} $, & if $a_{i}>0$.%
\end{tabular}%
\right.
\end{equation*}%
In words, if $i$ is inactive she can infer $y_{i}$ but has no clue about $%
x_{i}$, if she is active she obtains joint information about $y_{i}$ and $%
x_{i}$ that she cannot disentangle.

If there are \emph{no global externalities}, that is, if $\bar{y}_{i}=0$,
then being inactive reveals nothing, because $v_{i}\left( 0,x_{i}\right) =0$
independently of $x_{i}$, while being active reveals that%
\begin{equation*}
x_{i}=\frac{\hat{v}_{i}-\alpha _{i}a_{i}+\frac{1}{2}a_{i}^{2}}{a_{i}}=\frac{%
\hat{v}_{i}}{a_{i}}-\alpha _{i}+\frac{1}{2}a_{i}\text{.}
\end{equation*}%
With this assumption about feedback, the interactive situation is
represented by the mathematical structure%
\begin{equation*}
NG=\left \langle I,\mathcal{Z},\mathcal{G},\left(
A_{i},X_{i},Y_{i},v_{i},\ell _{i},g_{i}\right) _{i\in I}\right \rangle \text{%
,}
\end{equation*}%
determined by eq.s (\ref{eq:LQ}), (\ref{eq:L aggr}), and (\ref{eq:G aggr}),
which we call (parameterized) \textbf{linear-quadratic network game with
just observable payoffs, }or simply \textbf{network game.} This structure is
summarized in equation \eqref{eq:global_intro}.

To choose an action, a subjectively rational agent $i$ must have some
deterministic or probabilistic conjecture about the payoff state $x_{i}$.
Yet, her post-feedback update about $x_{i}$ depends on what she thinks about 
$y_{i}$, because she gets imperfect joint feedback about both. Therefore, we
model how $i$ forms conjectures about $x_{i}$ and $y_{i}$. We refer to
conjectures about the states $x_{i}$ and $y_{i}$ as \textbf{shallow
conjectures}, as opposed to\textbf{\ deep conjectures}, which concern the
specific network topology $\mathbf{Z}$, the global externality parameter $%
\gamma $ (when present), and the actions of other players $\mathbf{a}_{-i}$.
In our equilibrium analysis, given the continuity of the best-reply function
and the connectedness of $X_{i}$ and $Y_{i}$, it is sufficient to focus on 
\emph{deterministic shallow conjectures.} Indeed, for each $i\in I$ and
every probabilistic conjecture $\mu _{i}\in \Delta \left( X_{i}\times
Y_{i}\right) $, there exists a corresponding deterministic conjecture $%
\left( \hat{x}_{i},\hat{y}_{i}\right) \in X_{i}\times Y_{i}$ that justifies
the same action $a_{i}^{\ast }$ as the unique best reply.\footnote{%
See the analysis in Appendix \ref{app:SCEconj}} Deep conjectures are
relevant for the analysis of strategic thinking based on common belief in
rationality {(see Appendix \ref{ration})}, but our
equilibrium concept does not rely on strategic thinking.

\subsection{Selfconfirming equilibrium}

We analyze a notion of equilibrium that characterizes the steady states of
learning dynamics and therefore relaxes the mutual-best-reply condition of
the Nash equilibrium concept. Recall that our approach allows for the
possibility of agents being unaware of many aspects of the game. In
equilibrium, agents best respond to (deterministic)\footnote{%
{Without essential loss  of generality.}} shallow conjectures
consistent with the feedback that they receive given the true parameter
values $\left( \mathbf{Z},\gamma \right) $. 

\begin{definition}
\label{Def: SCE}A profile $\left( a_{i}^{\ast },\hat{x}_{i},\hat{y}%
_{i}\right) _{i\in I}\in \times _{i\in I}\left( A_{i}\times X_{i}\times
Y_{i}\right) $ of actions and (shallow) deterministic conjectures is a 
\textbf{selfconfirming equilibrium (SCE) at} $\left( \mathbf{Z},\gamma
\right) $ if, for each $i\in I$,

\begin{enumerate}
\item \emph{(subjective rationality)} $a_{i}^{\ast }=r_{i}\left( \hat{x}%
_{i}\right) $,

\item \emph{(confirmed conjecture)} $v_{i}\left( a_{i}^{\ast },\hat{x}_{i},%
\hat{y}_{i}\right) =v_{i}\left( a_{i}^{\ast },\ell _{i}\left( \mathbf{a}%
_{-i}^{\ast },\mathbf{Z}\right) ,g_{i}\left( \mathbf{a}_{-i}^{\ast },\gamma
\right) \right) $.
\end{enumerate}
\end{definition}

The two conditions require that: 1) each agent best responds to her own
conjecture; 2) the conjecture in equilibrium must belong to the ex post
information set, so that the expected payoff (feedback) coincides with the
realized payoff (feedback) given $a_{i}^{\ast }$, $x_{i}=\ell _{i}\left( 
\mathbf{a}_{-i}^{\ast },\mathbf{Z}\right) $, and $y_{i}=g_{i}\left( \mathbf{a%
}_{-i}^{\ast },\gamma \right) $. We say that $\mathbf{a}^{\ast }=\left(
a_{i}^{\ast }\right) _{i\in I}$ is a \textbf{\ selfconfirming action profile}
at $\left( \mathbf{Z},\gamma \right) $ if there exists a corresponding
profile of conjectures $\left( \hat{x}_{i},\hat{y}_{i}\right) _{i\in I}$
such that $\left( a_{i}^{\ast },\hat{x}_{i},\hat{y}_{i}\right) _{i\in I}$ is
a selfconfirming equilibrium at $\left( \mathbf{Z},\gamma \right) $, and we
let $\mathbf{A}_{\mathbf{Z},\gamma }^{SCE}$ denote the set of such action
profiles; in the special case of no global externalities, we write $\mathbf{A%
}_{\mathbf{Z}}^{SCE}$ to ease notation. Also, for any $\mathbf{Z}\in 
\mathcal{Z}$, we denote by $\mathbf{A}_{\mathbf{Z}}^{NE}$ the set of (pure)
Nash equilibria of the game determined by $\mathbf{Z}$ neglecting the
non-strategic global externalities, that is,%
\begin{equation*}
\mathbf{A}_{\mathbf{Z}}^{NE}:=\left \{ \mathbf{a}^{\ast }\in \times _{i\in
I}A_{i}:\forall i\in I,a_{i}^{\ast }=r_{i}\left( \ell _{i}\left( \mathbf{a}%
_{-i}^{\ast },\mathbf{Z}\right) \right) \right \} \text{.}
\end{equation*}%
%
%
%
%
%
%
%
%
%
%
%
%
%
%
%
%
%
%
%
%
%
%
%
%
%
%
%
%
%
%
%
%
%
%
%
%
%
%
%
%
%
Since, for each $\mathbf{Z}$, the joint best-reply function $\mathbf{a}%
^{\ast }\mapsto \left( r_{i}\left( \ell _{i}\left( \mathbf{a}_{-i}^{\ast },%
\mathbf{Z}\right) \right) \right) _{i\in I}$ is a continuous self-map on the
compact and convex subset $\times _{i\in I}\left[ 0,\bar{a}_{i}\right]
\subseteq \mathbb{R}^{I}$, Brower Fixed Point Theorem implies that a Nash
equilibrium exists. Hence, we obtain the existence of selfconfirming
equilibria for each $\left( \mathbf{Z},\gamma \right) \in \mathcal{Z\times B}
$. Indeed, a Nash equilibrium $\mathbf{a}^{\ast }$ corresponds to a
selfconfirming equilibrium with correct conjectures $\left( a_{i}^{\ast },%
\hat{x}_{i},\hat{y}_{i}\right) _{i\in I}=\left( a_{i}^{\ast },\ell
_{i}\left( \mathbf{a}_{-i}^{\ast },\mathbf{Z}\right) ,g_{i}\left( \mathbf{a}%
_{-i}^{\ast },\gamma \right) \right) _{i\in I}$. To summarize:

\begin{remark}
\label{Rem: NE is SCE}For every $\mathbf{Z}\in \mathcal{Z}$ and $\gamma \in 
\mathcal{G}$, there is at least one Nash equilibrium at $\mathbf{Z}$, and
every Nash equilibrium at $\mathbf{Z}$ is a selfconfirming action profile at 
$\left( \mathbf{Z},\gamma \right) $:%
\begin{equation*}
\forall \  \left( \mathbf{Z},\gamma \right) \in \mathcal{Z}\times \mathcal{G}%
\text{, }\emptyset \neq \mathbf{A}_{\mathbf{Z}}^{NE}\subseteq \mathbf{A}_{%
\mathbf{Z},\gamma }^{SCE}.
\end{equation*}
\end{remark}

In the next sections we study selfconfirming equilibria and learning, first
when there are only local externalities, and then when also global
externalities are considered.

\section{Local externalities}

\label{sec:local}

In this section, we analyze the set of selfconfirming equilibria and the
learning paths in linear-quadratic network games with just observable
payoffs and \emph{without global externalities}. Several proofs are derived
from the results in Appendix \ref{app:SCE}, which refers to the case of
generic network games with feedback, and from the results in Appendix \ref%
{app:interior}. The proofs themselves are collected in Appendix \ref%
{app:proofs}. In subsection \ref{Subsec:SCEstructure} we relate
selfconfirming equilibria to the Nash equilibria of auxiliary reduced games
and we classify equilibria according to the set of active agents. In
subsection \ref{Subsec:RelUniqueness} we provide properties of $\mathbf{Z}$
that imply uniqueness of active agents' equilibrium actions. In subsection %
\ref{sec:learning} we analyze learning paths.

\subsection{Nash equilibrium and structure of the SCE set\label%
{Subsec:SCEstructure}}

Let $I_{0}$ denote the \textbf{set of players for whom being inactive is
justifiable }(that is, undominated):\footnote{%
This definition is motivated by Lemma \ref{Lemma:nice BR} in Appendix \ref%
{app:SCE}, 
in which we analyze also the more general case of probabilistic conjectures
and we explain why restricting attention to deterministic conjectures is
without loss of generality.} 
\begin{equation*}
I_{0}:=\left \{ i\in I:\exists ~x_{i}\in X_{i},\ r_{i}\left( x_{i}\right)
=0\right \} =\left \{ i\in I:\alpha _{i}+\underline{x}_{i}\leq 0\right \} 
\text{.}
\end{equation*}%
Also, for each $\mathbf{Z}\in \mathcal{Z}$ and non-empty subset of players $%
J\subseteq I$, let $\mathbf{A}_{J,\mathbf{Z}}^{NE}$ denote the set of Nash
equilibria of the auxiliary game with player set $J$ obtained by imposing $%
a_{i}=0$ for each $i\in I\backslash J$, that is,%
\begin{equation*}
\mathbf{A}_{J,\mathbf{Z}}^{NE}=\left \{ \mathbf{a}_{J}^{\ast }\in \times
_{j\in J}A_{j}:\forall j\in J,a_{j}^{\ast }=r_{j}\left( \ell _{j}\left( 
\mathbf{a}_{J\backslash \{j\}}^{\ast },\mathbf{0}_{I\backslash J},\mathbf{Z}%
\right) \right) \right \} \text{,}
\end{equation*}%
where $\mathbf{0}_{I\backslash J}\in \mathbb{R}^{I\backslash J}$ is the
profile that assigns $0$ to each $i\in I\backslash J$. If $J=\emptyset $,
let $\mathbf{A}_{J,\mathbf{Z}}^{NE}=\left \{ \varnothing \right \} $ by
convention, where $\varnothing $ is the pseudo-action profile such that $%
\left( \varnothing ,\mathbf{0}_{I}\right) =\mathbf{0}_{I}$.\footnote{%
As we do in set theory with the empty set, when we consider functions whose
domain is a subset $J$ of some index set $I$, it is convenient to have a
symbol for the pseudo-function with empty domain. For example, if $%
J\subseteq I=\mathbb{N}$, such functions are (finite and countably infinite)
sequences and $\varnothing $ denotes the empty sequence.} We relate the set
of selfconfirming equilibria to the sets of Nash equilibria of such
auxiliary games.

\begin{proposition}
\label{Prop: SCE if LQ-OiffA text}In a linear-quadratic network game with
just observable payoffs, for each $\mathbf{Z}\in \mathcal{Z}$, the set of
selfconfirming action profiles is%
\begin{equation*}
\mathbf{A}_{\mathbf{Z}}^{SCE}=\bigcup_{J:I\backslash J\subseteq I_{0}}%
\mathbf{A}_{J,\mathbf{Z}}^{NE}\times \left \{ \mathbf{0}_{I\backslash
J}\right \} \text{,}
\end{equation*}%
that is, in each selfconfirming action profile $\mathbf{a}^{\ast }$, a
subset $I\backslash J$ of players for whom being inactive is justifiable
choose $0$, and every other player chooses the best reply to the actions of
her co-players. Therefore, in each selfconfirming action profile $\mathbf{a}%
^{\ast }$ and for each player $i\in I$, 
\begin{eqnarray}
a_{i}^{\ast } &=&0\Rightarrow \underline{x}_{i}\leq -\alpha _{i}\text{,} 
\notag \\
a_{i}^{\ast } &>&0\Rightarrow \left( \alpha _{i}+\sum_{j\in
I}z_{ij}a_{j}^{\ast }>0\text{ }\wedge a_{i}^{\ast }=\min \left \{ \bar{a}%
_{i},\alpha _{i}+\sum_{j\in I}z_{ij}a_{j}^{\ast }\right \} \right) \text{.}
\label{eq:network_payoffs}
\end{eqnarray}
\end{proposition}

Suppose for simplicity that, in every restricted auxiliary game with player
set $J$, Nash equilibrium actions are strictly positive (Proposition \ref%
{prop:interiorSCE} below provides sufficient conditions). Then in every SCE
we can partition the set of agents in two subsets. Agents in $J\subseteq I$
are active, 
while agents in $I\setminus J$ choose the null action. Start by considering
the latter group of agents. They must belong to the set of agents for whom
inactivity is justifiable; as such, they choose $0$ as a best reply to a
possibly wrong conjecture, and get null payoff independently of others'
actions. Since every conjecture is consistent with this payoff, their
conjecture is (trivially) consistent with their feedback. As for agents in $%
J $, since they choose a strictly positive action, they receive a message
that enables them to infer the true payoff state; with this, they
necessarily choose the objective best reply to their neighbors' actions,
whether or not they are aware of them. Note that, if being inactive is
justifiable for every agent ($I_{0}=I$), then $\mathbf{0}_{I}\in \mathbf{A}_{%
\mathbf{Z}}^{SCE}$ for every $\mathbf{Z}\in \mathcal{Z}$. In the polar
opposite case, being inactive is unjustifiable for every agent ($%
I_{0}=\emptyset $) and SCE coincides with Nash equilibrium. For example,
assume that $\mathbf{Z}=w\mathbf{Z}_{0}$, with $w>0$ and that $\mathbf{Z}%
_{0}\in \{0,1\}^{I\times I}$. In this context it is natural to also assume
that $\min X_{i}\geq 0$, which implies that being inactive is unjustifiable
(recall, $\alpha _{i}>0$). This represents the standard case of local
complementarities studied by \cite{ballester2006s}. If $w\left( n-1\right)
<1 $, there is a unique Nash equilibrium which is also interior and
coincides with the unique SCE action profile.

Thus, the SCE set can be characterized by means of the Nash equilibria of
the auxiliary games in which only active agents are considered. 
If, for example, for every given set $J\subseteq I$ there is a unique Nash
equilibrium of the corresponding auxiliary game (Proposition \ref%
{prop:interiorSCE} provides sufficient conditions), then $|\mathbf{A}_{%
\mathbf{Z}}^{SCE}|=2^{|I_{0}|}$, because for each $J$ with $I\backslash
J\subseteq I_{0}$ there is exactly one SCE where the set of active agents is 
$J$. Since each auxiliary game has at least one Nash equilibrium (see Remark %
\ref{Rem: NE is SCE}), $2^{|I_{0}|}$ is a lower bound on the number of
SCE's. If we assume strategic substitutes, then the Nash equilibria for each
auxiliary game in which only agents in $J\subseteq I$ may be active, can be
characterized as in \cite{bramoulle2014strategic}. Note that in this case,
some of the agents in $J$ can be active and some inactive. Appendix \ref%
{app:SCEchar} discusses the equilibrium characterization for the general
case of non linear-quadratic network games.

\begin{example}
\label{ex:net1} {Consider Figure \ref{fig:net1}, representing a
network with 4 nodes/players. We set $\alpha _{i}=0.1$ for every $i$. First
assume that each arrow represents a positive externality of $0.2$ (and
arrows point to the source of the externality), but we allow agents to
believe that links may also be a source of negative externality. Then,
agents may find it justifiable to be inactive. In this case we have one Nash
equilibrium (NE)\footnote{%
Note that with positive externalities the unique Nash equilibrium is the
only rationalizable action profile, i.e., the only one consistent with
common knowledge of the game, rationality, and common belief in rationality.}%
, but 16 possible SCE's, one for each subset of the players that we allow to
be active.} {Table \ref{table:1} reports the actions of players
in each case (we omit redundant doubletons and singletons). Note that player 
$3$, when active, always plays the same action $a_{3}=0.1$, because she is
not affected by any externality. Other players, instead, when active, play
differently according to who else is active.}

\begin{figure}[h]
\begin{center}
\includegraphics[scale=0.5]{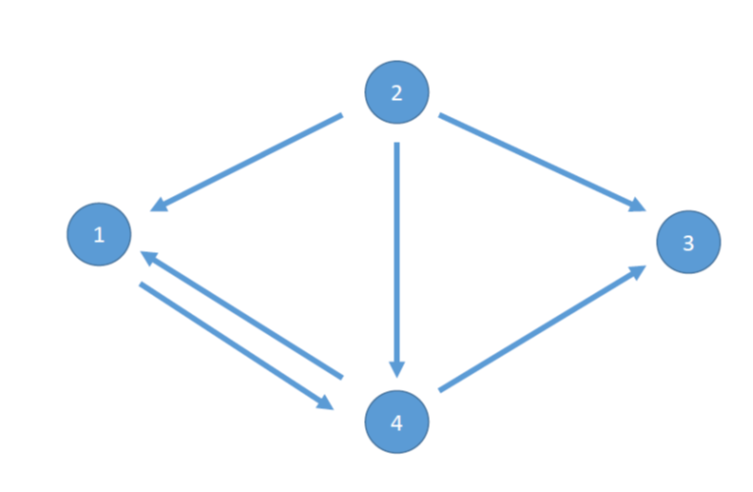} 
\end{center}
\caption{\textit{A network with 4 nodes.} Every arrow identifies an
externality of equal magnitude and sign.}
\label{fig:net1}
\end{figure}

\begin{table}[h!]
\centering
\begin{tabular}{||l|llllllllll||}
\hline
& \textbf{All} & $\{ 1,2,3\}$ & $\{ 1,2,4\}$ & $\{ 1,3, 4 \}$ & $\{ 2,3, 4
\} $ & $\{ 1,2\}$ & $\{ 1,3\}$ & $\{ 1,4\}$ & \dots & $\emptyset$ \\%
[0.5ex] \hline \hline
$a_1$ & \textbf{0.1292} & 0.1 & 0.125 & 0.1292 & 0 & 0.1 & 0.1 & 0.125 &  & 0
\\ 
$a_2$ & \textbf{0.1750} & 0.14 & 0.15 & 0 & 0.144 & 0.12 & 0 & 0 &  & 0 \\ 
$a_3$ & \textbf{0.1} & 0.1 & 0 & 0.1 & 0.1 & 0 & 0.1 & 0 &  & 0 \\ 
$a_4$ & \textbf{0.1458} & 0 & 0.125 & 0.1458 & 0.12 & 0 & 0 & 0.125 &  & 0
\\ \hline
\end{tabular}%
\caption{Selfconfirming equilibria of the network from Figure \protect \ref%
{fig:net1}, with all positive externalities of $0.2$. Columns are for the
subsets of active players. The unique Nash Equilibrium is in bold.}
\label{table:1}
\end{table}

{\noindent Consider now the same network, but assume that each
arrow represents a negative externality of $-0.6$. In this case we have more
NE's (there is no NE where all players are active, but there are 3 NE's),
but less than 16 SCE's (there are 13), because for some subset $J$ of
players (such as $J=I=\{1,2,3,4\}$) there is no SCE in which all its agents
are active. Table \ref{table:2} reports the actions of players in each case
(we omit redundant doubletons and singletons).} 
\end{example}

\begin{table}[h!]
\centering
\begin{tabular}{||l|lllllll||}
\hline
& $\mathbf{\{ 1,2,4\}}$ & $\mathbf{\{ 2,3, 4 \}}$ & $\{ 1,2\}$ & $\mathbf{\{
1,3\}}$ & $\{ 1,4\}$ & \dots & $\emptyset$ \\[0.5ex] \hline \hline
$a_1$ & \textbf{0.0625} & \textbf{0} & 0.1 & \textbf{0.1} & 0.0625 &  & 0 \\ 
$a_2$ & \textbf{0.025} & \textbf{0.016} & 0.04 & \textbf{0} & 0 &  & 0 \\ 
$a_3$ & \textbf{0} & \textbf{0.1} & 0 & \textbf{0.1} & 0 &  & 0 \\ 
$a_4$ & \textbf{0.0625} & \textbf{0.04} & 0 & \textbf{0} & 0.0625 &  & 0 \\ 
\hline
\end{tabular}%
\caption{Selfconfirming equilibria of the network from Figure \protect \ref%
{fig:net1}, with all negative externalities of $-0.6$. Columns are for the
subsets of active players. Nash Equilibria are in bold.}
\label{table:2}
\end{table}

\subsection{Relative uniqueness\label{Subsec:RelUniqueness}}

We list below some properties of the weighted adjacency matrix $\mathbf{Z}$
that will be used throughout the text but are \emph{not} maintained
assumptions.\footnote{%
That is, they appear explicitly among the hypotheses of some of the
subsequent propositions.} In what follows, we will assume some of these
properties to retrieve sufficient conditions for the existence and stability
of selfconfirming equilibria. In particular, they imply the uniqueness of
selfconfirming equilibrium actions relative to any given set $J$ of active
players. We refer to Appendix \ref{app:interior} for a deeper discussion on
these assumptions and their implications. 

\begin{assumption}
\label{ass:bound} Matrix $\mathbf{Z}$ of size $n$ has \emph{bounded values},
i.e.,~for each $i,j\in I$, $|z_{ij}|<\frac{1}{n}$.
\end{assumption}

\begin{assumption}
\label{ass:sign} Matrix $\mathbf{Z}$ has the \emph{same sign property},
i.e.,~for each $i,j\in I$, $sign(z_{ij})=sign(z_{ji})$, where the $sign$
function can have values $-1$, $0$ or $1$.\footnote{%
The sign condition is the one used in \cite{bervoets2016learning} to prove
convergence to Nash equilibria in network games, under a particular form of
learning.}
\end{assumption}

\begin{assumption}
\label{ass:negative} Matrix $\mathbf{Z}$ is \emph{negative}, i.e.,~for each $%
i,j\in I$, $z_{ij}<0$.
\end{assumption}

We recall here that the spectral radius $\rho(\mathbf{Z})$ of $\mathbf{Z}$
is the largest absolute value of its eigenvalues.

\begin{assumption}
\label{ass:limited} Matrix $\mathbf{Z}$ is \emph{limited}, i.e.,~$\rho(%
\mathbf{Z}) < 1$.
\end{assumption}


In some cases, we can write $\mathbf{Z}=\mathbf{WZ}_{0}$, where $\mathbf{W}$
is a diagonal matrix, and $\mathbf{Z}_{0}\in \{0,1\}^{I\times I}$ is the
basic underlying topology of the network. Whenever this is the case, matrix $%
\mathbf{Z}$ represents a basic network combined with an additional
idiosyncratic effect by which every agent $i$ weights the effects of others
on her. These effects are modeled by the parameter $w_{i}$.\footnote{%
Then the payoff of $i\in I$ at a given profile $\mathbf{a}$ of the original
game is%
\begin{equation*}
u_{i}\left( \mathbf{a},\mathbf{Z}\right) =\alpha _{i}a_{i}-\frac{1}{2}%
a_{i}^{2}+a_{i}w_{i}\sum_{j\in I}z_{0,ij}a_{j}=\alpha _{i}a_{i}-\frac{1}{2}%
a_{i}^{2}+a_{i}\sum_{j\in I}z_{ij}a_{j}\ .
\end{equation*}%
} The next assumption adds a symmetry condition on $\mathbf{Z}_{0}$.

\begin{assumption}
\label{ass:symmetrizable} Matrix $\mathbf{Z}$ is \emph{symmetrizable},
i.e.,~it can be written as $\mathbf{Z=W Z}_{0}$, with $\mathbf{W}$ diagonal
and $\mathbf{Z}_{0}$ symmetric. Moreover, $\mathbf{W}$ has all strictly
positive entries in the diagonal.
\end{assumption}

Note that if $\mathbf{Z}$ is symmetrizable then all its eigenvalues are
real. Moreover, since $\mathbf{W}$ has all strictly positive entries,
Assumption \ref{ass:symmetrizable} implies that the sign condition
(Assumption \ref{ass:sign}) holds. \newline
Our final assumption is discussed in \cite{bramoulle2014strategic} and
combines Assumptions \ref{ass:limited} and \ref{ass:symmetrizable} above.

\begin{assumption}
\label{ass:symmetrizable_limited} $\mathbf{Z}=\mathbf{W Z}_{0}$ is \emph{%
symmetrizable-limited}, i.e., $\mathbf{Z}$ is symmetrizable and the matrix $%
\bar{\mathbf{Z}}$, whose entries are defined, for each $i,j\in I$, as $\bar{z%
}_{ij}= z_{0,ij} \sqrt{w_i w_j}$, is limited. 
\end{assumption}

Our previous results, about the characterization of selfconfirming
equilibria, state that we can choose any subset $J\subseteq I_{0}$ of agents
and have them inactive in an SCE. However, we cannot ensure that the other
agents are active, because their best response in the reduced game could be
to stay inactive, since the Nash equilibrium of the reduced game in which
only agents in $I\backslash J$ are considered may have both active and
inactive agents. The next result goes in the direction of specifying under
what sufficient conditions this does not happen. Given the matrix $\mathbf{Z}
$, and given $J\subseteq I$, we call $\mathbf{Z}_{J}$ the submatrix which
has only rows and columns corresponding to the elements of $J$.

\begin{proposition}
\label{prop:interiorSCE} Consider a linear-quadratic network game and a
subset of players $J\subseteq I$, such that $I\backslash J\subseteq I_{0}$
(that is, $\alpha _{i}+\underline{x}_{i}\leq 0$ for each $i\notin J\,$).
Suppose that $\mathbf{Z}_{J}$ satisfies at least one of the three conditions
below:

\begin{enumerate}
\item it has bounded values (Assumption \ref{ass:bound});

\item it is negative and limited (Assumptions \ref{ass:negative} and \ref%
{ass:limited});

\item it is symmetrizable--limited (Assumption \ref%
{ass:symmetrizable_limited}).
\end{enumerate}

Then, we have the following two results:

\begin{itemize}
\item the auxiliary game with player set $J$ has a unique and strictly
positive Nash equilibrium: $\mathbf{A}_{\mathbf{Z}_{J}}^{NE}=\left \{ 
\mathbf{a}_{J}^{NE}\right \} $ with ${a}_{j}^{NE}>0$ for all $j\in J$;

\item $(\mathbf{a}_{J}^{NE},\mathbf{0}_{I\backslash J})$ is a selfconfirming
equilibrium at $\mathbf{Z}$.
\end{itemize}
\end{proposition}


Proposition \ref{prop:interiorSCE} provides sufficient conditions to have
arbitrary sets of active and inactive players in a selfconfirming
equilibrium. In particular, if any of the three conditions is satisfied for
every subset of $I$, and if for all players being inactive is justifiable ($%
I_{0}=I$), then the set of SCE's has the same cardinality as the power set $%
2^{I}$, that is $2^{n}$. The first sufficient condition about (sub)matrix $%
\mathbf{Z}_{J}$ is novel, while the other two were obtained respectively by 
\cite{ballester2006s} and \cite{stanczak2006resource}, and by \cite%
{bramoulle2014strategic}.

{We provide here below two examples, one with all positive
externalities, the other with mixed externalities.}

\begin{example}
{Consider $n$ players, and a randomly generated network between
them, of the type $\mathbf{Z=WZ}_{0}$, obtained from the following
generating process. $\mathbf{Z}_{0}$ is undirected, generated by an \cite%
{erdos1960evolution} process for which each link is i.i.d., and such that
its expected number of overall links (i.e., counted in both directions) is $%
k\cdot n$, for some $k\in \mathbb{R}_{+}$. This means that the expected
number of links for each player is $k$. It is well known that this model
predicts, as $n$ goes to infinity, that $\mathbf{\ Z}_{0}$ will have null
clustering and, with $k\geq 2$, a connected giant component. }

$\mathbf{W}$ is a diagonal matrix, such that each element $w_{i}$ in the
diagonal is strictly positive and is generated by some i.i.d.~random process
with mean $\mu $ and variance $\sigma ^{2}$. In this case, \cite%
{furedi1981eigenvalues} prove that the expected highest eigenvalue of $%
\mathbf{Z}$, as $n$ grows, is 
\begin{equation*}
\mathbb{E}[\lambda _{1}]=k\mu +\frac{\sigma ^{2}}{\mu }+O\left( \frac{1}{%
\sqrt{n}}\right) \ .
\end{equation*}%
Under Assumption \ref{ass:symmetrizable_limited}, as $n$ tends to infinity, $%
\mathbf{Z}$ is symmetrizable--limited if $\mathbb{E}[\lambda _{1}]<1$, which
is equivalent to 
\begin{equation*}
\frac{\mu -\sigma ^{2}}{\mu ^{2}}>k\ .
\end{equation*}%
Clearly, a necessary condition for the previous inequality is that $\mu
>\sigma ^{2}$. When this is the case, as $n$ grows to infinity, there always
exists a unique NE of the game where all players are active, as stated by
Proposition \ref{prop:interiorSCE}. \newline
Note that, since the expected clustering of $\mathbf{Z}_{0}$ goes to $0$,
this limiting result excludes the possibility that there is a subset $J$ of
players forming a dense sub--network, and a high realization of $w_{i}$'s,
such that there does not exist $\mathbf{a}^{\ast }\in \mathbf{A}_{\mathbf{Z}%
}^{SCE}$, for which $\mathbf{a}^{\ast }=\left \{ \mathbf{a}%
_{J}^{NE}\right
\} \times \left \{ \mathbf{0}_{I\backslash J}\right \} 
\text{.}$ In fact, if this were the case, since there are only positive
externalities, we would not have an all-active equilibrium for the whole
population of $n$ agents. 
\end{example}

\begin{figure}[h]
\begin{center}
\includegraphics[scale=0.7]{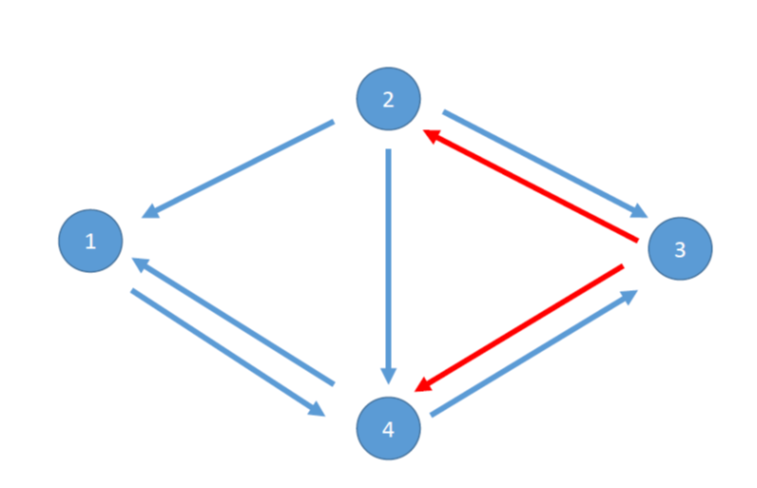}
\end{center}
\caption{\textit{A network with 4 nodes.} Blue (resp., red) arrows represent
positive (resp., negative) externalities.}
\label{fig:net2}
\end{figure}

\begin{example}
Proposition \ref{prop:interiorSCE} provides alternative \emph{sufficient}
conditions for an interior NE in the auxiliary game with player set $J$.
Figure \ref{fig:net2} provides an example of game that does not satisfy any
of them, but still has a unique interior NE. We set $\alpha _{i}=0.1$ for
each player $i$. Every blue arrow represents a positive externality of
intensity $0.2$ (so, the blue arrows represent the first case from Example %
\ref{ex:net1}). The two red arrows represent negative externalities of
intensity $-0.2$. This network game has a unique NE, and 16 SCE's. Table \ref%
{table:3} shows them all (redundant doubletons and singletons are omitted). 
\begin{table}[h]
\centering
\begin{tabular}{||l|lllllllllll||}
\hline
& \textbf{All} & $\{ 1,2,3\}$ & $\{ 1,2,4\}$ & $\{ 1,3, 4 \}$ & $\{ 2,3, 4
\} $ & $\{ 1,2\}$ & $\{ 1,3\}$ & $\{ 1,4\}$ & $\{ 2,3\}$ & \dots & $%
\emptyset $ \\[0.5ex] \hline \hline
$a_1$ & \textbf{0.1257} & 0.1 & 0.125 & 0.128 & 0 & 0.1 & 0.1 & 0.125 & 0 & 
& 0 \\ 
$a_2$ & \textbf{0.1603} & 0.1346 & 0.15 & 0 & 0.144 & 0.12 & 0 & 0 & 0.1154
&  & 0 \\ 
$a_3$ & \textbf{0.0412} & 0.731 & 0 & 0.720 & 0.1 & 0 & 0.1 & 0 & 0.0729 & 
& 0 \\ 
$a_4$ & \textbf{0.1336} & 0 & 0.125 & 0.14 & 0.12 & 0 & 0 & 0.125 & 0 &  & 0
\\ \hline
\end{tabular}%
\caption{Selfconfirming equilibria of the network from Figure \protect \ref%
{fig:net2}, with positive (resp., negative) externalities of intensity $0.2$
(resp.,$-0.2$). Columns correspond to subsets of active players. The unique
Nash Equilibrium is in bold.}
\label{table:3}
\end{table}
\end{example}

\subsection{Learning paths\label{sec:learning}}


Definition \ref{Def: SCE} of selfconfirming equilibrium and the
characterization stated in Proposition \ref{Prop: SCE if LQ-OiffA text}
identify steady states: if agents' conjectures are confirmed (not
contradicted) by the feedback they receive, these conjectures will not
change in the next interactions. However, we may wonder how agents get to
play SCE action profiles and if these profiles are stable.\footnote{%
Throughout all our analysis, players perform adaptive learning given an
exogenously fixed (but possibly unknown) network. For models in which
players adaptively change also their links, with a quadratic payoff function
analogous to ours, and the overall network evolve endogenously, see \cite%
{konig2011network} and \cite{konig2014nestedness}.}

We first point out that SCE has solid learning foundations.\footnote{%
See, for example, \cite{BGM92rie}, \cite{BFLM2016}, \cite{fk95geb}, and the
references therein.} The following result is specifically relevant for this
paper (see \citealp{Gilli99red} and Chapter 7 of \citealp{battigalli2018}).
Consider a temporal sequence (path) of action profiles $\left( \mathbf{a}%
_{t}\right) _{t=0}^{\infty }$. Then, \emph{if }$\left( \mathbf{a}_{t}\right)
_{t=0}^{\infty }$\emph{\ is consistent with adaptive learning\footnote{%
In a \emph{finite }game, a path of play $\left( \mathbf{a}_{t}\right)
_{t=0}^{\infty }$ is consistent with adaptive learning if for every $\hat{t}$%
, there exists some $T$ such that, for every $t>\hat{t}+T$ and $i\in I$, $%
a_{i,t}$ is a best reply to some \emph{deep} conjecture $\mu _{i}$ that
assigns probability $1$ to the set of action profiles $\mathbf{a}_{-i}$
consistent with the feedback received from $\hat{t}\, \ $through $t-1$. \
The definition for compact-continuous games is a bit more complex (see %
\citealp{milgrom1990rationalizability}, who assume perfect feedback).} and }$%
\mathbf{a}_{t}\rightarrow \mathbf{a}^{\ast }$\emph{, it follows that }$%
\mathbf{a}^{\ast }$\emph{\ must be a selfconfirming action profile.}


To ease the analysis we consider conjectural best-reply paths for shallow
conjectures. For each network $\mathbf{Z}$, each period $t\in \mathbb{N}$,
and each agent $i\in I$, $a_{i,t}=r_{i}\left( \hat{x}_{i,t}\right) $ is the
best reply to $\hat{x}_{i,t}$. After actions are chosen, given the feedback
received, agents update their conjectures. If conjectures are confirmed then
an agent keeps her previous conjecture, otherwise she updates it using as
new conjecture the one that would have been correct in the previous period.
Thus, 
\begin{equation}
\hat{x}_{i,t+1}=\left \{ 
\begin{array}{cc}
\hat{x}_{i,t} & \text{if }a_{i,t}=0, \\ 
\ell _{i}\left( \mathbf{a}_{-i,t},\mathbf{Z}\right) & \text{if }a_{i,t}>0,%
\end{array}%
\right.  \label{eq:learning}
\end{equation}%
and, from \eqref{eq:BR to x if LQ} we obtain

\begin{equation*}
a_{i,t+1}=r_{i}\left( \hat{x}_{i,t+1}\right) =\left \{ 
\begin{tabular}{ll}
$0$, & if $\hat{x}_{i,t}\leq -\alpha _{i}$, \\ 
$\bar{a}_i$, & if $\hat{x}_{i,t+1} \geq \bar{a}_i - \alpha_i$, \\ 
$\alpha _{i}+\hat{x}_{i,t+1}$, & otherwise.%
\end{tabular}%
\right.
\end{equation*}%
We will consider the possibility that the upper bound $\bar{a}_{i}$ is
reached only in the analysis of diverging dynamics. Given our assumptions
about feedback, \textit{being inactive is an absorbing state}: if an agent
is inactive at time $t$ she will remain so also at time $t+1$. If instead
the agent is active ($a_{i,t}>0$), feedback is such that the agent can
perfectly infer the payoff state $x_{i,t}=\ell _{i}\left( \mathbf{a}_{-i,t},%
\mathbf{Z}\right) $, and so she updates conjectures according to (\ref%
{eq:learning}), which becomes the updated conjectures. This is a conjectural
best-reply path. The result cited above implies that if the path described
above converges, then it must converge to a selfconfirming equilibrium,
i.e., a rest point where players keep repeating their choices.

In this subsection, we analyze the local stability of such rest points %
\citep[cf.][]{bramoulle2007public}.

\begin{definition}[Conjectural best-reply paths]
\label{def:learning} A sequence of profiles of actions and shallow
deterministic conjectures $(\mathbf{a}_{t},\mathbf{\hat{x}}_{t})_{t\in 
\mathbb{N}_{0}}$ is a \textbf{conjectural best-reply path} if it has the
following features:

\begin{enumerate}
\item Each player $i\in I$ starts at time $0$ with a belief, and beliefs are
represented by a profile of shallow deterministic conjectures $\mathbf{\hat{x%
}}_{0}=\left( \hat{x}_{i,0}\right) _{i\in I}$.

\item In each period \thinspace $t$, players best reply to their
conjectures: for each $i\in I$, $a_{i,t}=\max \{ \alpha _{i}+\hat{x}%
_{i,t},0\} $.

\item At the beginning of each period $t+1$, each player $i$ keeps her
period--$t$ shallow conjecture if she was inactive, and updates her
conjecture to period--$t$ revealed payoff state if she was active, that is, $%
\hat{x}_{i,t+1}=\frac{u_{i}\left( \mathbf{a}_{t}, \mathbf{Z}\right) }{a_{i,t}%
}-\alpha _{i}+\frac{1}{2}a_{i,t}$.
\end{enumerate}
\end{definition}

Observe that the system is deterministic and the initial conditions
completely determine the paths. From conditions \eqref{eq:network_payoffs}
and \eqref{eq:learning}, the system is not linear because, for each $i\in I$
and $t\in \mathbb{N}_{0}$, 
\begin{equation*}
\hat{x}_{i,t+1}=\left \{ 
\begin{array}{ccc}
\hat{x}_{i,t} & \mbox{ if } & \hat{x}_{i,t}\leq -\alpha _{i}\ , \\ 
\sum_{j\in I}z_{ij}a_{j,t} & \mbox{ if } & \hat{x}_{i,t}>-\alpha _{i}\ .%
\end{array}%
\right.
\end{equation*}

Clearly an SCE of the game is always a rest point of these learning paths.
Indeed, every SCE $\left( \mathbf{a}^{\ast },\mathbf{\hat{x}}\right) $
is---trivially---the limit of the constant conjectural best-reply path
starting at $(\mathbf{a}_{0},\mathbf{\hat{x}}_{0})=\left( \mathbf{a}^{\ast },%
\mathbf{\hat{x}}\right) $. Furthermore, the set of inactive agents in a
conjectural best-reply path can only increase: 
\begin{equation*}
I_{0}\left( \mathbf{\hat{x}}_{t}\right) \subseteq I_{0}\left( \mathbf{\hat{x}%
}_{t+1}\right) \ ,
\end{equation*}%
where $I_{0}\left( \mathbf{\hat{x}}\right) $ denotes the set of inactive
agents given profile of conjectures $\mathbf{\hat{x}=}\left( \hat{x}%
_{i}\right) _{i\in I}$.

We now consider the stability of such rest points. Say that a profile of
conjectures $\mathbf{\hat{x}}$ \textbf{justifies }action profile\textbf{\ }$%
\mathbf{a}^{\ast }$ if, for each $i\in I $, $a_{i}^{\ast }=r_{i}\left( \hat{x%
}_{i}\right) $.


\begin{definition}
\label{def:stable} A profile $\mathbf{a}^{\ast }\in \mathbf{A}_{\mathbf{Z}%
}^{SCE}$ is locally stable if there exists a profile of conjectures $\mathbf{%
\hat{x}}$ such that $(\mathbf{a}^{\ast },\mathbf{\hat{x}})$ is a
selfconfirming equilibrium, and if there exists an $\epsilon >0$ such that,
for each $\mathbf{\hat{x}}_{0}$ with $\left \Vert \mathbf{\hat{x}}_{0}-%
\mathbf{\hat{x}}\right \Vert <\epsilon $ (where $\left \Vert \cdot
\right
\Vert $ is the Euclidean norm), the conjectural best-reply path,
starting at $\mathbf{\hat{x}}_{0}$, has a limit and it is such that $%
\lim_{t\rightarrow \infty} \mathbf{a}_t=\mathbf{a^*}$.
\end{definition}


Since $(\mathbf{a}_{t},\mathbf{\hat{x}}_{t})_{t\in \mathbb{N}_{0}}$ is
determined by the initial conjectures $\mathbf{\hat{x}}_{0}$, we analyze
stability with respect to perturbations of $\mathbf{\hat{x}}_{0}$. Our
notion of stability with respect to conjectures relates to the standard
notion of stability with respect to actions in the following way. First of
all, since played actions are justified by some conjectures, the only reason
for these actions to change is a perturbation of the justifying conjectures,
but this is not a sufficient condition. If all agents are active, the two
definitions have the same consequences in terms of stability, since a
perturbation with respect to actions happens if and only if every agent's
conjecture is perturbed. Indeed, each active agent $i$ has perfect feedback
about $x_i$, and always chooses the best reply to neighbors' actions in
previous time step. However, consider an SCE with inactive agents, who
choose the null action as a corner solution, that is, whose subjective
expected marginal utility for increasing activity is strictly negative. For
such agents a small perturbation of their conjectures would not change their
null subjective best reply. This is so because inactive agents have
imperfect feedback and cannot infer the value of the local externality
aggregator. This implies that if an action profile is locally stable with
respect to action perturbations, then it is also locally stable under
conjectures perturbations, but the converse does not hold. Specifically,
forcing inactive agents to be active may lead some of them to be active
forever. The two definitions would be equivalent under perfect feedback for
all agents. Note finally that a temporary perturbation of shallow
conjectures $\mathbf{\hat{x}}_0$ has the same effect of a temporary shock in
the parameter $\mathbf{\alpha}$. By looking at the first order conditions,
they both induce the same effect on agents' best reply and on payoffs.

\bigskip

Each SCE is characterized by a set of active agents. So, given an action
profile $\mathbf{a}=\left( a_{i}\right) _{i\in I}$, let $I_{\mathbf{a}%
}:=\{i\in I:a_{i}>0\}$ denote the set of active players at profile $\mathbf{a%
}$. 
Also let $I_{0}^{\ast }:=\left \{ i\in I:\alpha _{i}+\underline{x}%
_{i}<0\right \} $ (a subset of $I_{0}$) denote the set of agents for whom
being inactive is a \textquotedblleft corner solution\textquotedblright \
for a set of conjectures with nonempty interior. 
For each action profile $\mathbf{a}$, $\mathbf{Z}_{I_{\mathbf{a}}}$ denotes
the sub--matrix with rows and columns corresponding to players who are
active in $\mathbf{a}$. The following result provides sufficient conditions
for a selfconfirming equilibrium to be locally stable.

\begin{proposition}
\label{remark:radius} The action profile in a selfconfirming equilibrium $(%
\mathbf{a}^{\ast },\mathbf{\hat{x}})$ such that $\hat{x}^i\neq \alpha_i$ for
each $i\in I$, is locally stable if

\begin{itemize}
\item Assumption \ref{ass:limited} holds for matrix $\mathbf{Z}_{I_{\mathbf{a%
}^{\ast }}}$;

\item $I\backslash I_{\mathbf{a}^{\ast }}\subseteq I_{0}^{\ast }$.
\end{itemize}
\end{proposition}

Intuitively, consider a sufficiently small perturbation of players'
conjectures. The first condition ensures that active players keep being
active and their actions converge back to the unique Nash equilibrium of the
auxiliary game with player set $I_{\mathbf{a}^{\ast }}$. The second
condition ensures that inactive players keep being inactive. Next, we
provide alternative sufficient conditions that allow to find the subsets of
active agents associated to SCE's.

\begin{proposition}
\label{prop:stableSCE} Consider the action profile $\mathbf{a}^{\ast }$ in a
selfconfirming equilibrium $(\mathbf{a}^{\ast },\mathbf{\hat{x}})$ such that 
$I\backslash I_{\mathbf{a}^{\ast }}\subseteq I_{0}^{\ast }$ and $\hat{x}%
^i\neq \alpha_i$ for each $i\in I$. If $\mathbf{Z}_{I_{\mathbf{a}^{\ast }}}$
satisfies at least one of the three conditions below:

\begin{enumerate}
\item it has bounded values (Assumption \ref{ass:bound}),

\item it is negative and limited (Assumptions \ref{ass:negative} and \ref%
{ass:limited}),

\item it is limited and symmetrizable (Assumptions \ref{ass:limited} and \ref%
{ass:symmetrizable}),
\end{enumerate}

\noindent then $\mathbf{a}^{\ast }$ is locally stable. Moreover, for every $%
J\subseteq I_{\mathbf{a}^{\ast }}$such that $I\backslash J\subseteq
I_{0}^{\ast }$, $\mathbf{a}^{\ast \ast }=(\mathbf{a}_{J}^{NE},\mathbf{0}%
_{I\backslash J})$ is a locally stable SCE action profile, where $\mathbf{a}%
_{J}^{NE}$ is the unique and strictly positive Nash equilibrium action
profile of the auxiliary game restricted to player set $J$.
\end{proposition}


%
%

%
%


The proof is based on results from linear algebra. In fact, if an adjacency
matrix satisfies one of the conditions from Proposition \ref{prop:stableSCE}%
, then also every submatrix of that matrix satisfies that property.

We know that there may be SCE's that are not Nash equilibria, because some
agents are inactive even if this is not a best response to the actions of
others. 
Proposition \ref{prop:stableSCE} provides an additional observation. Under
the stated conditions, for any given SCE action profile $\mathbf{a}^{\ast }$
with set of active agents $I_{\mathbf{a}^{\ast }}$, any subset $J\subseteq
I_{\mathbf{a}^{\ast }}$ of those agents such that $I\backslash J\subseteq
I_{0}^{\ast }$ is associated to a stable SCE where all agents in $J$ are
active, and the other agents are inactive.

\bigskip

The following example shows that we can reach SCE's that are not NE's also
if the initial beliefs induce strictly positive actions for all agents at
the beginning of the learning paths.

\begin{figure}[h]
\begin{center}
\includegraphics[height=4.3cm]{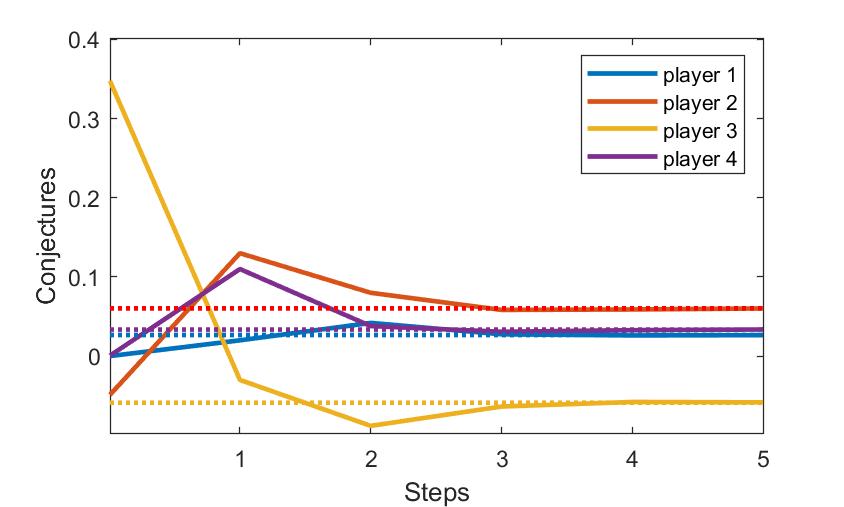} %
\includegraphics[height=4.3cm]{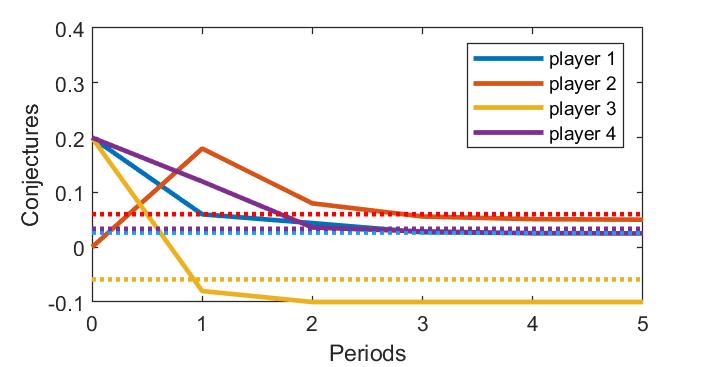}
\end{center}
\caption{\textit{Positive and negative externalities.} Starting from
different conjectures, given the same network (from Figure \protect \ref%
{fig:net2}), the learning process may converge to the unique Nash
equilibrium (left panel -- dotted lines are the Nash equilibrium) or to an
SCE which is not a Nash equilibrium (right panel). For active players,
actions are just an upward shift of conjectures of amount $\protect \alpha_i$%
. In the right panel, for the inactive player $3$ the action is $0$ from
step $2$ on. }
\label{fig:netex3}
\end{figure}

\begin{example}
Consider the case of $4$ players, with the network matrix $\mathbf{Z}\in
\{-0.2,0,0.2\}^{I\times I}$ shown in Figure \ref{fig:net2}, and, for every $%
i $, $\alpha _{i}=0.1$. This is a case of externalities that can be positive
or negative. Figure \ref{fig:netex3} shows the learning paths of actions
that start from different initial conditions. In one case (left panel) the
path converges to the unique Nash equilibrium of this game (the dotted
lines), in the other (right panel) the path makes a player inactive after
two rounds and converges to a selfconfirming equilibrium which is not Nash. 
\end{example}

{The next example (which does not satisfy the local stability
conditions of Proposition \ref{prop:stableSCE}) shows that convergence may
not occur even in a simple case of positive externalities.%
}

{\color{red} }

\begin{figure}[h]
\begin{center}
\includegraphics[height=6cm]{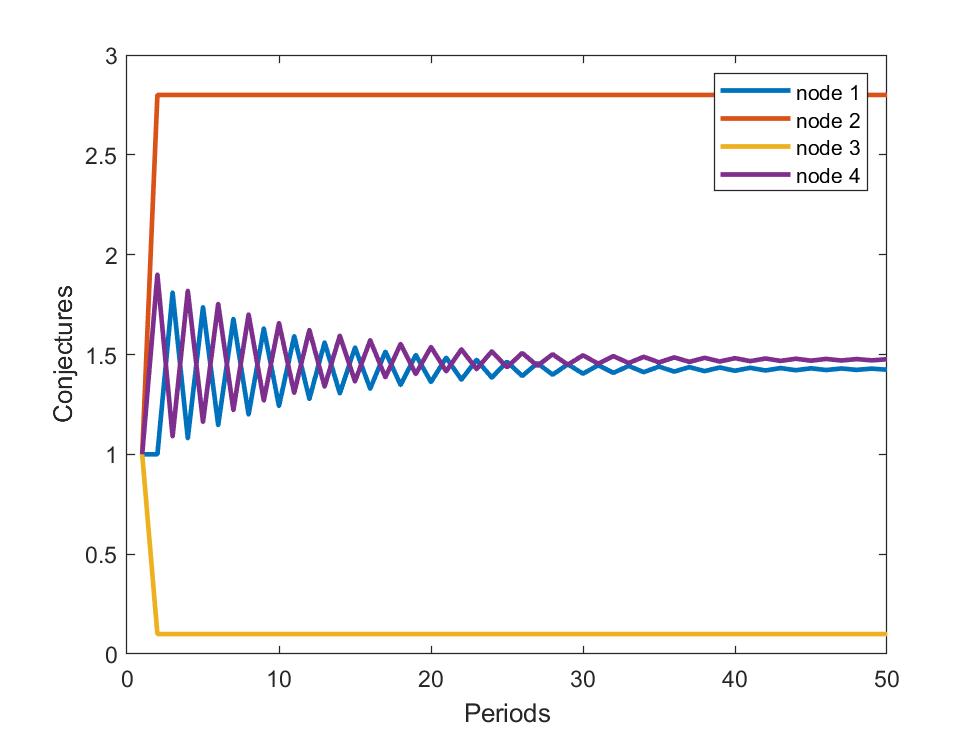} %
\includegraphics[height=6cm]{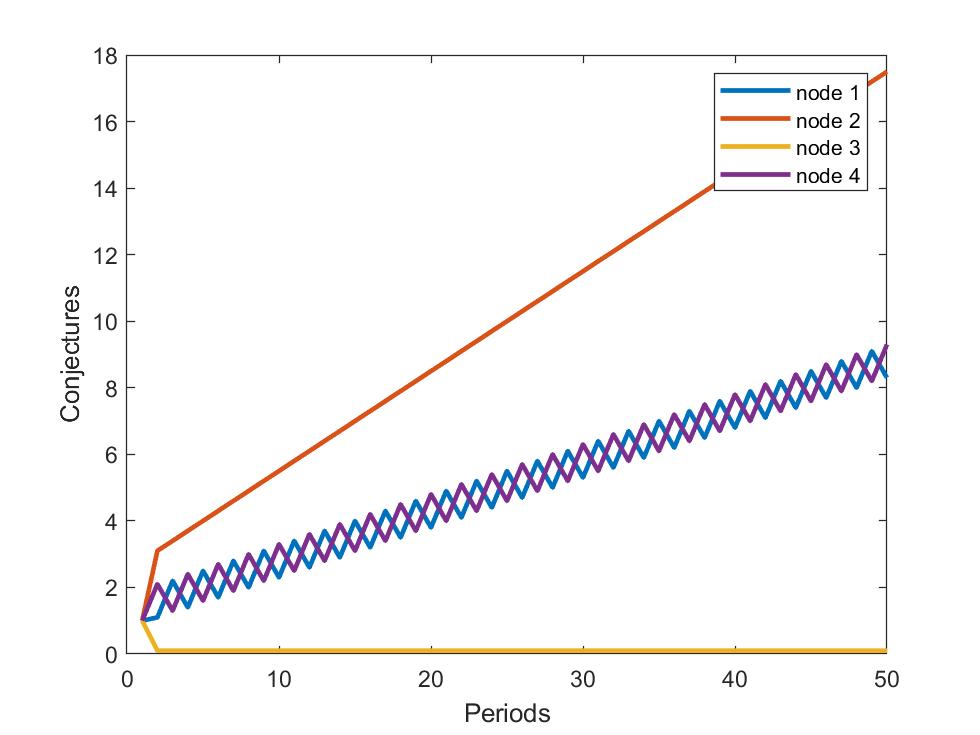}
\end{center}
\caption{\textit{Positive externalities only.} Given a network structure
(from Figure \protect \ref{fig:net1}), starting from the same conjectures,
the learning path may converge or not depending on the size of $w$: $w = 0.9$
in the left panel; $w=1$ in the right panel. Actions are an upward shift of
conjectures, of amount $\protect \alpha_i$. }
\label{fig:netex5}
\end{figure}

\begin{example}
\label{ex:non_conv} Go back to the 4-node network of Example \ref{ex:net1}
(Figure \ref{fig:net1}). Even if there are only positive externalities,
convergence depends on the magnitude of $w$. If $w<1$, there is convergence.
If instead $w\geq 1$, there is divergence. Figure \ref{fig:netex5} shows two
cases, with $w=0.9$ and $w=1$ respectively, starting from the same initial
beliefs. Note that the actions of nodes/players 1 and 4 reinforce each
other's beliefs, and this gives rise to an oscillating path of their
beliefs. The case of $w=1$, where the amplitude of oscillations remains
constant, is actually non--generic. 
\end{example}

\clearpage

\section{Local and global externalities}

\label{sec:local_global}

In many applications the feedback that active players receive is not enough
to find out the objectively optimal response. Users of online platforms may
not understand \emph{ex post} the objective best response to others' activity%
{and a firm in a complex market may not be able to infer
optimal investment plans by just observing prices}. In our context, this
means that perfect feedback may \emph{not} hold even for active players. 
In particular, this is the case if players just observe their realized
payoffs, but there are global externalities, which introduce a confounder.
This implies there may be other equilibria besides those analyzed above.
Assuming that local externalities are positive, the following analysis
yields two important observations. First, players may be more active if they
think that they are more linked in the network than they actually are, and
this can be welfare improving for the whole society. Second, agents with
excessive perceived connectedness may have the effect of preventing the
convergence of conjectural best-reply paths to non-corner solutions. Recall
Definition \ref{Def: SCE} (of selfconfirming equilibrium), based on general
linear-quadratic network games with just observable payoff (see equations %
\eqref{eq:global_intro}-\eqref{eq:G aggr}). We can characterize the set of
SCE's as follows:

\begin{proposition}
\label{Prop: SCEglob} A profile of actions and conjectures $\left(
a_{i}^{\ast },\hat{x}_{i},\hat{y}_{i}\right) _{i\in I}\in \times _{i\in
I}\left( A_{i}\times X_{i}\times Y_{i}\right) $ in a linear--quadratic
network game with just observable payoffs and global externalities is a
selfconfirming equilibrium at $\left( \mathbf{Z},\gamma \right) $ if and
only if, for every $i\in I$,

\begin{enumerate}
\item $a_{i}^{\ast }=0$ implies $\hat{x}_{i}\in \left[ \underline{x}%
_{i},-\alpha _{i}\right] $ and $\hat{y}_{i}=\gamma \sum_{j\neq i}a_{j}^{\ast
}$;

\item $a_{i}^{\ast }>0$ implies $a_{i}^{\ast }=\min \{ \alpha _{i}+\hat{x}%
_{i},\bar{a}_{i}\}$ and $\hat{y}_{i}=\gamma \sum_{k\neq i}a_{k}^{\ast
}+a_{i}^{\ast }\left( \sum_{j\neq i}z_{ij}a_{j}^{\ast }-\hat{x}_{i}\right) $.
\end{enumerate}
\end{proposition}

We discuss how the presence of the global externality term in the utility
function changes the characterization of selfconfirming equilibria. Although
we maintain the assumption of just observable payoffs, with global
externalities it is not anymore the case that active players have perfect
feedback about the payoff state. 
Indeed, for all $i\in I$ and for all pairs of realized externalities $\left(
x_{i},y_{i}\right) $, $v_{i}\left( 0,x_{i},y_{i}\right) =y_{i}$; thus,
inactive players have correct conjectures about the global externality, but
may have incorrect conjectures about the local externality. Active players,
on the other hand, are not able to determine precisely the relative
magnitude of the local effects with respect to the global effects. Given any
strictly positive action $a_{i}^{\ast }$, the confirmed conjectures
condition yields $(\hat{y}_{i}-y_{i})=a_{i}^{\ast }\left( x_{i}-\hat{x}%
_{i}\right) $. Then, in equilibrium, if agent $i$ overestimates
(underestimates) the local externality, she must compensate this error by
underestimating (overestimating) the global externality. Compared to the
case of only local externalities, we have that: $(i)$ active agents choose a
best response to a (possibly) wrong conjecture about the payoff state; thus, 
$(ii)$ it is not possible to completely characterize the set of SCE's by
means of Nash equilibria of the auxiliary games restricted to the active
players.

Yet, the analysis of Section \ref{sec:local} allows to identify a subset of
selfconfirming equilibria, those where agents have correct (shallow)
conjectures about the global payoff state.

\begin{remark}
\label{Rem:SCE local-global}Fix $\mathbf{Z}$ and $\gamma $. The set SCE
action profiles of the network game with only local externalities is
contained in the set of SCE action profiles of the game with local and
global externalities, that is, $\mathbf{A}_{\mathbf{Z}}^{SCE}\subseteq 
\mathbf{A}_{\mathbf{Z},\gamma }^{SCE}$. Specifically, if $\left( a_{i}^{\ast
},\hat{x}_{i}\right) _{i\in I}$ is an SCE of the game with only local
externalities, then $\left( a_{i}^{\ast },\hat{x}_{i},\hat{y}_{i}\right)
_{i\in I}$ with $\hat{y}_{i}=\gamma \sum_{k\neq i}a_{k}^{\ast }$ for each $%
i\in I$ is an SCE of the game with local and global externalities.
\end{remark}

Indeed, by Proposition \ref{Prop: SCE if LQ-OiffA text}, in profile $\left(
a_{i}^{\ast },\hat{x}_{i}\right) _{i\in I}$ each inactive player has a
(trivially) confirmed conjecture that makes her choose $0$, and each active
player must have a correct conjecture about the local externality. In
profile $\left( a_{i}^{\ast },\hat{x}_{i},\hat{y}_{i}\right) _{i\in I}$
conjectures $\left( \hat{y}_{i}\right) _{i\in I}$ about the global
externalities are correct by assumption. Thus, by Proposition \ref{Prop:
SCEglob}, $\left( a_{i}^{\ast },\hat{x}_{i},\hat{y}_{i}\right) _{i\in I}$ is
an SCE.

To ease the following analysis, in the remainder of this whole Section, we
assume that (i) each agent $i$ has \emph{the same stand-alone parameter} $%
\alpha >0$ \emph{and upper bound} $\bar{a}$, and (ii) $\gamma >0$. We assume
also that (iii) each matrix $\mathbf{Z}\in \mathcal{Z}$ is non--negative,
and (iv) either condition 1.~or 3.~of Proposition \ref{prop:interiorSCE} is
satisfied, so that there exists a unique NE. 
Finally, (v) we assume that the admissible range of possible best replies
for any player has no negative elements and does contain the upper bound $%
\bar{a}$.


Understanding how conjectures are shaped in a SCE also allows us to shed
some light on the efficiency properties of the SCE's. First of all note that
the problem of finding a maximizer of the sum of the utilities is a concave
quadratic problem and there exists a bliss point. The presence of positive
externalities makes the unique NE Pareto-dominated by other actions
profiles. Moreover the presence of a bliss point makes an arbitrary increase
of agents' actions not always welfare improving. Let us analyze these issues
in detail.\newline
Given the presence of global externalities, it is straightforward to see
that the Nash equilibrium is inefficient. Now, consider an SCE action
profile $\mathbf{a}^{SCE}$ (possibly $\mathbf{a}^{NE}$). This action profile
is justified by some profile of confirmed conjectures $(\hat{x}_{i},\hat{y}%
_{i})_{i\in I}$. Then, we can find another SCE, $\mathbf{a^{\prime }}%
^{SCE}\geq \mathbf{a}^{SCE}$, such that $\mathbf{a^{\prime }}^{SCE}$ yields
a higher aggregate payoff than $\mathbf{a}^{SCE}$. A possible way to find
such an equilibrium is to decrease, for each $i\in I$, the global
externality (shallow) conjecture $\hat{y}_{i}$. To keep the confirmation
condition, it is necessary to increase the local (shallow) conjectures $%
\left( \hat{x}_{i}\right) _{i\in I}$, and thus to increase the best-reply
actions. This, in turn, makes the local and global externalities increase.
However, this makes 
it necessary that the local conjectures are further increased, which induces
another increase in actions, and so on. The following proposition imposes a
condition for the existence of an interior SCE.

\begin{proposition}
\label{prop:inequality_SCE} If, for every pair of agents $(i,j)$ and for
every profile of local conjectures $\mathbf{\hat{x}}$, the following
inequality is satisfied 
\begin{equation}
\displaystyle \sum_{k\in I\backslash \left \{ i,j\right \} }z_{ik}\left(
\alpha +\hat{x}_{k}\right) -z_{ij}\displaystyle \sum_{h\in I\backslash \left
\{ i,j\right \} }\left( \alpha +\hat{x}_{h}\right) \alpha \geq 0\  \ ,
\end{equation}%
then, for every profile of global conjectures $\mathbf{\hat{y}}$ with $\hat{y%
}_{i}<\bar{a}\bigl(\alpha \displaystyle \sum_{k\in I\backslash \left \{
i\right \} }z_{ij}+\gamma n\bigr)$ for every $i$, there exists a unique SCE
with local conjectures $\mathbf{\hat{x}}$ and action profile $\mathbf{a}%
^{\ast }$, with $a_{i}^{\ast }<\bar{a}$.
\end{proposition}

The condition of the proposition imposes concavity on some fixed point
equations derived from the best replies functions, and then ensure existence
and uniqueness of this fixed point. Note that the condition is always
satisfied if $\alpha \leq 1$ and $\mathbf{Z}=w\mathbf{Z}_{0}$, with $w>0$,
that is: every strictly positive $z_{ij}$ has the same value for each pair
of agents $i$ and $j$ in $I$. 
Otherwise, the larger the number of agents, the more likely it is that the
condition is violated for some pair $(i,j)$ for which $z_{ij}$ is high. If
the network is composed of just two agents, this condition is always
satisfied. {The following example illustrates some of the issues
just analyzed. }

\begin{example}
\label{ex:twoagents} Consider a simple network composed of two agents. Let $%
z_{ij}z_{ji}=\delta$. For simplicity we assume that $\alpha=\gamma=%
\delta=0.1 $. Figure \ref{figure:ex2nodes} represents several features of
this examples. On the axes we report $\hat{x}_1$ and $\hat{x}_2$,
respectively. The curve $\hat{y}_1=0$ represents all the possible $(\hat{x}%
_1, \hat{x}_2)$, such that agent 1 thinks about a null global externalities.
Since we know that, in a SCE, $\hat{y}_1\geq0$, then all the feasible
conjectures are on the left of this curve, since on the right of $\hat{y}%
_1=0 $, we would have negative conjectured global externalities. For the
very same reason, only pairs $(\hat{x}_1, \hat{x}_2)$ below $\hat{y}_2=0$
are consistent with positive conjectured global externality for agent 2. As
a result, in a SCE only pairs $(\hat{x}_1, \hat{x}_2)$ between the two
curves can be observed. The dashed lines show the NE conjectures. As is easy
to observe, SCE allows for much higher (and lower too) conjectures, so that
larger actions are allowed.\newline
The dashed line 
represents all the pairs of conjectures delivering the same welfare as the
NE. Above this dashed line the welfare is larger than in NE, below this line
it is smaller. In this example the SCE with the highest welfare is the
top-right kink with the highest possible conjectures (note that, in this
case, the bliss point for the welfare is $\hat{x}_1=\hat{x}_2=0.275$, that
is out of the confirmed conjectures area. 
\begin{figure}[h]
\begin{center}
\includegraphics[scale=0.45]{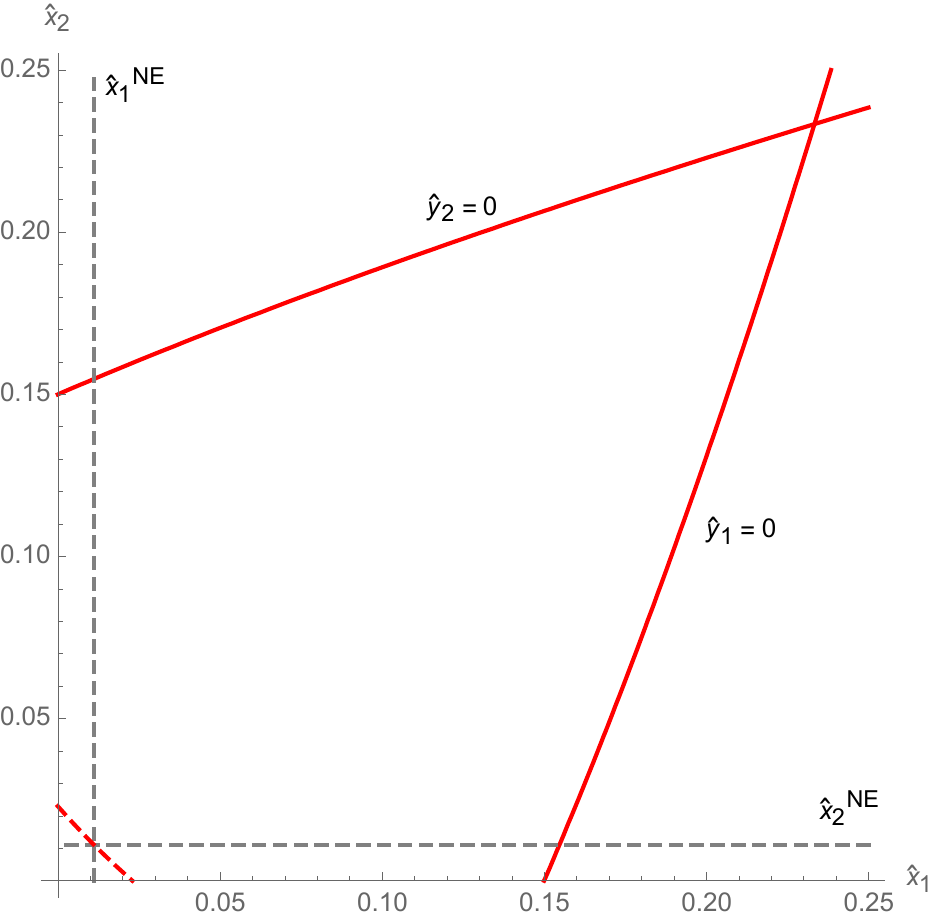}
\end{center}
\caption{SCE's for a network of two agents. Parameterization $\protect \alpha=%
\protect \gamma=\protect \delta=0.1$.}
\label{figure:ex2nodes}
\end{figure}
\end{example}

%
\bigskip

To better understand the structure of the equilibrium set, we introduce
additional assumptions about what agents know or think they know about the
strategic environment. This is a way to restrict their conjectures. We
provide some insights along two different dimensions: $i)$ what happens if
agents know something about the magnitude of the externalities ? $ii)$ What
happens if agents have definite beliefs about the relative size of local
with respect to global externality? {This last case, that we
call \textit{perceived centrality} will be crucial for the learning dynamics.%
}

\subsection{Knowledge of externalities parameters}

\label{sec:knwoledgeglobal} We assume that $\mathbf{Z}=w\mathbf{Z_{0}}$,
where $w>0$, and 
$\mathbf{Z_{0}\in }\left \{ 0,1\right \} ^{I\times I}$ is the unweighted
network. This means that there is a homogeneous positive externality $w$
between all connected players, so that equation \eqref{eq:global_intro}
becomes: 
\begin{equation}
u_{i}(a_{i},\mathbf{a}_{-i},\mathbf{Z})=\alpha _{i}a_{i}-\frac{1}{2}%
a_{i}^{2}+a_{i}w\sum_{j\in I\backslash \left \{ i\right \}
}z_{0,ij}a_{j}+\gamma \displaystyle \sum_{k\in I\backslash \left \{ i\right \}
}a_{k}\  \ .  \label{eq:global_game_beta}
\end{equation}

We do not impose any further restriction over the network structure $\mathbf{%
Z_{0}}$, but we assume that all agents understand they interact in a network
and \emph{know $w$ and $\gamma $}. Given these assumptions, we need to
slightly modify what aggregators and conjectures are. In detail, aggregators
about local and global externalities do not internalize $w$ and $\gamma $,
respectively, and the conjectures concern the aggregate actions of the
neighbors (local) and of all other players (global).

Consider the case in which $\mathbf{Z}=w\mathbf{Z}_{0}^{c}$, where $\mathbf{Z%
}_{0}^{c}$ is the matrix of the complete basic network (i.e., $z_{0,ij}=1$
for all non-diagonal entries). Note that if the agents conjecture that the
network is a complete one, then, for each $i\in I$, $\hat{x}_{i}=\hat{y}_{i}$%
, and this ensures uniqueness of the SCE. Then the SCE can just be indexed
by the conjecture about the local externality.\footnote{%
The discussion below about conjectured ratios will make this point clear.}
Given $(w,\gamma )$, let $(a_{i}^{c}(w,\gamma ),\hat{x}_{i}^{c}(w,\gamma
))_{i\in I}$ denote the unique SCE in which, for each $i\in I$, $\hat{x}%
_{i}^{c}(w,\gamma )$ is the (confirmed) shallow conjecture induced by $\bar{%
\mu}_{i}^{c}\in \left \{ \mathbf{Z}_{0}^{c}\right \} \times \mathbf{A}_{-i}$,
that is, a (confirmed) deep conjecture in which $i$ thinks she belongs to a
complete network.

\begin{proposition}
\label{prop:beta} Consider a linear quadratic network game with global
externalities, with $0<w<\frac{1}{n-1}$, and where all
agents know $w$ and $\gamma $. Let $\mathbf{a}_{\mathbf{Z}_{0}}^{NE}$ and $%
\mathbf{a}_{\mathbf{Z}_{0}^{c}}^{NE}$ be the unique Nash equilibria of the
game played on $(w\mathbf{Z_{0}},\gamma )$ and $(w\mathbf{Z}_{0}^{c},\gamma )
$, respectively. Then, (1) $a_{i}^{c}(w,\gamma )$ is increasing in the ratio 
$\frac{\gamma }{w}$; (2) $\lim_{\frac{\gamma }{w}\rightarrow 0 }%
\mathbf{a}^{c}(w,\gamma )=\mathbf{a}_{%
\mathbf{Z}_{0}}^{NE}$; and (3) $\lim_{\frac{\gamma }{w}\rightarrow \infty }%
\mathbf{a}^{c}(w,\gamma )=\mathbf{a}_{\mathbf{Z}_{0}^{c}}^{NE}$.
\end{proposition}

%
%

So, independently of the basic network $\mathbf{Z}_{0}$, if all players
believe to be more linked than they actually are and $\frac{\gamma}{w} $ is 
\emph{large}, then the action profile approaches what they would choose in
the NE of the game played on the complete network, where every player is
linked to every other player.

As it will be clear from Section \ref{Subsec:Learn global}, this result
implies that the learning paths are self--reinforcing. Players maintain
wrong conjectures about the network structure and they infer $\ell
_{i}\left( \mathbf{a}_{-i}^{\ast },\mathbf{Z}\right) $ from the payoff that
they receive as feedback, using \eqref{eq:global_game_beta}. This implies
that, converging to an SCE, as they increase their own action they infer a
higher $\ell _{i}\left( \mathbf{a}_{-i}^{\ast },\mathbf{Z}\right) $ and a
lower $g_{i}\left( \mathbf{a}_{-i}^{\ast },\gamma \right) $, to which they
will respond with an even higher action. Nevertheless, this process does not
diverge to hit the upper bounds of the action profiles, and it reaches the
NE on the complete network.

\begin{figure}[h]
\begin{center}
\includegraphics[scale=0.55]{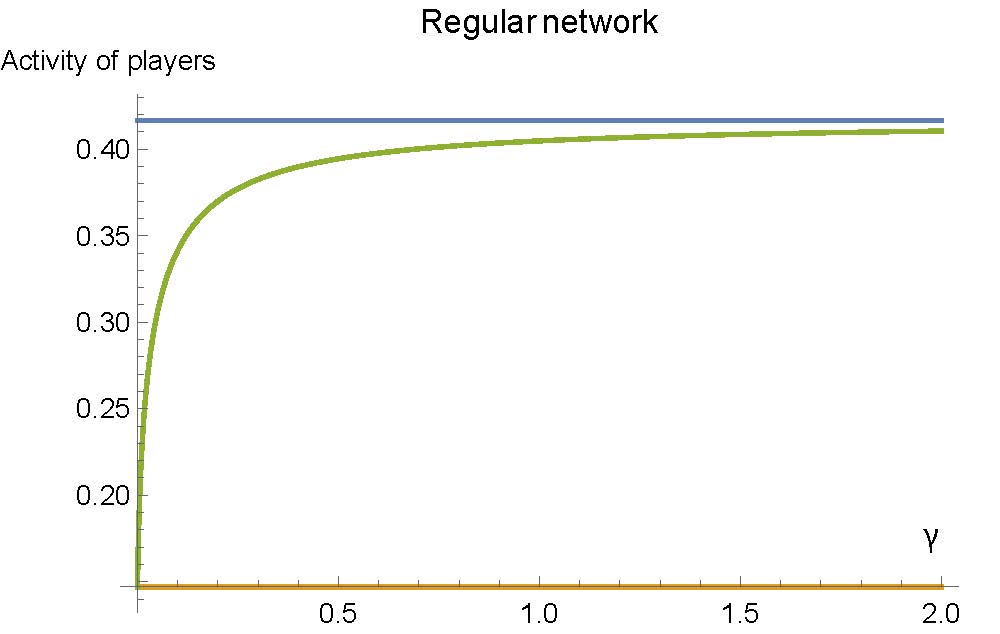} %
\includegraphics[scale=0.55]{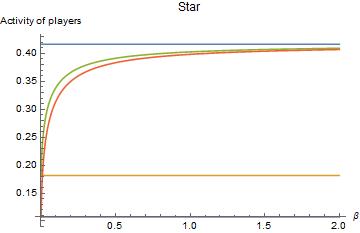}
\end{center}
\caption{The panels show the SCE common activity level as a function of
parameter $\protect \gamma$ when each agent thinks she is connected to every
other agent. Both cases have parameters $\protect \alpha=0.1$, $w=0.04$ and $%
n=20$. The left panel is for the regular network with common degree 8: in
blue we have the action that would be played in the NE of the complete
network; in yellow the NE of the regular network; in green the SCE action.
The right panel is for the star network: in blue we have the action that
would be played in the NE of the complete network; in yellow and purple the
NE action profiles for the center and the spokes, respectively, in the star
network; in green and red the SCE action profile for the center and the
spokes, respectively.}
\label{fig:regular_star}
\end{figure}

Proposition \ref{prop:beta} is a limiting result. However, for some networks
where NE's and SCE's can be easily computed analytically, we can show that
the actions of an SCE converge rapidly to the actions of the NE for the
complete network as $\gamma /w$ becomes large. Figure \ref{fig:regular_star}
shows how this happens when every player has the same number of links
(regular network) and when there is a central player and every other player
is linked only to her (star network).

In the Introduction we discussed the possible application of our model to
online social networks, where the provider may have the possibility to
affect the beliefs of the consumers. The previous result applies to the case
where consumers know the value of the parameters $w$ and $\gamma $, and
their overall number $n$. If we further assume that the profits of the
provider are positively correlated with the overall activity on the
platform, the provider may have an incentive to make people feel more
connected than they actually are. So, if $\frac{\gamma }{w}$ is large (that
is, in our interpretation, most of the payoff for the consumers is obtained
from using the platform \emph{per se}, and not from actual interaction), and
if these parameters are known to the users, companies make more profit by
letting players think that they have a lot of followers. With this
application in mind, in the end of this section we will extend the
discussion about the implications of biased beliefs on aggregate welfare. 

Proposition \ref{prop:beta} is based on the assumption that players know the
values of $\gamma $ and $w$. However, if they have wrong beliefs about $%
\gamma $, overestimating it, their actions would even exceed those of the NE
of the complete network. This is shown in the next example, where agents do
not know the true value of $\gamma $ and, overestimating the ratio between
local and global externalities, they play actions that are much above the
action that they would play in the NE of the complete network.\newline

%
%
%
%

\begin{example}
\label{ex:SCEglobal} Consider three agents in a star network (i.e., a line).
Let agent 2 be the center. Then, for every SCE, $\ell _{2}\left( \mathbf{a}%
_{-2}^{\ast },\mathbf{Z}\right) $ is proportional to $g_{2}\left( \mathbf{a}%
_{-2}^{\ast },\gamma \right) $, always with the same ratio $\frac{\gamma }{w}
$, while this is not true for agents $1$ and $3$. We assume that each agent
thinks that the network is complete, so every $i\in I$ thinks that $\ell
_{i}\left( \mathbf{a}_{-i}^{\ast },\mathbf{Z}\right) $ is proportional to $%
g_{i}\left( \mathbf{a}_{-i}^{\ast },\gamma \right) $. In this case agents $1$
and $3$ believe to be more linked than they actually are. Table \ref%
{table:star} provides the Nash equilibria for the actual network and for the
complete network, and the selfconfirming equilibrium actions for for some
specification of the parameters.

\begin{table}[h]
\centering
\begin{tabular}{||l|l|l|l||}
\hline
& Line NE & Complete Network NE & SCE \\[0.5ex] \hline \hline
$a_1$ & 0.130 & 0.167 & 1.569 \\ 
$a_2$ & 0.152 & 0.167 & 1.679 \\ 
$a_3$ & 0.130 & 0.167 & 1.569 \\ \hline
\end{tabular}%
\caption{Simulations for the case of $\protect \alpha =0.1$, $w=0.2$, and $%
\protect \gamma =1$. Columns refer to 1) Nash Equilibrium of the line
network; 2) Nash equilibrium of complete network; 3) SCE in the line network
in which each $i\in I$ believes that $\ell _{i}\left( \mathbf{a}_{-i}^{\ast
},\mathbf{Z}\right) =\frac{\protect \gamma }{w}g_{i }\left( \mathbf{a}%
_{-i}^{\ast },\protect \gamma \right) $. }
\label{table:star}
\end{table}
This numerical exercise shows that, when agents overestimate the impact of
local externalities, we get a \emph{multiplier} effect that makes SCE
actions increase at a level even larger than what would be predicted in a
complete network by Nash equilibrium. This follows from how agents
misinterpret their feedback. In particular, thinking to be in a complete
network makes agents $1$ and $3$ overestimate local externalities. Take for
instance agent $1$. Given any $\mathbf{a}_{-1}$, she chooses a subjective
best reply higher than the objective best reply since she overestimates the
local externality. This high action has the effect of increasing the global
externality term for agent $3$. Agent $3$, by overestimating the local
externality, partly attributes this higher global externality to the local
externality term, and chooses an action larger than predicted by Nash
equilibrium. The choice of agent $3$ increases in turn the global
externality perceived by agent $1$, and so on. At the same time agent $2$,
as neighbors choose higher actions, increases her own action level. This
effect goes on and gives rise to a multiplier effect. The limit of such a
conjectural best reply path is selfconfirming equilibrium in which actions
are almost ten times larger than the complete network NE actions 
\end{example}

We call $c_{i}:=\frac{\hat{x}_{i}}{\hat{y}_{i}}$ the \textbf{conjectured
ratio} of player $i$ with respect to local and global externalities. Then,
given a profile $(c_{i})_{i\in I}$, one can rewrite the SCE conditions as a
non-linear system of $n$ equations in $n$ unknowns solved either for $(\hat{x%
}_{i})_{i\in I}$ or $(\hat{y}_{i})_{i\in I}$, and characterize the set of
SCE's given the imposed restrictions. This is what we will use in the next
section when studying the learning paths.

{We can think of conjectured ratio $c_{i}$ as the \textbf{%
perceived centrality} of player $i$. For each player, this parameter
describes what she thinks to be the share of the activity in her
neighborhood with respect to the sum of all the actions of the population.
This perceived share has a strong relationship with the Bonacich centrality.
If there is a unique Nash equilibrium $\mathbf{a}^{\ast }$ of the game,
where all actions are strictly positive, we have, for each $i\in I$, 
\begin{equation*}
a_{i}^{\ast }=\alpha +x_{i}=\alpha +\sum_{j\in I\backslash \left \{ i\right
\} }z_{ij}a_{j}^{\ast }\ .
\end{equation*}%
The profile of \textbf{Bonacich centrality measures} $\mathbf{b}$ is the
unique solution of the linear system\footnote{%
{In general, independently of any game defined on the network, Bonacich
centrality is a network centrality measure that depends on parameter $\alpha >0$. It is defined exactly as the solution of that same linear
system. For a detailed discussion on this see \cite{DequiedtZenou17}.}} 
\begin{equation*}
\forall i\in I\text{, }b_{i}=\alpha +\sum_{j\in I\backslash \left \{ i\right
\} }z_{ij}b_{j}\ .
\end{equation*}%
So, when beliefs are correct, as in the Nash equilibrium, we have that, for
each $i\in I$, $b_{i}=a_{i}^{\ast }$, $y_i =\gamma \displaystyle \sum_{k\in
I\backslash \left \{ i\right \} }a^*_{k}$ and $c_{i}=\frac{b_{i}-\alpha }{%
y_{i}}$.\newline
Now, in the Nash equilibrium we have also that, for each $i$ and $j$, $\frac{%
1}{y_{i}}-\frac{1}{y_{j}}=\gamma \frac{a_{i}^{\ast }-a_{j}^{\ast }}{%
y_{i}y_{j}}$. If the number $n$ of players is large, for each $i$ and $j$, {%
\ $y_{i}$ and $y_j$ grow and the difference $\frac{1}{y_{i}}-\frac{1}{y_{j}}$
approaches $0$ faster than $\frac{1}{y_{i}}$ and than $\frac{1}{y_{j}}$. We
can express this writing $\frac{1}{y_{i}}\simeq \frac{1}{y_{j}}$, because as 
$n$ grows both $\frac{1}{y_{i}}$ and $\frac{1}{y_{j}}$ are of \emph{another
order of magnitude} with respect to $\frac{1}{y_{i}}-\frac{1}{y_{j}}$}, and
so every $c_{i}$ is roughly the same linear rescaling of $b_{i}$. {\ 
Our perceived centrality can then be interpreted, with a good approximation,
as the belief that player $i$ has, as a node in a large network, about her
Bonacich centrality.} }

\subsection{Learning with global externalities\label{Subsec:Learn global}}

We now study conjectural best reply paths with global externalities. To
simplify the analysis we assume, for each agent, a \emph{fixed conjectured
ratio}. Differently from Section \ref{sec:knwoledgeglobal}, we do not assume
agents to know anything about the parameters characterizing the strategic
environment.\newline
At each time, there are infinitely many profiles of feasible pairs $\left( 
\hat{x}_{i,t},\hat{y}_{i,t}\right) _{i\in I}$ consistent with feedback. For
each $i\in I$, and each time $t\in \mathbb{N}$, let $m_{i,t}=f_{i}\left(
a_{i,t},x_{i,t},y_{i,t}\right) =u_{i}(a_{i,t},\mathbf{a}_{-i,t}, \mathbf{Z},
\gamma)$ be the message agent $i$ receives. 
Then, given message $m_{i,t-1}$, and considering that agents perfectly
recall their past actions, $\hat{y}_{i,t}$ is uniquely determined as a
function of $\hat{x}_{i,t}$. In particular, if at each time period $t$ agent 
$i$'s conjectures $\hat{x}_{i,t}$ and $\hat{y}_{i,t}$ are consistent with
the message received at the previous period, we obtain 
\begin{equation*}
\hat{y}_{i,t+1}=m_{i,t}-\alpha a_{i,t}+\frac{1}{2}\left( a_{i,t}\right)
^{2}-a_{i,t}\hat{x}_{i,t+1}\ .
\end{equation*}%
Then, we can focus on the path of $\hat{x}_{i,t}$, given by

\begin{equation}
\hat{x}_{i,t+1}=\frac{ m_{i,t} -\hat{y}_{i,t+1}}{a_{i,t}}-\alpha +\frac{1}{2}%
a_{i,t} \  \ .  \label{eq:learningglobal}
\end{equation}

In this case, active agents do not have perfect feedback, because players
test a two--dimensional conjecture with a feedback, the payoff, that has a
single dimension. This brings also indeterminacy to the updating rule that
players use. To avoid bifurcations at each time period $t$, we need to use
simplifying assumptions on conjectures. We define for each $i\in I$ and
every $t\in \mathbb{N}_{0}$ 
\begin{equation}
c_{i,t}:=\frac{\hat{x}_{i,t}}{\hat{y}_{i,t}},\footnote{%
In doing so, we implicitly assume that players think there are active
co-players. This is a reasonable assumption, because under positive
externalities any best response should be at least $\alpha$.}
\end{equation}%
and in the following we assume that this \emph{conjectured ratio} is
constant along paths of learning dynamics for each player $i$.

\begin{assumption}
For each $i\in I$ and for each $t\in \mathbb{N}$, $c_{i,t}=c_{i,t+1}=c_{i}$. %
\label{ass:v}
\end{assumption}

From equation \eqref{eq:learningglobal}, and expressing the message as the
observed payoff, we get the following learning path, for each agent at each
time period: 
\begin{equation}
\hat{x}_{i,t+1}=x_{i,t}+\frac{y_{i,t}}{a_{i,t}}-\frac{\hat{y}_{i,t+1}}{%
a_{i,t}}\ ,  \label{learning1}
\end{equation}%
where $x_{i,t}$ and $y_{i,t}$ are the true realized values of the payoff
states. Plugging in $c_{i}=\frac{\hat{x}_{i,t}}{\hat{y}_{i,t}}$ we get, for
each $t $ and $i$, 
\begin{equation}
\hat{x}_{i,t+1}=\frac{c_{i}}{1+c_{i}a_{i,t}}\left(
a_{i,t}x_{i,t}+y_{i,t}\right) \ .  \label{eq:simulate_global}
\end{equation}


{\ Note that the true ratio of player $i$ at time $t$ is} 
\begin{equation*}
c_{i,t}^{\prime }:=\frac{x_{i,t}}{y_{i,t}}\  \ ,
\end{equation*}%
{\ with} $c_{i,t}^{\prime }\in \left[ 0,\frac{\sum_{j\neq i}z_{ij}}{\gamma }%
\right]$. For this reason, we also assume that the conjectured ratio of each
player $i$ is such that $c_{i}\in \left( 0,\frac{\sum_{j\neq i}z_{ij}}{%
\gamma }\right] $, and this specifies the set of all admissible conjectured
ratios.

The learning dynamic from \eqref{learning1}, then, can be written as 
\begin{equation}  \label{Hdynamics}
\hat{x}_{i,t+1}=c_{i}y_{i,t}\frac{a_{i,t}^{\ast }c_{i,t}^{\prime }+1}{%
a_{i,t}^{\ast}c_{i}+1}\  \ ,
\end{equation}%
which implies that the conjecture $\hat{x}_{i,t+1}$ is correct only when $%
c_{i}=c_{i,t}^{\prime }$.

We look at best responses $a_{i,t+1}=\alpha +\hat{x}_{i,t+1}$, and study the
existence and characterization of the steady state of this learning process.
Recall that $y_{i,t}=\gamma \sum_{j\neq i}a_{j,t}$. To find a fixed point we
look at the system of $n$ equations, one for each $i$, 
\begin{equation}
H_{i}(\mathbf{a}^{\ast },\mathbf{c},\gamma ,\mathbf{Z}):=\alpha +c_{i}\left(
\gamma \sum_{j\neq i}a_{j}^{\ast }\right) \frac{a_{i}^{\ast }c_{i}^{\prime
}+1}{a_{i}^{\ast }c_{i}+1}-a_{i}^{\ast }=0\  \ .  \label{system1}
\end{equation}%
For comparison, we also study the system of equations that provide the Nash
Equilibrium of this network game, that is, for each $i$: 
\begin{equation}
F_{i}(\mathbf{a}^{\ast },\mathbf{Z}):=\alpha +\sum_{j\neq
i}z_{ij}a_{j}^{\ast }-a_{i}^{\ast }=0\  \ .  \label{systemNE}
\end{equation}

\bigskip

Let $\mathcal{A}\subset \lbrack \alpha ,\infty )^{I}$ denote the set of the
solutions of system \eqref{system1}. We have the following result.

\begin{proposition}
\label{prop:homeo1} If the system defined by \eqref{systemNE} admits a
solution $\mathbf{a}^{\ast }$ with non-negative entries, then for each
profile $\mathbf{c} $ of conjectured ratios also the system defined by %
\eqref{system1} admits a solution. Moreover, there is a homeomorphism $\Phi$
between the set of all profiles $\mathbf{c}$ and $\mathcal{A}$. The
homeomorphism $\Phi$ is strictly monotone with respect to the lattice order
of the domain of all profiles $\mathbf{c}$ and the codomain $\mathcal{A}$.
\end{proposition}

The assumption of non-negative solutions implies a unique NE of the game,
and we refer to Proposition \ref{prop:interiorSCE} for sufficient conditions
for uniqueness. This result provides information only on the steady states
of our learning paths. It is important because it establishes a one--to--one
function between profiles of conjectured ratios and SCEs: there is one and
only one SCE strategy profile for each profile $\mathbf{c}$ but there may
SCEs that do not result from the hypothesized learning paths. The
homeomorphism also provides continuity on the initial parameters, as a
marginal change in the conjectured ratios will result in a marginal change
in the resulting SCE, even if this function may be highly non--linear, as
shown in the example below.

\begin{example}
\label{ex:line_complete} Under the conditions of Proposition \ref%
{prop:homeo2}, we use equation \eqref{eq:simulate_global} to express
learning paths converging to the SCE implicitly defined by \eqref{system1}.
This allows us to provide a graphical illustration of Proposition \ref%
{prop:homeo1}, for the case of three nodes. 
We do this for the case of a line network (where each of the two links is
bidirectional), and for the case of a complete network. We consider equation %
\eqref{eq:global_game_beta}, with $\gamma =1$ and $w=0.2$. Figure \ref%
{fig:hom} shows the results. We can start from any pattern of conjectured
ratios for the three nodes. The left panel shows the profile of conjectured
ratios when at least one node has maximal conjectured ratio (the three faces
of the cube have different colors, according to which node has the maximal
centrality). The central panel shows the corresponding SCE conjecture
profile $\vec{\hat{{x}}}$ when the network is a line (the node that has
conjectured ratio 1 in the red dots is the central node). The right panel
shows the corresponding SCE conjecture profile $\vec{\hat{{x}}}$ when the
network is a complete triangle. The figure suggests that homeomorphism $\Phi 
$ (from Proposition \ref{prop:homeo1}) is highly non--linear, because of the
self-reinforcement process in beliefs that we discussed in Example \ref%
{ex:SCEglobal}. The figure also shows that, as stated by Proposition \ref%
{prop:homeo1}, homeomorphism $\Phi $ respects the lattice order on the two
sets. 
\end{example}

\begin{figure}[h]
\begin{center}
\includegraphics[scale=0.25]{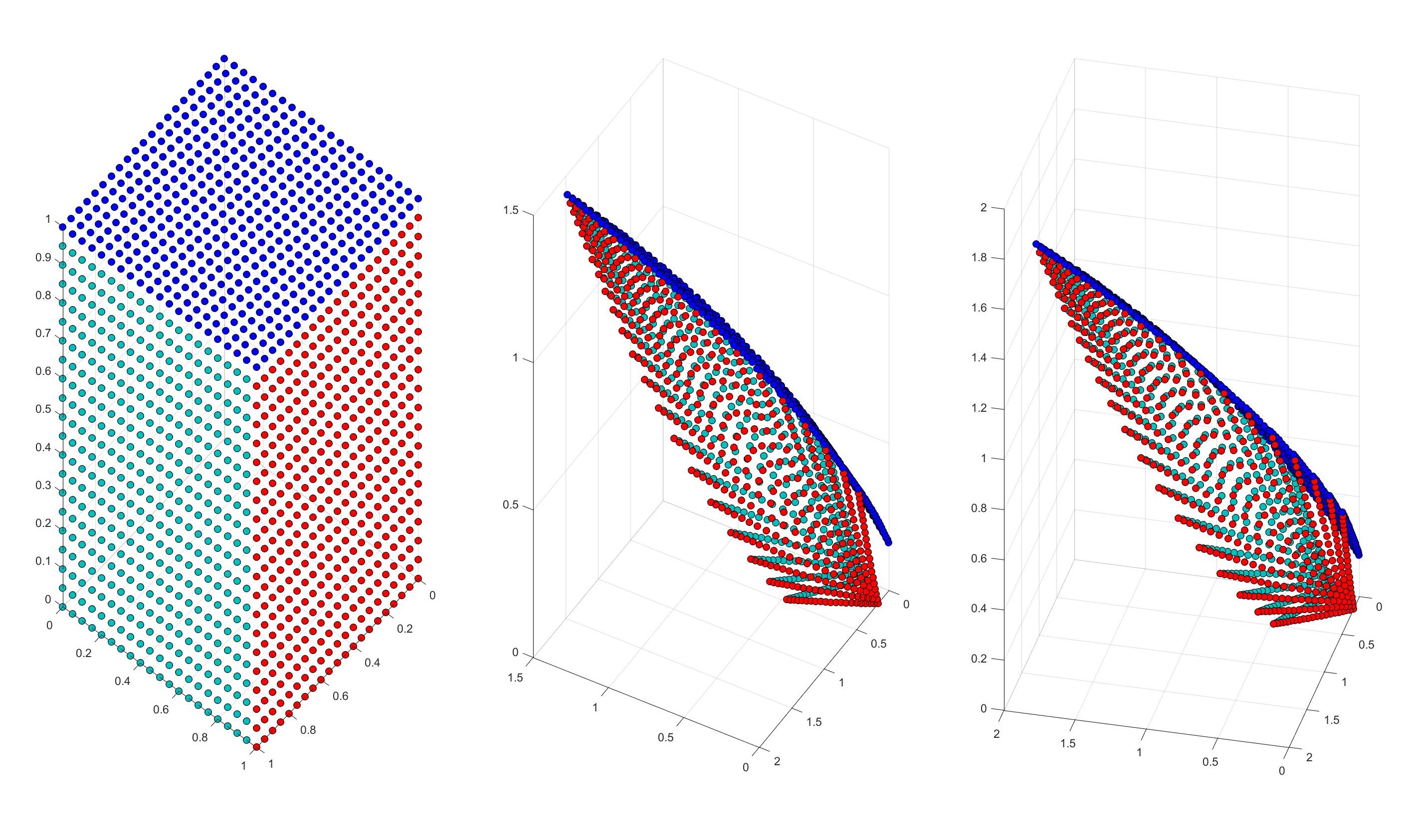}
\end{center}
\caption{Simulations showing the homeomorphism of Proposition \protect \ref%
{prop:homeo2} for the case of 3 nodes, as discussed in Example \protect \ref%
{ex:line_complete}. The left panel shows vectors of conjectured ratios. The
central panel shows the corresponding SCE conjecture profile $\hat{\mathbf{x}%
}$ when the network is a line (the node that has conjectured ratio 1 in the
red dots is the central node). The right panel shows the corresponding SCE
conjecture profile $\hat{\mathbf{x}}$ when the network is a complete
triangle. }
\label{fig:hom}
\end{figure}

Monotonicity implies that increasing the conjectured ratio of one player
will have a weakly monotonic effect on all the actions of that player and
other players in the corresponding SCE. A final \emph{caveat} to remember is
that the homeomorphism is implied by the particular learning path that we
are assuming, which is based on constant conjectured ratios. Considering the
paths in this special case, in the following proposition we show that if
local and global externalities are not too large, the learning paths always
converge.

\begin{proposition}
\label{prop:homeo2} If, for each player $i\in I$, $0<c_{i}\gamma
(n-1)<\sum_{j\neq i}z_{ij}<2$, then the paths defined by the learning paths %
\eqref{Hdynamics} always converge to the unique solution of \eqref{system1},
which is locally stable.\footnote{%
Definition \ref{def:stable} of local stability extends naturally to the case
of learning with global externalities with paths of the form $(\mathbf{a}%
_{t},\mathbf{\hat{x}}_{t},\mathbf{\hat{y}}_{t})_{t\in \mathbb{N}_{0}}$.}
\end{proposition}

It should be noted that, in a game with just local externalities, where $%
\gamma =0$, the assumptions of Proposition \ref{prop:homeo2} are more
general than assuming that $|\sum_{j\neq i}z_{ij}|<1$, which in turn implies
that Assumption \ref{ass:limited} holds and hence that the learning paths
converge. That is because we are focusing on a precise learning path in
which players act as if global externalities were present. Moreover, in a
game with $\gamma >0$, if for some players the conjectured ratios are too
high, the learning paths defined by \eqref{system1} may not converge to an
interior solution, but may hit instead the upper boundaries of the feasible
action profiles. 

Proposition \ref{prop:homeo1} tells us that a non-negative shift in each
conjectured ratio will always result in a non-negative shift of each agent's
action in the resulting SCE. However, Proposition \ref{prop:homeo2} gives an
implicit warning. Too high conjectured ratios may imply that the sufficient
conditions for stability are lost, and convergence to the corresponding SCE
may not occur. Note also that, summing up equation \eqref{eq:global_intro}
for all the players, the aggregate welfare is maximized if $\mathbf{a}^{\ast
}$ solves the following linear system of equalities%
\begin{equation*}
\forall i\in I\text{, \ }a_{i}^{\ast }=\alpha +(n-1)\gamma +\sum_{j\in
I\backslash \left \{ i\right \} }(z_{ij}+z_{ij})a_{j}^{\ast }\ .
\end{equation*}

To better understand this aspect, consider the online social networks
application we often referred to. The results of this last subsection apply
to the case where consumers do not know the parameters of the model and
their own total number, but have only a conjecture about the ratio of the
benefits from just using the platform, and from the actual strategic
interaction on the platform. Social platforms like Facebook and Twitter
often provide information to users about the activity of their peers. The
social platform Reddit does not show to users their followers, but only a
measure of popularity called \emph{karma}. A rationale for this marketing
strategy may be that these companies want to change the beliefs of players,
making them feel more important (i.e., more followed) in the social network.
Even a benevolent social planner may want to set the conjectured ratios to
the level for which the social optimum is achieved. However, according to
our model, if conjectured ratios are too high, the learning paths may
diverge. For example, in the context of the model and from the assumptions
of Proposition \ref{prop:homeo2} a conjectured ratio is \emph{too high} as
soon as $c_{i} \geq \frac{\sum_{j\neq i}z_{ij}}{\gamma (n-1)}$, because in
this case learning can lead to SCE where the activity of some player $i$
hits her upper bound $a_i$ and the strategy profile is inefficiently high
for the players.

This is shown in the following example.

\begin{figure}[h]
\begin{center}
\includegraphics[scale=0.25]{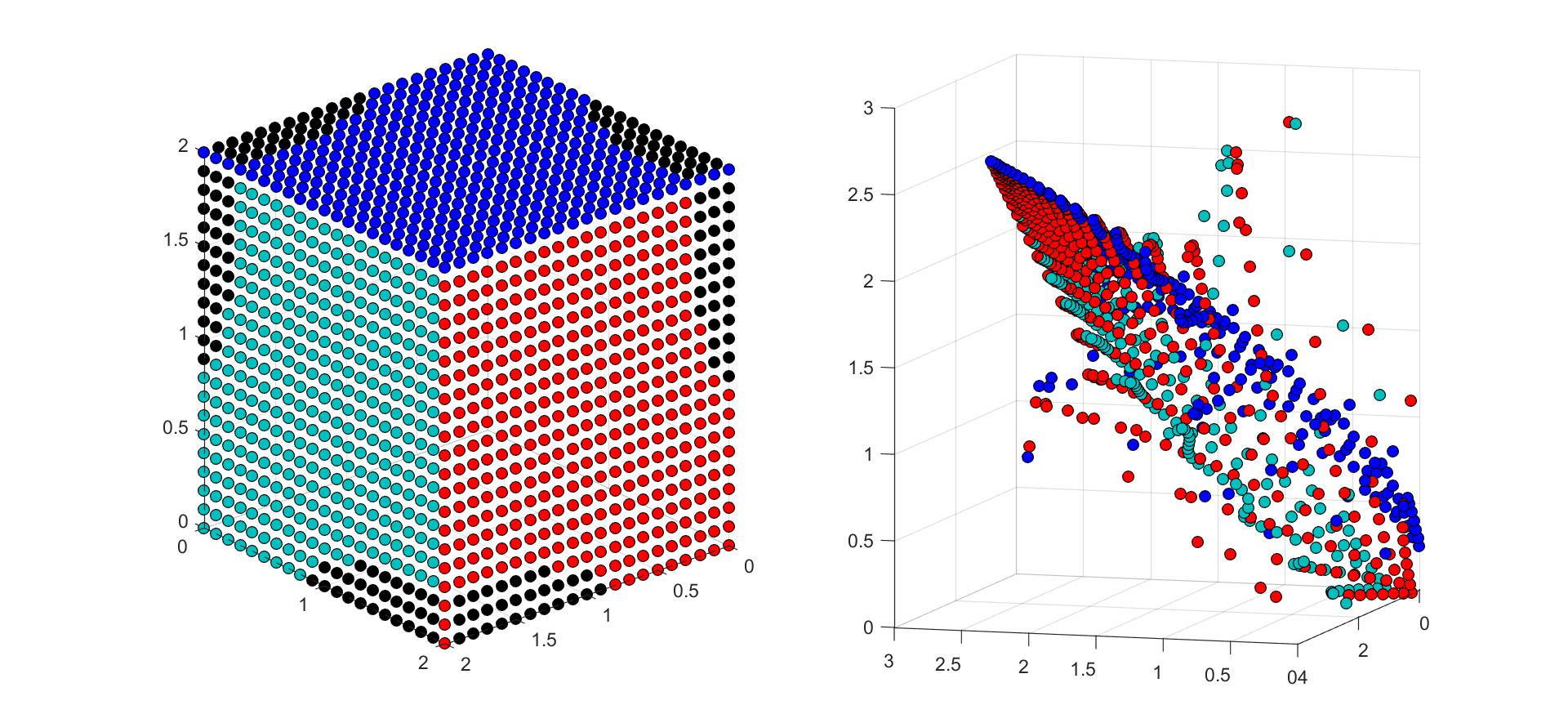}
\end{center}
\caption{Simulations showing the homeomorphism of Proposition \protect \ref%
{prop:homeo2} for the case of 3 nodes, as discussed in Example \protect \ref%
{ex:line_complete_blackpoint}. The left panel shows vectors of conjectured
ratios. With respect to Figure \protect \ref{fig:hom}, we allow for higher
values of conjectured ratios. Black dots represent cases for which the
learning dynamics diverge. The right panel shows the corresponding SCE
conjecture profile $\hat{\mathbf{x}}$ when the network is a complete
triangle, and when the learning dynamics are converging. }
\label{fig:hom2}
\end{figure}

\begin{example}
\label{ex:line_complete_blackpoint} We replicate the same exercise that we
did in Example \ref{ex:line_complete}, but only for the case of the complete
triangle. However we do it for a wider range of conjectured ratios. Figure %
\ref{fig:hom2} shows that in this case there may be combinations of
conjectured ratios that prevent convergence of the learning paths to
interior equilibria.
\end{example}

\section{Conclusion}

\label{sec:conclusion}

In this paper we offer a novel approach to network games. A key application
of network games is in modelling large societies with millions of nodes and
non regular distributions of connections. It is natural to assume that
players may ignore the complete structure of the network; this prevents them
from performing sophisticated strategic reasoning possibly leading to a Nash
equilibrium. Instead, they just best respond to some subjective beliefs
affected by the information feedback they receive. We analyze simple
conjectural best-reply paths and show that in some cases they converge to
stable Nash equilibria. However, we also characterize those situations in
which stable action profiles are not Nash equilibria, but rather
selfconfirming equilibrium action profiles in which some (if not \emph{all})
players have wrong beliefs and yet the feedback they receive is consistent
with such beliefs. We also show that, in the presence of global
externalities, simple biases in the perception of position in the network
may lead players to play action profiles that are far from the Nash
equilibria of the game.

One natural application of this approach is to directed online social
platforms like {\ Twitter {\ and Instagram}, where links need not to be
reciprocated. Using a linear--quadratic structure for the payoff function we
have also laid the ground for a tractable welfare analysis of the model.
However, policy implications are not straightforward if we want to consider
the long run benefits of connections and not only the instantaneous payoffs
of the users of those platforms. }

The analysis in {Appendix} \ref{ration} also provides a first
account of the strategic reasoning that agents can perform given some
commonly known features of the network. For example, we use known results
about rationalizability to show that, if the network game has strategic
complementarities, there is common knowledge of the game, and the Nash
equilibrium is unique, then sophisticated strategic reasoning leads to the
unique NE, whereas the results differ when actions are strategic substitutes.%
\footnote{%
On rationalizability in nice games with strategic complementarities see,
e.g., Chapter 5 of \cite{battigalli2018} and the references therein.} 
%
%
%

\newpage

\appendix

\section{Selfconfirming equilibria in parameterized nice games with
aggregators}

\label{app:SCE}


In this section we develop a more general analysis of selfconfirming
equilibria in a class of games that contains the linear-quadratic network
games with just observable payoffs studied in the main text. To ease
reading, we make this section self-contained, repeating some definitions
from the main text. We write this section focusing on local externalities.
This because the analysis that follows mainly concerns best-replys, that are
not affected by the presence of global externalities, so that all the
considerations about best-replys in this section also apply to the case of
the presence of global externalities.

A \textbf{parameterized nice game with aggregators and feedback} is a
structure%
\begin{equation*}
G=\left \langle I,\mathcal{Z} ,\left( A_{i},\ell _{i},v_{i},f_{i}\right)
_{i\in I}\right \rangle
\end{equation*}%
where

\begin{itemize}
\item $I$ is the finite \textbf{players} \textbf{set}, with cardinality $%
n=\left \vert I\right \vert $ and generic element $i$.

\item $\mathcal{Z} \subseteq \mathbb{R}^{m}$ is a \emph{compact} \textbf{%
parameter space}.

\item $A_{i}=\left[ 0,\bar{a}_{i}\right] \subseteq \mathbb{R}_{+}$, a \emph{%
compact interval}, is the \textbf{action space} of player $i$ with generic
element $a_{i}\in A_{i}$.

\item $X_{i}=\left[ \underline{x}_{i},\bar{x}_{i}\right] \subseteq \mathbb{R}
$, a \emph{compact interval}, is the \textbf{space} of \textbf{payoff states}
for $i$.

\item $\ell _{i}:\mathbf{A}_{-i}\times \mathcal{Z}\rightarrow X_{i}$ (where $%
\mathbf{A}_{-i}=\times _{j\in I\backslash \{i\}}A_{j}$) is a \emph{continuous%
} parameterized \textbf{aggregator} of the actions of $i$'s co-players such
that its \emph{range} $\ell _{i}\left( \mathbf{A}_{-i}\times \mathcal{Z}%
\right) $ is \emph{connected}.\footnote{%
Since the range of each section $\ell _{i,\mathbf{Z}}$ must be a compact
interval, we require that the union of the compact intervals $\ell _{i,%
\mathbf{Z}}\left( \mathbf{A_{-i}}\right) $ ($\mathbf{Z}\in \mathcal{Z}$) is
also an interval, which must be compact because $\mathcal{Z}$ is compact and 
$\ell _{i}$ continuous.} 

\item $v_{i}:A_{i}\times X_{i}\rightarrow \mathbb{R}$ is the \textbf{utility
function} of player $i$, which is \emph{strictly quasi-concave }in $a_{i}$
and \emph{continuous,}\footnote{%
That is, $v_{i}$ is jointly continuous in $\left( a_{i},x_{i}\right) $ and,
for each $x_{i}\in \left[ \underline{x}_{i},\bar{x}_{i}\right] $, the
section $v_{i,x_{i}}:\left[ 0,\bar{a}_{i}\right] \rightarrow \mathbb{R}$ has
a unique maximizer $a_{i}^{\ast }$ (that typically depends on $x_{i}$), it
is strictly increasing on $\left[ 0,a_{i}^{\ast }\right] $, and it is
strictly decreasing on $\left[ a_{i}^{\ast },\bar{a}_{i}\right] $. Of
course, the monotonicity requirement holds vacuously when the relevant
sub--interval is a singleton.} and from which we derive the \textbf{%
parameterized payoff function}%
\begin{equation*}
\begin{tabular}{llll}
$u_{i}:$ & $A_{i}\times \mathbf{A}_{-i}\times \mathcal{Z}$ & $\rightarrow $
& $\mathbb{R}$, \\ 
& $\  \  \  \left( a_{i},\mathbf{a}_{-i},\mathbf{Z}\right) $ & $\mapsto $ & $%
v_{i}\left( a_{i},\ell _{i}\left( \mathbf{a}_{-i},\mathbf{Z}\right) \right) $%
.%
\end{tabular}%
\end{equation*}%
Thus, $x_{i}=\ell _{i}\left( \mathbf{a}_{-i},\mathbf{Z}\right) $ is the
payoff--relevant state that $i$ has to guess in order to choose a
subjectively optimal action. With this, for each $\mathbf{Z}\in \mathcal{Z}$%
, $\left \langle I,\left( A_{i},u_{i,\mathbf{Z}}\right) _{i\in
I}\right
\rangle $ is a nice game (cf. \citealt{moulin1984dominance}), and $%
\left
\langle I,\mathcal{Z},\left( A_{i},u_{i}\right) _{i\in
I}\right
\rangle $ is a \textbf{parameterized nice game}. We let 
\begin{equation*}
\begin{tabular}{llll}
$r_{i}:$ & $X_{i}$ & $\rightarrow $ & $A_{i}$ \\ 
& $x_{i}$ & $\mapsto $ & $\arg \displaystyle \max_{a_{i}\in
A_{i}}v_{i}\left( a_{i},x_{i}\right) $%
\end{tabular}%
\end{equation*}%
denote the \textbf{best-reply function} of player $i$. The Maximum theorem
implies that $r_{i}$ is continuous.

\item Let $M\subseteq \mathbb{R}$ be a set of \textquotedblleft
messages,\textquotedblright \ $f_{i}:A_{i}\times X_{i}\rightarrow M$ is a 
\emph{continuous} \textbf{feedback function} that describes what $i$
observes (a \textquotedblleft message,\textquotedblright \ e.g., a monetary
outcome) after taking any action $a_{i}$ given any payoff state $x_{i}$.
\end{itemize}

On top of the formal assumptions stated above, we maintain the following 
\emph{minimal} \emph{informal assumption} about players' knowledge of the
game:

\begin{itemize}
\item Each player $i$ \emph{knows} $v_{i}$ \emph{and} $f_{i}$.
\end{itemize}

Unless we explicitly say otherwise, we instead do not assume that $i$
necessarily knows $\mathbf{Z}$, or function $\ell _{i}$, or even that $i$
understands that her payoff is affected by the actions of other players.
However, since $i$ knows the feedback function $f_{i}:A_{i}\times
X_{i}\rightarrow M$ and the action she takes, what $i$ infers about the
payoff state $x_{i}$ after she has taken action $a_{i}$ and observed message 
$m$ is that 
\begin{equation*}
x_{i}\in f_{i,a_{i}}^{-1}\left( m\right) :=\left \{ x_{i}^{\prime
}:f_{i}\left( a_{i},x_{i}^{\prime }\right) =m\right \} .
\end{equation*}

\subsection{Conjectures}

\label{app:SCEconj}

If player $i$ only knows the feedback function $f_{i}$, but does not know
how the payoff state $x_{i}$ is determined, then she just forms a conjecture
about $x_{i}$. If instead $i$ knows that $x_{i}$ is determined by the
actions of others given parameter $\mathbf{Z}$ through the aggregator $\ell
_{i}$, then $i$ forms a conjecture about $\left( \mathbf{a}_{-i},\mathbf{Z}%
\right) $.

\begin{definition}
\label{def:conj} A \textbf{shallow conjecture }for $i\in I$ is a probability
measure $\mu _{i}\in \Delta \left( X_{i}\right) $. A \textbf{deep} \textbf{%
conjecture} for $i$ is a probability measure $\bar{\mu}_{i}\in \Delta \left( 
\mathbf{A}_{-i}\times \mathcal{Z}\right) $. An action $a_{i}^{\ast }$ is 
\textbf{justifiable }if there exists a shallow conjecture $\mu _{i}$ such
that%
\begin{equation*}
a_{i}^{\ast }\in \argmax_{a_{i}\in A_{i}}\int_{X_{i}}v_{i}\left(
a_{i},x_{i}\right) \mu _{i}\left( \mathrm{d}x_{i}\right) \text{;}
\end{equation*}%
in this case we say that $\mu _{i}$ \textbf{justifies }$a_{i}^{\ast }$.
Similarly, we say that deep conjecture $\bar{\mu}_{i}\in \Delta \left( 
\mathbf{A}_{-i}\times \mathcal{Z}\right) $ \textbf{justifies} $a_{i}^{\ast }$
if the shallow conjecture induced by $\bar{\mu}_{i}$ ($\mu _{i}=\bar{\mu}%
_{i}\circ \ell _{i}^{-1}\in \Delta \left( X_{i}\right) $) justifies $%
a_{i}^{\ast }$.
\end{definition}

The following lemma summarizes well known results about nice games (see,
e.g., \citealp{battigalli2018}) and some straightforward consequences for
the more structured class of nice games with aggregators considered here. We
include the proof to make the exposition self-contained.

\begin{lemma}
\label{Lemma:nice BR}The best-reply function $r_{i}:X_{i}\rightarrow A_{i}$
is continuous, hence its range $r_{i}\left( X_{i}\right) $ is a compact
interval, just like $X_{i}$. Furthermore, for each $a_{i}^{\ast }\in A_{i}$,
the following are equivalent:

\begin{itemize}
\item $a_{i}^{\ast }$ is justifiable,

\item $a_{i}^{\ast }\in r_{i}\left( X_{i}\right) $ (that is, $a_{i}^{\ast }$
is justified by a deterministic shallow conjecture),

\item there is no $a_{i}$ such that $v_{i}\left( a_{i}^{\ast },x_{i}\right)
<v_{i}\left( a_{i},x_{i}\right) $ for all $x_{i}\in X_{i}$ (that is, $%
a_{i}^{\ast }$ is not dominated by any other pure action).
\end{itemize}
\end{lemma}

\noindent \textbf{Proof.} With a slight abuse of notation, we let $%
r_{i}\left( \mu _{i}\right) $ denote the set of best replies to (shallow)
conjecture $\mu _{i}$:%
\begin{equation*}
r_{i}\left( \mu _{i}\right) :=\arg \max_{a_{i}\in
A_{i}}\int_{X_{i}}v_{i}\left( a_{i},x_{i}\right) \mu _{i}\left( \mathrm{d}%
x_{i}\right) \text{.}
\end{equation*}%
By the Maximum theorem $\mu _{i}\mapsto r_{i}\left( \mu _{i}\right) $ has a
closed graph, which---under the stated assumptions---is equivalent to upper
hemi-continuity. By strict quasi-concavity, the restriction of the
best-reply correspondence to the domain $X_{i}$ of deterministic conjectures
is single-valued; hence, it must be a continuous function.

Fix any \emph{closed }(hence, compact)\emph{\ }sub-\emph{interval} $%
C\subseteq X_{i}$. Let $ND_{i,p}\left( C\right) $ denote the set of \textbf{%
actions that are not strictly dominated by other pure actions.} By
inspection of the definitions, it holds that%
\begin{equation*}
r_{i}\left( C\right) \subseteq r_{i}\left( \Delta \left( C\right) \right)
\subseteq ND_{i,p}\left( C\right) \text{.}
\end{equation*}%
We prove that $ND_{i,p}\left( C\right) \subseteq r_{i}\left( C\right) $,
that is, $A_{i}\backslash r_{i}\left( C\right) \subseteq A_{i}\backslash
ND_{i,p}\left( C\right) $, which therefore implies the thesis. Since $r_{i}$
is a continuous function on $X_{i}\supseteq C$ and $C$ is compact and
connected, $r_{i}\left( C\right) $ is compact and connected as well, hence,
it is a compact interval. Therefore, it is enough to show that all the
actions below $\min r_{i}\left( C\right) $ or above $\max r_{i}\left(
C\right) $ are dominated. Fix any $a_{i}<\min r_{i}\left( C\right) $, by
strict quasi-concavity,%
\begin{equation*}
\forall x_{i}\in C\text{, }v_{i}\left( a_{i},x_{i}\right) <v_{i}\left( \min
r_{i}\left( C\right) ,x_{i}\right) \leq v_{i}\left( r_{i}\left( x_{i}\right)
,x_{i}\right) \text{.}
\end{equation*}%
Therefore, every $a_{i}<\min r_{i}\left( C\right) $ is strictly dominated by 
$\min $ $r_{i}\left( C\right) $. A similar argument shows that every $%
a_{i}>\max r_{i}\left( C\right) $ is strictly dominated by $\max r_{i}\left(
C\right) $. Since there are no other actions outside $r_{i}\left( C\right) $%
, this concludes the proof.\hfill $\blacksquare $

\begin{corollary}
\label{Cor: J by shallow vs deep}Suppose that the aggregator $\ell _{i}$ is 
\emph{onto}. Then, an action of player $i$ is justifiable if an only if it
is justified by a deterministic (Dirac) deep conjecture.
\end{corollary}

\noindent \textbf{Proof.} The \textquotedblleft if\textquotedblright \ part
is trivial. For the \textquotedblleft only if\textquotedblright \ part, fix
a justifiable action $a_{i}^{\ast }$ arbitrarily. By Lemma \ref{Lemma:nice
BR}, there is some $x_{i}\in X_{i}$ such that $a_{i}^{\ast }=r_{i}\left(
x_{i}\right) $. Since the aggregator $\ell_{i}$ is onto, there is some $%
\left( \mathbf{a}_{-i},\mathbf{Z} \right) \in \ell_{i}^{-1}\left(
x_{i}\right) $ such that%
\begin{equation*}
a_{i}^{\ast }\in \arg \max_{a_{i}\in A_{i}}u_{i}\left( a_{i},\mathbf{a}_{-i},%
\mathbf{Z} \right) \text{.}
\end{equation*}%
Hence $a_{i}^{\ast }$ is justified by the deep conjecture $\delta _{\left( 
\mathbf{a}_{-i},\mathbf{Z} \right) }$, that is, the Dirac measure supported
by $\left( \mathbf{a}_{-i},\mathbf{Z} \right) $. \hfill $\blacksquare $

\bigskip

With this, from now on we mostly restrict our attention to (shallow, or
deep) \emph{deterministic conjectures}.

\subsection{Feedback properties}

\label{app:SCEfeed}

\begin{definition}
\label{Def: OP}Feedback $f_{i}$ satisfies \textbf{observable payoffs }(OP)
relative to $v_{i}$ if there is a function $\bar{v}_{i}:A_{i}\times
M\rightarrow \mathbb{R}$ such that%
\begin{equation*}
v_{i}\left( a_{i},x_{i}\right) =\bar{v}_{i}\left( a_{i},f_{i}\left(
a_{i},x_{i}\right) \right)
\end{equation*}%
for all $\left( a_{i},x_{i}\right) \in A_{i}\times X_{i}$; if the section $%
\bar{v}_{i,a_{i}}$ is injective for each $a_{i}\in A_{i}$, then we say that $%
f_{i}$ satisfies \textbf{just observable payoffs }(JOP) relative to $v_{i}$.
Game $G$ satisfies (just) observable payoffs if, for each player $i\in I$,
feedback $f_{i}$ satisfies (J)OP relative to $v_{i}$.
\end{definition}

If $f_{i}$ satisfies JOP, we may assume without loss of generality that $%
f_{i}=v_{i}$, because, for each action $a_{i}$, the partitions of $X_{i}$
induced by the preimages of $v_{i,a_{i}}$ and $f_{i,a_{i}}$ coincide:

\begin{remark}
\label{Rem: IOP is v=f}Feedback $f_{i}$ satisfies JOP relative to $v_{i}$ if
and only if%
\begin{equation}
\forall a_{i}\in A_{i}\text{, }\left \{ v_{i,a_{i}}^{-1}\left( u\right)
\right \} _{u\in v_{i,a_{i}}\left( X_{i}\right) }=\left \{
f_{i,a_{i}}^{-1}\left( m\right) \right \} _{m\in f_{i,a_{i}}\left(
X_{i}\right) }\text{.}  \label{eq:same part}
\end{equation}
\end{remark}

\noindent \textbf{Proof. }(Only if) Fix $a_{i}\in A_{i}$. Since $f_{i}$
satisfies JOP relative to $v_{i}$, $v_{i,a_{i}}\left( X_{i}\right) =\left( 
\bar{v}_{i,a_{i}}\circ f_{i,a_{i}}\right) \left( X_{i}\right) $ (by OP), for
each $u\in v_{i,a_{i}}\left( X_{i}\right) $ there is a unique message $%
m_{a_{i},u}=\bar{v}_{i,a_{i}}^{-1}\left( u\right) $ (by injectivity of $\bar{%
v}_{i,a_{i}} $), and 
\begin{eqnarray*}
v_{i,a_{i}}^{-1}\left( u\right) &=&\left \{ x_{i}\in X_{i}:v_{i}\left(
a_{i},x_{i}\right) =u\right \} \\
&=&\left \{ x_{i}\in X_{i}:\bar{v}_{i}\left( a_{i},f_{i}\left(
a_{i},x_{i}\right) \right) =u\right \} \\
&=&\left \{ x_{i}\in X_{i}:f_{i}\left( a_{i},x_{i}\right) =m_{a_{i},u}\right
\} =f_{i,a_{i}}^{-1}\left( m_{a_{i},u}\right) \text{,}
\end{eqnarray*}%
which implies eq. (\ref{eq:same part}).

(If) Suppose that eq. (\ref{eq:same part}) holds. For every $a_{i}\in A_{i}$
and $m\in f_{i,a_{i}}\left( X_{i}\right) $ select some $\xi _{i}\left(
a_{i},m\right) \in f_{i,a_{i}}^{-1}\left( m\right) $. Let 
\begin{equation*}
D:=\bigcup_{a_{i}\in A_{i}}\left \{ a_{i}\right \} \times f_{i,a_{i}}\left(
X_{i}\right).
\end{equation*}%
With this,%
\begin{equation*}
\xi _{i}:D\rightarrow X_{i}
\end{equation*}%
is a well defined function. Domain $D$ is the set of action-message pairs
for which the definition of $\bar{v}_{i}$ matters. Define $\bar{v}_{i}$ as
follows:%
\begin{equation*}
\bar{v}_{i}\left( a_{i},m\right) =\left \{ 
\begin{tabular}{ll}
$v_{i}\left( a_{i},\xi _{i}\left( a_{i},m\right) \right) $ & if $\left(
a_{i},m\right) \in D$, \\ 
$0$ & otherwise.%
\end{tabular}%
\right.
\end{equation*}%
By construction, eq. (\ref{eq:same part}) implies that 
\begin{equation*}
\forall \left( a_{i},x_{i}\right) \in A_{i}\times X_{i}\text{, }\bar{v}%
_{i}\left( a_{i},f_{i}\left( a_{i},x_{i}\right) \right) =v_{i}\left(
a_{i},x_{i}\right) \text{.}
\end{equation*}%
Hence, OP holds. Furthermore, for all $a_{i}\in A_{i}$, $m^{\prime
},m^{\prime \prime }\in f_{a_{i}}\left( X_{i}\right) $,%
\begin{eqnarray*}
m^{\prime } \neq m^{\prime \prime }\Rightarrow &\xi _{i}\left(
a_{i},m^{\prime }\right) \neq \xi _{i}\left( a_{i},m^{\prime \prime }\right)
\\
\Rightarrow & v_{i}\left( a_{i},\xi _{i}\left( a_{i},m^{\prime }\right)
\right) \neq v_{i}\left( a_{i},\xi _{i}\left( a_{i},m^{\prime \prime
}\right) \right) \\
\Rightarrow & \bar{v}_{i}\left( a_{i},m^{\prime }\right) \neq \bar{v}%
_{i}\left( a_{i},m^{\prime \prime }\right)
\end{eqnarray*}%
where the first and the second implications follow from eq. (\ref{eq:same
part}) ($\xi _{i}\left( a_{i},m^{\prime }\right) $ and $\xi _{i}\left(
a_{i},m^{\prime \prime}\right) $ belong to different cells of the coincident
partitions, hence yield different utilities), and the third holds by
construction. Therefore, $\bar{v}_{i,a_{i}}$ is injective for every $a_{i}$,
which means that JOP holds.\hfill $\blacksquare $

\begin{definition}
\label{Def: OA}Feedback $f_{i}$ satisfies \textbf{observability if and only
if }$i$ \textbf{is active} (OiffA) if section $f_{i,a_{i}}$ is injective for
each $a_{i}>0$ and constant for $a_{i}=0$. Game $G$ satisfies \textbf{%
observability by active players} if OiffA holds for each $i$.
\end{definition}

\begin{remark}
\label{Rem: LQ JOP imply OAP}If a network game is \emph{linear-quadratic}
and satisfies \emph{just observable payoffs}, then it satisfies
observability by active players.
\end{remark}

\noindent \textbf{Proof.} By Remark \ref{Rem: IOP is v=f} JOP implies that,
for each $a_{i}\in A_{i}$,%
\begin{equation*}
\left \{ v_{i,a_{i}}^{-1}\left( u\right) \right \} _{u\in v_{i,a_{i}}\left(
X_{i}\right) }=\left \{ f_{i,a_{i}}^{-1}\left( m\right) \right \} _{m\in
f_{i,a_{i}}\left( X_{i}\right) }\text{.}
\end{equation*}%
The linear-quadratic form of $v_{i}$ implies that, for every $x_{i}\in X_{i}$%
,%
\begin{equation*}
v_{i,0}^{-1}\left( v_{i,0}\left( x_{i}\right) \right) =X_{i} \ ,
\end{equation*}
\begin{equation*}
\forall a_{i}>0,v_{i,a_i}^{-1}\left( v_{i,a_i}\left( x_{i}\right) \right)
=\left \{ x_{i}\right \} \text{.}
\end{equation*}%
These equalities imply that $f_{i,0}$ is constant and $f_{i,a_{i}}$ is
injective for $a_{i}>0$, that is, $NG$ satisfies observability by active
players.\hfill $\blacksquare $

\begin{definition}
\label{Def: OAI}Feedback $f_{i}$ satisfies \textbf{own-action independence}
(OAI) of feedback about the state if, for all \emph{justifiable} actions $%
a_{i}^{\ast },a_{i}^{o}$ and all payoff states $\hat{x}_{i},x_{i}$,%
\begin{equation*}
f_{i}\left( a_{i}^{\ast },\hat{x}_{i}\right) =f_{i}\left( a_{i}^{\ast
},x_{i}\right) \Rightarrow f_{i}\left( a_{i}^{o},\hat{x}_{i}\right)
=f_{i}\left( a_{i}^{o},x_{i}\right) \text{.}
\end{equation*}%
Game $G$ satisfies own-action independence of feedback about the state if,
for each player $i\in I$, feedback $f_{i}$ satisfies OAI.
\end{definition}

In other words, OAI says that if player $i$ cannot distinguish between two
payoff states $\hat{x}_{i}$ and $x_{i}$ when she chooses some given
justifiable action $a_{i}^{\ast }$, then she cannot distinguish between
these two states when he chooses any other justifiable action $a_{i}^{o}$.
This is equivalent to requiring that the partitions of $X_{i}$ of the form $%
\left \{ f_{i,a_{i}}^{-1}\left( m\right) \right \} _{m\in f_{i,a_{i}}\left(
X_{i}\right) }$ coincide across justifiable actions, i.e.~across actions $%
a_{i}\in r_{i}\left( X_{i}\right) $ (see Lemma \ref{Lemma:nice BR}).

The following lemma---which holds for any game, not just nice games---states
that, under payoff observability and own-action independence, an action is
justified by a confirmed conjecture if and only if it is a best reply to the
actual payoff state:

\begin{lemma}
\label{Lemma: subj-obj BR}If $f_{i}$ satisfies\emph{\ observable payoffs}
relative to $v_{i}$ and \emph{own-action independence} of feedback about the
state, then for all $\left( a_{i}^{\ast },x_{i}\right) \in A_{i}\times X_{i}$
the following are equivalent:

\begin{enumerate}
\item there is some $\hat{x}_{i}\in X_{i}$ such that $a_{i}^{\ast }\in \arg
\max_{a_{i}\in A_{i}}v_{i}\left( a_{i},\hat{x}_{i}\right) $ and $f_{i}\left(
a_{i}^{\ast },\hat{x}_{i}\right) =f_{i}\left( a_{i}^{\ast },x_{i}\right) $,

\item $a_{i}^{\ast }\in \arg \max_{a_{i}\in A_{i}}v_{i}\left(
a_{i},x_{i}\right) $.
\end{enumerate}
\end{lemma}

\noindent \textbf{Proof. }(Cf. \citealp{battigalli2015self}, %
\citealp{battigalli2018}) It is obvious that 2 implies 1 independently of
the properties of $f_{i}$. To prove that 1 implies 2 under the stated
assumptions, suppose that $f_{i}$ satisfies OP-OAI and let $\hat{x}_{i}$ be
such that 1 holds. Let $a_{i}^{o}$ be a best reply to the actual state $%
x_{i} $. We must show that also $a_{i}^{\ast }$ is a best reply to $x_{i}$.
Note that both $a_{i}^{\ast }$ and $a_{i}^{o}$ are justifiable; hence, by
OAI, $f_{i}\left( a_{i}^{\ast },\hat{x}_{i}\right) =f_{i}\left( a_{i}^{\ast
},x_{i}\right) $ implies $f_{i}\left( a_{i}^{o},\hat{x}_{i}\right)
=f_{i}\left( a_{i}^{o},x_{i}\right) $. Using OP, condition 1, and OAI as
shown in the following chain of equalities and inequalities, we obtain%
\begin{equation*}
v_{i}\left( a_{i}^{\ast },x_{i}\right) \overset{(\mathrm{OP})}{=}\bar{v}%
_{i}\left( a_{i}^{\ast },f_{i}\left( a_{i}^{\ast },x_{i}\right) \right) 
\overset{(\mathrm{1})}{=}\bar{v}_{i}\left( a_{i}^{\ast },f_{i}\left(
a_{i}^{\ast },\hat{x}_{i}\right) \right) \overset{(\mathrm{OP})}{=}%
v_{i}\left( a_{i}^{\ast },\hat{x}_{i}\right) \overset{(\mathrm{1})}{\geq }
\end{equation*}%
\begin{equation*}
v_{i}\left( a_{i}^{o},\hat{x}_{i}\right) \overset{(\mathrm{OP})}{=}\bar{v}%
_{i}\left( a_{i}^{o},f_{i}\left( a_{i}^{o},\hat{x}_{i}\right) \right) 
\overset{(\mathrm{1,OAI})}{=}\bar{v}_{i}\left( a_{i}^{o},f_{i}\left(
a_{i}^{o},x_{i}\right) \right) \overset{(\mathrm{OP})}{=}v_{i}\left(
a_{i}^{o},x_{i}\right) \text{.}
\end{equation*}%
Since $a^{o}$ is a best reply to $x_{i}$ and $v_{i}\left( a_{i}^{\ast
},x_{i}\right) \geq v_{i}\left( a_{i}^{o},x_{i}\right) $, it must be the
case that also $a_{i}^{\ast }$ is a best reply to $x_{i}$.\hfill $%
\blacksquare $

\begin{corollary}
\label{Cor: SCE is NE}Suppose that the parameterized nice game with
aggregators and feedback $G$ satisfies \emph{observable payoffs} and \emph{%
own-action independence} of feedback about the state, then the sets of
selfconfirming action profiles and Nash equilibrium action profiles coincide
for each $\mathbf{Z}$:%
\begin{equation*}
\forall \mathbf{Z}\in \mathcal{Z}\text{, }\mathbf{A}_{\mathbf{Z}}^{SCE}=%
\mathbf{A}_{\mathbf{Z}}^{NE}\text{.}
\end{equation*}
\end{corollary}

\noindent \textbf{Proof} By Remark \ref{Rem: NE is SCE}, we only have to
show that $\mathbf{A}_{\mathbf{Z} }^{SCE}\subseteq \mathbf{A}_{\mathbf{Z}
}^{NE}$. Fix any $\mathbf{a}^{\ast }=\left( a_{i}^{\ast }\right) _{i\in
I}\in \mathbf{A}_{\mathbf{Z} }^{SCE}$ and any player $i$. By definition of
SCE and by Lemma \ref{Lemma:nice BR}, there is some $\hat{x}_{i}\in X_{i}$
such that $a_{i}^{\ast }\in r_{i}\left( \hat{x}_{i} \right) $ and $%
f_{i}\left( a_{i}^{\ast },\hat{x}_{i}\right) =f_{i}\left( a_{i}^{\ast
},\ell_{i}\left( \mathbf{a}_{-i}^{\ast },\mathbf{Z} \right) \right) $. By
Lemma \ref{Lemma: subj-obj BR} $a_{i}^{\ast }\in r_{i}\left( \ell_{i}\left( 
\mathbf{a}_{-i}^{\ast },\mathbf{Z} \right) \right) $. This holds for each $i$%
, hence $\mathbf{a}^{\ast }\in \mathbf{A}_{\mathbf{Z} }^{NE}$.\hfill $%
\blacksquare $

\bigskip

Corollary \ref{Cor: SCE is NE} provides sufficient conditions for the
equivalence between SCE and NE action profiles. Next, we give sufficient
conditions that allow a characterization of $\mathbf{A}_{\mathbf{Z} }^{SCE}$
by means of Nash equilibria of auxiliary games. %

\subsection{Equilibrium characterization}

\label{app:SCEchar}

If $a_{i}\in \lbrack 0,\bar{a}_{i}]$ is interpreted as an activity level
(e.g.,~effort) by player $i$, then it makes sense to say that $i$ is \textbf{%
active} if $a_{i}>0$ and \textbf{inactive }otherwise. Let $I_{0}$ denote the 
\textbf{set of players for whom being inactive is justifiable}. Note that,
by Lemma \ref{Lemma:nice BR},%
\begin{equation*}
I_{0}=\left \{ i\in I:\min r_{i}\left( X_{i}\right) =0\right \} \text{.}
\end{equation*}%
Also, for each $\mathbf{Z}\in \mathcal{Z}$ and non-empty subset of players $%
J\subseteq I$, let $\mathbf{A}_{J,\mathbf{Z}}^{NE}$ denote the set of Nash
equilibria of the auxiliary game with players set $J$ obtained by letting $%
a_{i}=0$ for each $i\in I\backslash J$, that is,%
\begin{equation*}
\mathbf{A}_{J,\mathbf{Z}}^{NE}=\left \{ \mathbf{a}_{J}^{\ast }\in \times
_{j\in J}A_{j}:\forall j\in J,a_{j}^{\ast }=r_{j}\left( \ell _{j}\left( 
\mathbf{a}_{J\backslash \{j\}}^{\ast },\mathbf{0}_{I\backslash J},\mathbf{Z}%
\right) \right) \right \} \text{,}
\end{equation*}%
where $\mathbf{0}_{I\backslash J}\in \mathbb{R}^{I\backslash J}$ is the
profile that assigns $0$ to each $i\in I\backslash J$. If $J=\emptyset $,
let $\mathbf{A}_{J,\mathbf{Z}}^{NE}=\left \{ \varnothing \right \} $ by
convention, where $\varnothing $ is the pseudo-action profile such that $%
\left( \varnothing ,\mathbf{0}_{I}\right) =\mathbf{0}_{I}$.

\begin{lemma}
\label{Lemma:SCE if OiffA}Suppose that the parameterized nice game with
aggregators and feedback $G$ satisfies \emph{observability by active players}%
. Then, for each $\mathbf{Z} $, the set of selfconfirming action profiles is%
\begin{equation*}
\mathbf{A}_{\mathbf{Z} }^{SCE}= \bigcup_{J:I\backslash J\subseteq I_{0}}%
\mathbf{A}_{J,\mathbf{Z} }^{NE}\times \left \{ \mathbf{0}_{I\backslash
J}\right \} \text{.}
\end{equation*}
\end{lemma}

\noindent \textbf{Proof} \ Fix $\mathbf{a}^{\ast }$ and let $J$ be the set
of players $i$ such that $a_{i}^{\ast }>0$. Fix $\mathbf{Z}\in \mathcal{Z}$
arbitrarily. Suppose that $\mathbf{a}^{\ast }\in $ $\mathbf{A}_{\mathbf{Z}%
}^{SCE}$ and fix any $i\in I$. If $a_{i}^{\ast }=0$, then $0$ is justifiable
for $i$, that is $i\in I_{0}$. If $a_{i}^{\ast }>0$, observability by active
players implies that $f_{i,a_{i}^{\ast }}$ is injective, that is, action $%
a_{i}^{\ast }$ reveals the payoff state, which implies that the (shallow)
conjecture justifying $a_{i}^{\ast }$ is correct: $a_{i}^{\ast }=r_{i}\left(
\ell _{i}\left( \mathbf{a}_{-i}^{\ast },\mathbf{Z}\right) \right) $. Hence $%
\mathbf{a}^{\ast}_J\in \mathbf{A}^{NE}_{J,\mathbf{Z}}$. Thus, $\mathbf{a}%
^{\ast }=\left( \mathbf{a}_{J}^{\ast },\mathbf{a}_{I\backslash J}^{\ast
}\right) $ is such that $a_{i}^{\ast }=0$ for each $i\in I\backslash
J\subseteq I_{0}$, and $a_{j}^{\ast }=r_{j}\left( \ell _{j}\left( \mathbf{a}%
_{J\backslash \{j\}}^{\ast },\mathbf{0}_{I\backslash J},\mathbf{Z}\right)
\right) >0$ for each $j\in J$. Hence,%
\begin{equation*}
\mathbf{a}^{\ast }=\left( \mathbf{a}_{J}^{\ast },\mathbf{a}_{I\backslash
J}^{\ast }\right) \in \mathbf{A}_{J,\mathbf{Z}}^{NE}\times \left \{ \mathbf{0%
}_{I\backslash J}\right \} \text{ with }I\backslash J\subseteq I_{0}\text{.}
\end{equation*}

Let $I\backslash J\subseteq I_{0}$ and $\left( \mathbf{a}_{J}^{\ast },%
\mathbf{a}_{I\backslash J}^{\ast }\right) \in \mathbf{A}_{J,\mathbf{Z}%
}^{NE}\times \left \{ \mathbf{0}_{I\backslash J}\right \} $. Since $G$
satisfies observability by active players, for each $i\in I\backslash J$,
any conjecture justifying $a_{i}^{\ast }=0$ (any $\hat{x}_{i}\in
r_{i}^{-1}\left( 0\right) $) is trivially confirmed. For each $j\in J$, $%
a_{j}^{\ast }>0$ is by assumption the best reply to the correct, hence
confirmed, shallow conjecture $\hat{x}_{j}=\ell _{i}\left( \mathbf{a}%
_{J\backslash \{j\}}^{\ast },\mathbf{0}_{I\backslash J},\mathbf{Z}\right) $.
Hence, $\left( \mathbf{a}_{J}^{\ast },\mathbf{a}_{I\backslash J}^{\ast
}\right) =\left( \mathbf{a}_{J}^{\ast },\mathbf{0}_{I\backslash J}\right)
\in \mathbf{A}_{\mathbf{Z}}^{SCE}$.\hfill $\blacksquare $


{\color{red} }

\section{Knowledge of the network and strategic reasoning}

\label{ration}The SCE concept does not rely, either explicitly or
implicitly, on strategic reasoning. Thus, some SCE's may be supported by
confirmed conjectures that are inconsistent with the assumption that other
agents are rational and think strategically. In this section we consider the
behavioral consequences of agents in a network game with feedback using the
common information they have about the network to reason strategically, thus
forming (deep) conjectures about the relevant unknowns, that is, actions of
others and parameters. We model strategic reasoning by means of the
assumption of common belief in rationality. Thus, we analyze which SCE's are
consistent with common belief in rationality, which may help in selecting
some SCE's when there is a multiplicity of equilibria (see Battigalli 1987,
Battigalli and Guaitoli 1997, Battigalli and Bordoli 2022). More
specifically, when agents have some information about the network, it is
reasonable to assume that they use it to determine how they should act.
Indeed, using one step of reasoning, every agent may try to infer which
actions her direct neighbors may play, restricting own conjectures
accordingly. Depending on the knowledge about the strategic interactions she
is exposed to, the agent determines the set of her own actions that are best
replies to conjectures satisfying said restrictions, i.e., consistent with
neighbors' rationality. Going further, she can take into account that her
neighbors actions should be best replies to conjectures consistent with the
rationality of her neighbors' neighbors, and so on. This yields a notion of
rationalizability of conjectures, and a corresponding definition of \textbf{%
selfconfirming equilibrium with rationalizable conjectures}, which is the
object of our analysis in this section. We obtain results for the cases
analyzed in the previous sections of the paper, that is positive local
externalities, arbitrary local externalities, and positive local
externalities joint with positive global externalities.

We can distinguish among different elements that can be the object of
knowledge: (i) the pure topological structure of the network (who is linked
with whom); (ii) the kind of local interaction (positive or negative local
externality) that operates on each link; (iii) the intensity of this
interaction. Here we focus on two extreme cases, common knowledge of the
network $\mathbf{Z}$, or common knowledge of only some aspects of the
network captured by the common exogenous uncertainty space $\mathbf{\mathcal{%
Z}}$, e.g., whether there are positive local externalities. Thus, we ignore
other intermediate cases that could be analyzed within our framework. In
particular, \emph{we ignore the possibility that agents have private
information about the network}, which simplifies the analysis. 

\paragraph{A characterization of SCE}

\label{sec:SCE} 
Here we simply characterize the set $\mathbf{A}_{\mathbf{Z}%
}^{SCE}$ of selfconfirming equilibrium action profiles at $\mathbf{Z}$. 

\begin{proposition}
\label{prop:SCE=NE} Consider a network game such that, for every $i\in I$
and for every $\hat{x}_{i}\in X_{i}$, $r_{i}\left( \hat{x}_{i}\right) >0$.
Then, for each $\mathbf{Z}\in \mathcal{Z}$, $\mathbf{A}_{\mathbf{Z}}^{SCE}=%
\mathbf{A}_{\mathbf{Z}}^{NE}$.\footnote{%
Given the stated assumptions about feedback, the same result holds also for
non--linear and continuous aggregators $\ell _{i}$ and continuous strictly
quasi-concave utility functions $v_{i}$.}
\end{proposition}

\textbf{Proof.} Since $NG$ is \emph{linear-quadratic} and satisfies \emph{%
just observable payoffs}, then it satisfies observability by active players.
Since being inactive is unjustifiable (dominated) for every player, \textbf{%
observability by active players} implies \emph{own-action independence of
the feedback about the state}. Then, the result follows from Corollary \ref%
{Cor: SCE is NE} in Appendix \ref{app:SCE}. \hfill $\blacksquare $

\paragraph{Knowledge and deep conjectures}

As defined in the previous sections, $\mathcal{Z}\subseteq \lbrack 
\underline{w},\bar{w}]^{I\times I}$ is the set of possible weighted
networks. Formally, we assume that the compact set $\mathcal{Z}$ is also 
\emph{connected}. Informally, we assume that $\mathcal{Z}$ \emph{is common
knowledge}, and that there is \emph{common knowledge of the payoff functions
parametrized by }$\mathbf{Z}$\emph{.} For the purposes of this analysis, we
consider two possible cases: i) $\mathcal{Z}=\left \{ \mathbf{Z}\right \} $,
so that the network is common knowledge; ii) $\mathcal{Z}\subseteq \lbrack 0,%
\bar{w}]^{I\times I}$, so that the network $\mathbf{Z}$ may be unknown, but
it is common knowledge that links are positive and bounded by $\bar{w}$.
Besides common knowledge of $\mathcal{Z}$, we have to consider \textbf{deep
conjectures}, that is, conjectures about the network $\mathbf{Z}$ and the
actions of other agents in the network. For each agent $i\in I$, deep
conjectures are defined as probability measures $\bar{\mu}_{i}\in \Delta (%
\mathbf{A}_{-i}\times \mathcal{Z})$ (see definition \ref{def:conj} in \ref%
{app:SCEconj}). Notice that, if $\mathcal{Z}$ is a singleton, the only
uncertainty agents have is about actions of others.

\paragraph{Rationalizability}

Given common knowledge of the parameterized game $\left \langle I,\mathcal{Z}%
,\left( A_{i},u_{i}\right) _{i\in I}\right \rangle $, we can characterize
the behavioral implications of \emph{rationality and common belief in
rationality (RCBR)}, i.e.,~the set of action profiles consistent with these
(so called) \textbf{epistemic assumptions}. A formal expression of these
epistemic assumptions and a characterization of their behavioral
implications in a class of games that contains those considered here is
given, for example, in \cite{battigalli2018interactive} 
\citep[see also][for
a more intuitive explanation]{battigalli2018}. In our setting, an action
profile is consistent with RCBR if and only if, given $\mathcal{Z}$, for
every $i\in I$, it survives the following procedure of iterated elimination
of non-best replies:

\begin{itemize}
\item $A_{i}^{0}=A_{i}$,

\item $A_{i}^{n+1}=\left \{ a_{i}^{\ast }\in A_{i}:\exists \bar{\mu} _{i}\in
\Delta (\mathbf{A}_{-i}^{n}\times \mathcal{Z}),\ a_{i}^{\ast }\in %
\displaystyle \arg \max_{a_{i}\in A_{i}}\mathbb{E}_{\bar{\mu} _{i}}\left[
u_{i}(a_{i},\cdot, \cdot )\right] \right \} $,

\item $A_{i}^{\infty }=\bigcap \limits_{n\in \mathbb{N}}A_{i}^{n}$.
\end{itemize}

\begin{definition}
\label{Def:R-zability}An action $a_{i}$ of player $i$ is\emph{\
rationalizable} if $a_{i}\in A_{i}^{\infty }$. A deep conjecture $\bar{\mu}%
_{i}$ of player $i$ is\emph{\ rationalizable} if $\bar{\mu}_{i}\in \Delta (%
\mathbf{A}_{-i}^{\infty }\times \mathcal{Z})$.
\end{definition}

\begin{remark}
For every $\mathbf{Z} \in \mathcal{Z}$ every Nash equilibrium at $\mathbf{Z}$
(every $\mathbf{a}^{*} \in \mathbf{A}_\mathbf{Z}^{NE}$) is a profile of
rationalizable actions.
\end{remark}

A compactness-continuity argument yields the following:

\begin{remark}
\label{Rem: rat action conj}An action is rationalizable if and only if it is
a best reply to a rationalizable conjecture.
\end{remark}

As we did for the case of shallow conjectures, for each agent $i\in I$, we
can restrict our attention to \emph{deterministic} deep conjectures $(%
\mathbf{\hat{a}}_{-i},\hat{\mathbf{Z}_{i}})\in \mathbf{A}_{-i}\times 
\mathcal{Z}$. We are allowed to use deterministic deep conjectures because,
for each $i\in I$, $\mathbf{A}_{-i}$ and $\mathcal{Z}$ are compact and
connected and thus, given the continuity of $u_{i}$ and strict
quasi-concavity of each section $u_{i,\mathbf{a}_{-i},\mathbf{Z}}$, for
every probabilistic deep conjecture there exists a deterministic deep
conjecture that delivers the same best reply (see Appendix \ref{app:SCE} and %
\citealp{battigalli2018}). This implies that if $\mathbf{A}_{-i}^{n}$ is
compact and connected, then $A_{i}^{n+1}$ is the compact interval of best
replies to deterministic conjectures over $\mathbf{A}_{-i}^{n}$ (see Lemma %
\ref{Lemma:nice BR}). With this, the following result follows from Lemma \ref%
{Lemma:nice BR} and a straightforward induction argument: each set $\mathbf{A%
}^{n}$ of $n$-rationalizable action profiles is a \textquotedblleft
box,\textquotedblright \ or \emph{order-interval}:

\begin{theorem}
\label{Th:R-zability nice}In a parametrized nice game with aggregators (with 
$\mathcal{Z}$ connected)%
\begin{equation*}
\mathbf{A}^{n}=\times _{i\in I}\left[ \min A_{i}^{n},\max A_{i}^{n}\right]
\end{equation*}%
for all $n\in \mathbb{N\cup }\left \{ \infty \right \} $.
\end{theorem}

\paragraph{Selfconfirming equilibrium with rationalizable conjectures}

Focusing on the case considered in the main body of the paper, \emph{%
linear-quadratic} network games with \emph{just observable payoffs}, we
refine the definition of selfconfirming equilibrium adding the requirement
of rationalizability of conjectures.

\begin{definition}
\label{def:WRSCE} A profile $\left( {a}_{i}^{\ast },\hat{\mathbf{a}}_{-i},%
\hat{\mathbf{Z}_{i}}\right) _{i\in I}\in \times _{i\in I}\left( A_{i}\times 
\mathbf{A}_{-i}\times \mathcal{Z}\right) $ of actions and deterministic deep
conjectures is a \textbf{selfconfirming equilibrium at} $\mathbf{Z}$ \textbf{%
with rationalizable conjectures} (SCER) of a game $G$ with \emph{just
observable payoffs} if, for each player $i\in I$,

\begin{enumerate}
\item \emph{(best reply)} ${\ a}_{i}^{\ast }\in r_{i}\left(\ell \left( \hat{%
\mathbf{a}}_{-i},\hat{\mathbf{Z}_{i}}\right) \right) $;

\item \emph{(confirmed conjectures, given just observable payoffs)} $%
u_{i}\left( {a}_{i}^{\ast },\hat{\mathbf{a}}_{-i},\hat{\mathbf{Z}_{i}}%
\right) =u_{i}\left( {a}_{i}^{\ast },\mathbf{\mathbf{a}}_{-i}^{\ast },%
\mathbf{Z}\right) $;

\item \emph{(rationalizable conjectures)} $(\hat{\mathbf{a}}_{-i},\hat{%
\mathbf{Z}_{i}} )\in \mathbf{A}_{-i}^{\infty }\times \mathcal{Z}$.
\end{enumerate}
\end{definition}

We denote by $\mathbf{A}_{\mathbf{Z,}\mathcal{Z}}^{SCER}$ the set of SCE
actions profiles at $\mathbf{Z}$ justified by rationalizable confirmed
conjectures, given the commonly known parameter space $\mathcal{Z}$.
Similarly, if $\mathcal{Z=}\left \{ \mathbf{Z}\right \} $, we let $\mathbf{A}%
_{\mathbf{Z}}^{SCER}$ denote the set of SCE actions profiles justified by
rationalizable confirmed conjectures, given the commonly known network $%
\mathbf{Z}$. Note, this is the set of action profiles consistent with the
following assumptions: (a) players are rational, (b) players' conjectures
are confirmed (given $\mathbf{Z}$), and (c) there is common belief of (a). A
stronger notion of \textquotedblleft rationalizable selfconfirming
equilibrium\textquotedblright \ for games with complete information %
\citep[due to][]{rubinstein1994rationalizable} is based on the following
assumptions: (a) players are rational, (b) players' conjectures are
confirmed, and (c*) there is common belief of (a) and (b).\footnote{%
See Esponda (2013) for games with incomplete information.} We limit our
analysis to the weaker SCER concept for two reasons: (i) it is simpler; (ii)
to our knowledge, there is no learning foundation of rationalizable SCE \`{a}
la Rubinstein and Wolinsky, whereas one can justify our concept by
considering learning paths like those analyzed in this paper, assuming that
players always hold rationalizable conjectures because there is common
belief in rationality. Note that such belief cannot ever be falsified by
what players observe, given that they best respond to rationalizable
conjectures, and therefore always choose rationalizable actions 
\citep[see][]{BB22}.

We now discuss how SCER actions are shaped depending on the type of
strategic interaction in the given network.

\paragraph{Positive local externalities}

The first case analyzed in the previous sections of the paper is when there
are local complementarities or mild substitutabilities. For simplicity of
exposition, we only consider the case of positive local externalities: if
the actual network is unknown, then $\mathcal{Z}\subseteq \lbrack 0,\bar{w}%
]^{I\times I}$ is not a singleton, if it is commonly known then $\mathcal{Z}%
=\left \{ \mathbf{Z}\right \} $ with $\mathbf{Z}\in \lbrack 0,\bar{w}%
]^{I\times I}$. Common knowledge of $\mathcal{Z}$ implies that $X_{i}=\ell
_{i}\left( \mathbf{A}_{-i}\times \mathcal{Z}\right) $ for each $i$. Thus,
the hypotheses of Proposition \ref{prop:SCE=NE} are satisfied, because $%
\underline{x}_{i}=0$ and $\min r_{i}\left( X_{i}\right) =r_{i}\left(
0\right) =\alpha _{i}>0$ for each $i$. Given positive local externalities,
it follows that the set of SCE action profiles at $\mathbf{Z}$ coincides
with the set of Nash equilibrium profiles at $\mathbf{Z}$ (Proposition \ref%
{prop:interiorSCE} provides sufficient conditions for uniqueness).
Consequently, adding rationalizability on top of the SCE requirements does
not change the result, because every Nash equilibrium action profile is
rationalizable.

\begin{corollary}
For every linear-quadratic network game with just observable payoffs, if $%
\mathbf{Z}\in \lbrack 0,\bar{w}]^{I\times I}$, then $\mathbf{A}_{\mathbf{Z}%
}^{SCE}=\mathbf{A}_{\mathbf{Z}}^{SCER}=\mathbf{A}_{\mathbf{Z,}[0,\bar{w}%
]^{I\times I}}^{SCER}=\mathbf{A}_{\mathbf{Z}}^{NE}$.\footnote{%
As we noted for Proposition \ref{prop:SCE=NE}, the same result holds also
for a non--linear and continuous aggregator $\ell _{i}$ and a generic
continuous and strictly quasi-concave utility function $v_{i}$.}
\end{corollary}

Even if, with positive local externalities, adding rationalizability does
not change the set of SCE's, it is still interesting to understand how
rationalizability works in a linear--quadratic network game, and more
generally in nice games with strategic complementarities.\newline
Given the finite index set $K$, the vector space $\mathbb{R}^{K}$ is endowed
with the standard partial order: $\mathbf{v}^{\prime }\leq \mathbf{v}%
^{\prime \prime }$ if and only if $v_{k}^{\prime }\leq v_{k}^{\prime \prime
} $ for each $k\in K$. With this, our assumptions imply that $\mathcal{Z}%
\subseteq \mathbb{R}^{I\times I}$ is a \emph{complete lattice}, which
implies that also $\mathbf{A}\times \mathcal{Z}$ is a complete lattice. We
let $\underline{\mathbf{Z}}$ and $\bar{\mathbf{Z}}$ respectively denote the
smallest and largest elements of $\mathcal{Z}$. Let $X_{i}=\ell _{i}\left( 
\mathbf{A}_{-i}\times \mathcal{Z}\right) $. A function $v_{i}:A_{i}\times
X_{i}\rightarrow \mathbb{R}$ has \textbf{increasing differences }if, for all 
$a_{i}^{\prime },a_{i}^{\prime \prime }\in A_{i}$, $x_{i}^{\prime
},x_{i}^{\prime \prime }\in X_{i}$ such that $a_{i}^{\prime }\leq
a_{i}^{\prime \prime }$ and $x_{i}^{\prime }\leq x_{i}^{\prime \prime }$%
\begin{equation*}
v_{i}\left( a_{i}^{\prime \prime },x_{i}^{\prime }\right) -v_{i}\left(
a_{i}^{\prime },x_{i}^{\prime }\right) \leq v_{i}\left( a_{i}^{\prime \prime
},x_{i}^{\prime \prime }\right) -v_{i}\left( a_{i}^{\prime },x_{i}^{\prime
\prime }\right) \text{.}
\end{equation*}

\begin{definition}
A linear-quadratic network game $NG$ has \textbf{strategic complementarities}
if $\mathcal{Z}\subseteq \lbrack 0,\bar{w}]^{I\times I}$ is a \emph{complete
lattice} and, for each $i\in I$, $v_{i}$ has increasing differences.
\end{definition}

\begin{remark}
If a linear-quadratic network game $NG$ has strategic complementarities,
then each game $\left \langle I,\left( A_{i},u_{i,\mathbf{Z}}\right) _{i\in
I}\right \rangle $ with $\mathbf{Z}\in \mathcal{Z}$ is supermodular.
\end{remark}

It is well known that the set of Nash equilibria of a supermodular game is a
complete lattice \citep[e.g.,][]{milgrom1990rationalizability}. With this,
for any linear-quadratic network game with strategic complementarities, we
let $\underline{\mathbf{a}}_{\underline{\mathbf{Z}}}^{NE}$ and $\mathbf{\bar{%
a}}_{\bar{\mathbf{Z}}}^{NE}$ respectively denote the smallest Nash
equilibrium of game $\left \langle I,\left( A_{i},u_{i,\underline{\mathbf{Z}}%
}\right) _{i\in I}\right \rangle $ and the largest Nash equilibrium of game $%
\left \langle I,\left( A_{i},u_{i,\bar{\mathbf{Z}}}\right) _{i\in
I}\right
\rangle $. The \textquotedblleft box,\textquotedblright \ or
order-interval in $\mathbb{R}^{I}$ determined by $\underline{\mathbf{a}}_{%
\underline{\mathbf{Z}}}^{NE}$ and $\mathbf{\bar{a}}_{\bar{\mathbf{Z}}}^{NE}$
is 
\begin{equation*}
\left[ \underline{\mathbf{a}}_{\underline{\mathbf{Z}}}^{NE},\mathbf{\bar{a}}%
_{\bar{\mathbf{Z}}}^{NE}\right] :=\times _{i\in I}\left[ \underline{a}_{i,%
\underline{\mathbf{Z}}}^{NE},\bar{a}_{i,\bar{\mathbf{Z}}}^{NE}\right] \text{.%
}
\end{equation*}

\begin{proposition}
\label{Th:R-bility in NG}Consider a linear-quadratic network game $NG$ with
strategic complementarities. The set of rationalizable action profiles is $%
\mathbf{A}^{\infty }=\left[ \underline{\mathbf{a}}_{\underline{\mathbf{Z}}%
}^{NE},\mathbf{\bar{a}}_{\bar{\mathbf{Z}}}^{NE}\right] $, that is, the set
of rationalizable actions of each player is the interval between the
smallest Nash equilibrium action in the game determined by the smallest
parameter $\underline{\mathbf{Z}}$ and the largest Nash equilibrium action
in the game determined by the largest parameter $\bar{\mathbf{Z}}$.
\end{proposition}

\noindent \textbf{Proof} \ Consider an auxiliary game $\hat{G}$ where an 
\emph{indifferent pseudo-player} chooses $\mathbf{Z}\in \mathcal{Z}$, and
the action sets and payoff functions of each $i\in I$ are those specified in
the network game $NG$ given $\mathbf{Z}$. It is easy to verify that the
auxiliary game $\hat{G}$ is supermodular and every $\mathbf{Z}\in \mathcal{Z}
$ is a Nash equilibrium action for the indifferent pseudo-player, that is,
the set of Nash equilibria of $\hat{G}$ is%
\begin{equation*}
\bigcup \limits_{\mathbf{Z}\in \mathcal{Z}}\mathbf{A}_{\mathbf{Z}%
}^{NE}\times \left \{ \mathbf{Z}\right \} \text{.}
\end{equation*}%
It is also easy to check that the set of rationalizable profiles of $\hat{G}$
is $\mathbf{A}^{\infty }\times \mathcal{Z}$, and Theorem \ref{Th:R-zability
nice} implies that $\mathbf{A}^{\infty }$ is an order-interval. Finally,
Theorem 5 in \cite{milgrom1990rationalizability} implies that the smallest
element of $\mathbf{A}^{\infty }\times \mathcal{Z}$ is $\left( \underline{%
\mathbf{a}}_{\underline{\mathbf{Z}}}^{NE},\underline{\mathbf{Z}}\right) $
and the largest element of $\mathbf{A}^{\infty }\times \mathcal{Z}$ is $%
\left( \mathbf{\bar{a}}_{\bar{\mathbf{Z}}}^{NE},\bar{\mathbf{Z}}\right) $;
therefore, $\mathbf{A}^{\infty }=\left[ \underline{\mathbf{a}}_{\underline{%
\mathbf{Z}}}^{NE},\mathbf{\bar{a}}_{\bar{\mathbf{Z}}}^{NE}\right] $. \hfill $%
\blacksquare $

\bigskip

Proposition \ref{Th:R-bility in NG} characterizes the set of rationalizable
action profiles for a generic complete lattice $\mathcal{Z}$. It is
straightforward to see that if the network $\mathbf{Z}$ is common knowledge,
i.e., $\bar{\mathbf{Z}}=\underline{\mathbf{Z}}$, and there is a unique Nash
equilibrium $\mathbf{a}_{\mathbf{Z}}^{NE}$ (e.g., if the assumptions of
Proposition \ref{prop:interiorSCE} hold), then $\underline{\mathbf{a}}_{%
\underline{\mathbf{Z}}}^{NE}=\mathbf{\bar{a}}_{\bar{\mathbf{Z}}}^{NE}=%
\mathbf{a}_{\mathbf{Z}}^{NE}$ and $\mathbf{A}^{\infty }=\left \{ \mathbf{a}_{%
\mathbf{Z}}^{NE}\right \} $. Thus, \emph{in this particular case, the Nash
equilibrium concept is justified by assuming that information is complete
and players are strategically sophisticated}, i.e., there is rationality and
common belief in rationality.

\paragraph{Positive and negative local externalities}

We consider now the case in which a network also allows for strictly
negative weights, so that $\mathcal{Z\subseteq }[\underline{w},\bar{w}%
]^{I\times I}$ with $\underline{w}<0$ and $\bar{w}>0$. The SCE analysis for
this case performed in Section \ref{sec:SCE} shows that a selfconfirming
equilibrium with shallow conjectures may allow any arbitrary set of agents
to be inactive as long as this is not dominated. Here we show that \emph{%
common knowledge of the network} and strategic reasoning may refine the SCE
set, even if one does not necessarily get rid of all the non-Nash SCE's.
Indeed, when negative local externalities (hence, strategic
substitutabilities) are at work, the set of rationalizable action profiles
is typically larger than the set of Nash equilibria. Here, we characterize
the set of SCE's with rationalizable conjectures. \newline

Consider the following two matrices. $\mathbf{Z}_{-}$, with all the negative
elements of $\mathbf{Z}$, is such that $z_{ij,-}<0$ if $z_{ij}<0$, and $%
z_{ij,-}=0$ otherwise. $\mathbf{Z}_{+}$, with all the positive elements of $%
\mathbf{Z}$, is such that $z_{ij,+}>0$ if $z_{ij}>0$, and $z_{ij,+}=0$
otherwise. Then, $\mathbf{Z}=\mathbf{Z}_{-}+\mathbf{Z}_{+}$. 
Define a sequence of pairs of action profiles $(\underline{\mathbf{a}}^{n},%
\bar{\mathbf{a}}^{n})_{n\in \mathbb{N}_{0}}$ as follows: 
$\underline{\mathbf{a}}^{0}=\mathbf{0}$, $\bar{\mathbf{a}}^{0}=\left( \bar{a}%
_{i}\right) _{i\in I}$ and, for every $n\in \mathbb{N}$, $\underline{\mathbf{%
a}}^{n}=\bm{\alpha}+\mathbf{Z}_{+}\underline{\mathbf{a}}^{n-1}+\mathbf{Z}_{-}%
\bar{\mathbf{a}}^{n-1}$ and $\bar{\mathbf{a}}^{n}=\bm{\alpha}+\mathbf{Z}_{+}%
\bar{\mathbf{a}}^{n-1}+\mathbf{Z}_{-}\underline{\mathbf{a}}^{n-1}$. Then, at
the $n^{th}$ step of iterated deletion of dominated strategies, the interval
of actions agent $i\in I$ can play is $A_{i}^{n}=[\underline{a}_{i}^{n},\bar{%
a}_{i}^{n}]$. Indeed, for each $i\in I$, $\underline{a}_{i}^{n}$ is the best
reply to the \textquotedblleft most pessimistic\textquotedblright \
conjecture consistent with $n-1$ steps of strategic reasoning, which is
given by (i) the \emph{largest} possible actions of neighbors towards whom $%
i $ experiences \emph{negative} externalities (strategic substitution) that
can be rationalized in $n-1$ steps, and (ii) the \emph{smallest} possible
actions that can be rationalized in $n-1$ steps of neighbors towards whom $i$
experiences \emph{positive} externalities (strategic complementarities).
Similarly, $\bar{a}_{i}^{n}$ is the best reply to the \textquotedblleft most
optimistic\textquotedblright \ conjecture consistent with $n-1$ steps of
strategic reasoning, which is given by (i) the largest possible actions that
can be rationalized in $n-1$ steps of neighbors towards whom $i$ experiences
positive externalities, and (ii) the smallest possible actions that can be
rationalized in $n-1$ steps of neighbors towards whom $i$ experiences
negative externalities. Let 
\begin{equation*}
I_{0}^{\infty }:=\{i\in I:\lim_{n\rightarrow \infty }\min
A_{i}^{n}=0\}=\{i\in I:\lim_{n\rightarrow \infty }\underline{a}_{i}^{n}=0\}
\end{equation*}%
denote the set of agents for whom being inactive is rationalizable. Relying
on Proposition \ref{prop:eq_SCERsub}, we can characterize $\mathbf{A}_{%
\mathbf{Z}}^{SCER}$ as the set of SCE's in which only players in $%
I_{0}^{\infty }$ can be inactive.

\begin{proposition}
\label{prop:eq_SCERsub} Consider a linear-quadratic network game with just
observable payoffs and common knowledge of the network ($\mathcal{Z=}%
\left
\{ \mathbf{Z}\right \} $). Then, 
\begin{equation*}
\mathbf{A}_{\mathbf{Z}}^{SCER}=\bigcup_{J:I\backslash J\subseteq
I_{0}^{\infty }}\mathbf{A}_{J,\mathbf{Z}}^{NE}\times \left \{ \mathbf{0}%
_{I\backslash J}\right \} .
\end{equation*}
\end{proposition}

\noindent \textbf{Proof. }Recall from Proposition \ref{Prop: SCE if LQ-OiffA
text} that%
\begin{equation*}
\mathbf{A}_{\mathbf{Z}}^{SCE}=\bigcup_{J:I\backslash J\subseteq I_{0}}%
\mathbf{A}_{J,\mathbf{Z}}^{NE}\times \left \{ \mathbf{0}_{I\backslash
J}\right \} \text{,}
\end{equation*}%
where $I_{0}\supseteq I_{0}^{\infty }$ denotes the set of players for whom
being inactive is undominated.

First we prove by induction that, if $I\backslash J\subseteq I_{0}^{\infty }$
then $\mathbf{A}_{J,\mathbf{Z}}^{NE}\times \left \{ \mathbf{0}_{I\backslash
J}\right \} \subseteq \mathbf{A}^{n}$ for every $n$; hence, $\mathbf{A}_{J,%
\mathbf{Z}}^{NE}\times \left \{ \mathbf{0}_{I\backslash J}\right \}
\subseteq \mathbf{A}^{\infty }$. Indeed, for each profile $\left( \mathbf{a}%
_{J},\mathbf{0}_{I\backslash J}\right) \in \mathbf{A}_{J,\mathbf{Z}%
}^{NE}\times \left \{ \mathbf{0}_{I\backslash J}\right \} $ and each player $%
i\in J$, action $a_{i}$ is the best reply to $\left( \mathbf{a}_{J\backslash
\left \{ i\right \} },\mathbf{0}_{I\backslash J}\right) $ and for each $i\in
I\backslash J$, action $a_{i}=0$ is rationalizable; thus, $\mathbf{A}_{J,%
\mathbf{Z}}^{NE}\times \left \{ \mathbf{0}_{I\backslash J}\right \}
\subseteq \mathbf{A}^{1}$. Suppose by way of induction that, for some $n\geq
2$, $\mathbf{A}_{J,\mathbf{Z}}^{NE}\times \left \{ \mathbf{0}_{I\backslash
J}\right \} \subseteq \mathbf{A}^{n-1}$. Then, for each profile $\left( 
\mathbf{a}_{J},\mathbf{0}_{I\backslash J}\right) \in \mathbf{A}_{J,\mathbf{Z}%
}^{NE}\times \left \{ \mathbf{0}_{I\backslash J}\right \} $ and each player $%
i\in J$, action $a_{i}$ is the best reply to $\left( \mathbf{a}_{J\backslash
\left \{ i\right \} },\mathbf{0}_{I\backslash J}\right) \in \mathbf{A}%
_{-i}^{n-1}$ and for each $i\in I\backslash J$, action $a_{i}=0$ is
rationalizable; thus, $\mathbf{A}_{J,\mathbf{Z}}^{NE}\times \left \{ \mathbf{%
0}_{I\backslash J}\right \} \subseteq \mathbf{A}^{n}$. With this, for each
action profile in $\mathbf{A}_{J,\mathbf{Z}}^{NE}\times \left \{ \mathbf{0}%
_{I\backslash J}\right \} $, each player $i\in J$ is best replying to the
co-players' actions, hence to a rationalizable confirmed conjecture, and
each player $i\in I\backslash J\subseteq I_{0}^{\infty }$ is rationalizably
inactive, hence, she is best replying to a rationalizable conjecture (Remark %
\ref{Rem: rat action conj}), which is trivially confirmed. Thus, $%
I\backslash J\subseteq I_{0}^{\infty }$ implies $\mathbf{A}_{J,\mathbf{Z}%
}^{NE}\times \left \{ \mathbf{0}_{I\backslash J}\right \} \subseteq \mathbf{A%
}_{\mathbf{Z}}^{SCER}$.

If instead $\left( I\backslash J\right) \nsubseteq I_{0}^{\infty }$, for
each action profile $\mathbf{a\in A}_{J,\mathbf{Z}}^{NE}\times \left \{ 
\mathbf{0}_{I\backslash J}\right \} $ there is some $i\in I\backslash J$
such that $a_{i}=0$ is not rationalizable, hence it is not a best reply to
any rationalizable conjecture (Remark \ref{Rem: rat action conj}). This
implies that $\mathbf{a\notin A}_{\mathbf{Z}}^{SCER}$.\hfill $\blacksquare $

\bigskip %

Finally, we note that even if the network is not common knowledge, but it is
common knowledge that local externalities are mild, then there is a unique
SCER, which---necessarily---coincides with the unique Nash equilibrium. This
is the case, for example, if $-\frac{\alpha }{(n-1)\bar{a}}<\underline{w}$
and $\bar{w}<\frac{\bar{a}-\alpha }{(n-1)\alpha }$, and this is common
knowledge, then rationalizability yields the unique interior Nash
equilibrium. One can get intermediate results by changing the threshold for
just one of $\underline{w}$ and $\bar{w}$.

\paragraph{Local and global externalities}

We consider now the case of both local and global externalities. As
discussed in Section \ref{sec:local_global}, we restrict our attention to
situations in which local externalities are positive. In this case, there is
a continuum of SCE's, one for each vector of conjectured ratios. We now
study whether strategic reasoning helps in selecting some SCE's. The main
result is that, if there is common knowledge of the network, strategic
reasoning expressed as common belief in rationality selects the unique Nash
equilibrium among the infinitely many SCE's.

\begin{proposition}
\label{prop:ratioglobalZ} Consider a linear-quadratic network game with just
observable payoffs, positive local externalities, global externalities,
common knowledge of the network ($\mathcal{Z}=\left \{ \mathbf{Z}\right \} $%
), and a unique Nash equilibrium. Then $\mathbf{A}_{\mathbf{Z}}^{SCER}=%
\mathbf{A}_{\mathbf{Z}}^{NE}$
\end{proposition}

\noindent \textbf{Proof. } The result follows from Proposition \ref%
{Th:R-bility in NG}. Indeed the game we are considering has strategic
complementarities. Then, $\mathbf{A}^{\infty }=\left[ \underline{\mathbf{a}}%
_{\underline{\mathbf{Z}}}^{NE},\mathbf{\bar{a}}_{\bar{\mathbf{Z}}}^{NE}%
\right] $. Since $\mathbf{Z}$ is common knowledge, and there exists a unique
Nash equilibrium, viz $\mathbf{a}_{\mathbf{Z}}^{NE}$, it follows that $%
\mathbf{A}^{\infty }=\left \{ \mathbf{a}_{\mathbf{Z}}^{NE}\right \} $. Then, 
$\mathbf{A}_{\mathbf{Z}}^{SCER}=\left \{ \mathbf{a}_{\mathbf{Z}%
}^{NE}\right
\} $. \hfill $\blacksquare $ \newline

We can alternatively prove this result by showing how step-by-step strategic
reasoning works in this case, assuming for simplicity that the unique Nash
equilibrium is \emph{interior}. Recall that if the network is common
knowledge and local externalities are positive, then agents can only have
positive justifiable actions. Consider $\underline{\mathbf{a}}^{0}=%
\boldsymbol{\alpha }=\left( \alpha _{i}\right) _{i\in I}$ and $\bar{\mathbf{a%
}}^{0}=\bar{\mathbf{a}}=\left( \bar{a}_{i}\right)_{i\in I} $. If the network
is common knowledge, then 
\begin{align}
\underline{\mathbf{a}}^{1}=& \boldsymbol{\alpha }+\mathbf{Z}\boldsymbol{%
\alpha },\  & \bar{\mathbf{a}}^{1}=& \boldsymbol{\alpha }+\mathbf{Z}\bar{%
\mathbf{a}}  \notag \\
\underline{\mathbf{a}}^{2}=& \boldsymbol{\alpha }+\mathbf{Z}(\boldsymbol{%
\alpha }+\mathbf{Z}\boldsymbol{\alpha }),\  & \bar{\mathbf{a}}^{1}=& 
\boldsymbol{\alpha }+\mathbf{Z}(\boldsymbol{\alpha }+\mathbf{Z}\bar{\mathbf{a%
}})  \notag \\
=& \boldsymbol{\alpha }+\mathbf{Z}\boldsymbol{\alpha }+\mathbf{Z}^{2}%
\boldsymbol{\alpha },\  & =& \boldsymbol{\alpha }+\mathbf{Z}\boldsymbol{%
\alpha }+\mathbf{Z}^{2}\bar{\mathbf{a}}  \notag \\
\underline{\mathbf{a}}^{3}=& \boldsymbol{\alpha }+\mathbf{Z}\boldsymbol{%
\alpha }+\mathbf{Z}^{2}\boldsymbol{\alpha }+\mathbf{Z}^{3}\boldsymbol{\alpha 
},\  & \bar{\mathbf{a}}^{3}=& \boldsymbol{\alpha }+\mathbf{Z}\boldsymbol{%
\alpha }+\mathbf{Z}^{2}\boldsymbol{\alpha }+\mathbf{Z}^{3}\bar{\mathbf{a}} 
\notag \\
\dots & \  & \dots &  \notag \\
\underline{\mathbf{a}}^{n}=& \alpha \sum_{t=0}^{n}\mathbf{Z}^{t},\  & \bar{%
\mathbf{a}}^{n}=& \alpha \sum_{t=0}^{n-1}\mathbf{Z}^{t}+\mathbf{Z}^{n}\bar{%
\mathbf{a}}  \notag
\end{align}%
Since the game is assumed to have a unique Nash equilibrium that is also
interior, then $\lim_{n\rightarrow \infty }\sum_{t=0}^{n}\mathbf{Z}^{t}$
exists and it is finite, and $\lim_{n\rightarrow \infty }\mathbf{Z}^{n}=%
\mathbf{0}$. Then $\underline{\mathbf{a}}^{\infty }=\bar{\mathbf{a}}^{\infty
}=\mathbf{a}^{NE}$. Since $\mathbf{A}_{\mathbf{Z}}^{\infty }=\mathbf{A}_{%
\mathbf{Z}}^{NE}=\{ \mathbf{a}^{NE}\} \supseteq \mathbf{A}_{\mathbf{Z}%
}^{SCE} $, it follows that $\mathbf{A}_{\mathbf{Z}}^{SCER}=\mathbf{A}_{%
\mathbf{Z}}^{\infty }\cap \mathbf{A}_{\mathbf{Z}}^{SCE}=\mathbf{A}_{\mathbf{Z%
}}^{NE}$.


\section{Interior Nash equilibria}

\label{app:interior}

Proposition \ref{Prop: SCE if LQ-OiffA text} shows
that, given our maintained assumptions about the network game with feedback,
selfconfirming action profiles can be characterized as Nash equilibria of
auxiliary games with a restricted set of players, which must include all
those for whom being inactive is unjustifiable (dominated), but may leave
out any player for whom inactivity is justifiable (undominated). We now
provide some results about existence of these SCE's that will be useful in
proving Proposition \ref{prop:interiorSCE}. We
first present sufficient conditions that are present in the literature for
the existence and uniqueness of interior Nash equilibria, then we provide
some original results. 

\bigskip

In this appendix we formulate the problem with the approach of linear
algebra. We consider a square matrix $\mathbf{Z} \in \mathbb{R}^{n \times n}$
such that $z_{ii}=0$ for all $i \in \{1, \dots,n\}$. We denote by $\mathbf{I}
$ the identity matrix, $\lambda_{max} (\mathbf{Z})$ the maximal eigenvalue
of $\mathbf{Z}$, $\rho(\mathbf{Z})$ the spectral radius of $\mathbf{Z}$
(i.e., the largest absolute value of its eigenvalues), $\vec{1} $ the vector
of all $1$'s, $\vec{0} $ the vector of all $0$'s, and $\gg$ the strict
partial ordering between vectors (meaning that all the entries in the first
vector are coordinatewise strictly greater than the entries in the second
vector). With this notation, the condition for the existence of a unique
Nash equilibrium which is also interior is $\left( \mathbf{I} - \mathbf{%
\mathbf{Z}} \right)^{-1} \cdot \vec{1} \gg \vec{0}$.

\begin{proposition}
\label{prop:BCZ_exist} Consider a square matrix $\mathbf{Z} \in \mathbb{R}%
^{n \times n}$ such that $(i)$ $\rho(\mathbf{Z})<1$, $(ii)$ for each $i\in I$%
, $z_{ii}=0$, and $(iii)$ for each $j \ne i$, $z_{ij}\leq0$. Then $\left( 
\mathbf{I} - \mathbf{\mathbf{Z}} \right)^{-1} \cdot \vec{1} \gg \vec{0}$.%
\footnote{%
This is Theorem 1 in \cite{ballester2006s}. The same result is in Appendix A
in \cite{stanczak2006resource}.}
\end{proposition}

Some results can be provided also when the sign of the externalities are
mixed. Recall that matrix $\mathbf{\mathbf{Z}}$ is symmetrizable if there
exists a diagonal matrix $\mathbf{W}$ and a symmetric matrix $\mathbf{Z}_{0}$
such that $\mathbf{Z}=\mathbf{WZ}_{0}$. Note that, if $\mathbf{Z}$ is
symmetrizable, then all its eigenvalues are real. If for all $i$, $z_{ii}=0$%
, and $\mathbf{Z}$ is symmetrizable, we define the symmetric matrix $\tilde{%
\mathbf{Z}}$ to be such that $\tilde{z}_{ij}=z_{ij}\sqrt{w_{i}w_{j}}$.

\begin{proposition}
\label{prop:BKD_exist} Consider a square matrix $\mathbf{Z} \in \mathbb{R}%
^{n \times n}$ such that $(i)$ for each $i\in I$, $z_{ii}=0$, $(ii)$ $%
\mathbf{Z}$ is symmetrizable, and $(iii)$ $| \lambda_{max} (\tilde{\mathbf{Z}%
}) | <1$. Then $\left( \mathbf{I} - \mathbf{Z} \right)^{-1} \cdot \vec{1}
\gg \vec{0}$.\footnote{%
See Section VI of \cite{bramoulle2014strategic}, generalizing Proposition 2
therein. Note that in their payoff specification externalities have a \emph{%
minus} sign, while in \eqref{eq:LQ} we have a \emph{plus} sign: this is why
we have a condition on the maximal eigenvalue and not on the minimal
eigenvalue.}
\end{proposition}

Finally, we provide below a novel alternative condition.

\begin{proposition}
\label{prop:our_exist} Consider a square matrix $\mathbf{Z} \in \mathbb{R}%
^{n \times n}$ such that $(i)$ for each $i\in I$, $z_{ii}=0$ and $(ii)$ for
each $i\neq j$, $|z_{ij}|< \frac{1}{n}$. Then $\left( \mathbf{I} - \mathbf{Z}
\right)^{-1} \cdot \vec{1} \gg \vec{0}$.
\end{proposition}

\noindent{\  \textbf{Proof:} Let $\mathbf{B} := \left( \mathbf{I} - \mathbf{Z}
\right)$. First of all, by Gershgorin circle theorem, 
$\mathbf{B}$ has all eigenvalues, possibly complex, with real part strictly
between $0$ and $2$, so $det(\mathbf{B}) \ne 0$. }

Consider the $n$ vectors $\vec{b}^1, \dots, \vec{b}^n$ given by the $n$ rows
of $\mathbf{B}$, and take the hyperplane in $\mathbb{R}^n$ passing by those $%
n$ points: 
\begin{equation*}
H := \{ \vec{h} \in \mathbb{R}^n : \exists \vec{\alpha} \in \mathbb{R}^n, 
\vec{\alpha}^{\prime }\cdot \vec{1} = 1 \wedge \vec{h} = \mathbf{B}^{\prime }%
\vec{\alpha} \} \ .
\end{equation*}
Now, consider the following vector 
\begin{equation*}
\vec{v} := \mathbf{B}^{-1} \vec{1} \ .
\end{equation*}
$v_i$ is exactly the sum of the entries in $i^{th}$ row of $\mathbf{B}^{-1}$%
. However, $\vec{v}$ is also a vector perpendicular to $H$. This is because
for any $\vec{h} \in H$ we have, for some $\vec{\alpha} \in \mathbb{R}^n$, 
\begin{eqnarray}
\vec{h} \cdot \vec{v} & = & \left( \mathbf{B}^{\prime }\vec{\alpha}
\right)^{\prime }\cdot \mathbf{B}^{-1} \vec{1}  \notag \\
& = & \vec{\alpha}^{\prime }\vec{1}  \notag \\
& = & \sum_{i=1}^n \alpha_i = 1 \ ,  \notag
\end{eqnarray}
which is a constant.

\bigskip

Now, we want to show that $H$ does not pass through the convex region of
vectors with all negative elements: $H\cap (-\infty ,0]^{n}=\emptyset $. In
fact, it is impossible to find $\vec{w}\in \mathbb{R}^{n}$, such that $\vec{w%
}^{\prime }\cdot \vec{1}=1$ and $\mathbf{B}^{\prime }\vec{w}\ll \vec{0} $.
Suppose, by way of contradiction, that such vector $\vec{w}$ exists. Let $%
k:=\arg \max_{i\in \{1,\dots ,n\}}\{w_{i}\}$ ($w_{k}>0$ because $%
\sum_{i=1}^{n}w_{i}=1$), then (calling $\vec{b}_{k}$ the $k^{th}$ row of
matrix $\mathbf{B}$) 
\begin{equation*}
\vec{b}_{k} \cdot \vec{w}=w_{k}+\sum_{j\neq k}w_{j}b_{jk}>w_{k}-\sum_{j\neq
k}|w_{j}||z_{jk}|>w_{k}\left( 1-\sum_{j\neq k}|z_{jk}|\right) >0\ ,
\end{equation*}%
which is a contradiction.

\bigskip

Finally, we show that if a hyperplane $H$ satisfies $H \cap (-\infty,0]^n =
\emptyset$, then its perpendicular vector from the origin has all strictly
positive entries, and this concludes the proof . \newline
We do so by induction on $n$.

\begin{enumerate}
\item $\mathbf{n=2}$: This is easy to show graphically. In the Cartesian
plane the hyperplane is a line. Not passing by $(-\infty,0]^2$, it will
cross both axes in their strictly positive part: call these intersection
points $A$ and $B$. So, the segment that from the origin crosses this line
perpendicularly will cross it in a point $C$ that lies on the line between $%
A $ and $B$.

\item \textbf{Induction hypothesis}: Suppose it is true for $n-1$.

\item \textbf{Inductive step}: a hyperplane $H \subset \mathbb{R}^n$ that
satisfies $H \cap (-\infty,0]^n = \emptyset$ does not pass through the
origin. So, it has an orthogonal vector $\vec{v}$ such that $\vec{v} \in H$.
By assumption on $H$, $\vec{v}$ cannot have all elements non strictly
positive. So, there exists $i \in \{1,\dots, n\}$ such that $v_i>0$. Let us
take $P_{\neg i} = \{ \vec{p} \in \mathbb{R}^n: p_i =0 \}$. Call $H_{\neg i}$
the intersection of $H$ with $P_{\neg i}$. Take the vector $\vec{v}_{\neg i} 
$ that is the projection of $\vec{v}$ on $P_{\neg i}$. This vector has all
entries equal to $\vec{v}$, except for entry $i$ which is null. Also, $\vec{v%
}_{\neg i}$ is perpendicular to $H_{\neg i}$.

By assumption on $H$, $H_{\neg i} \cap (-\infty,0]^{n-1} = \emptyset$.
Moreover, because of the induction hypothesis, $\vec{v}_{\neg i}$ has all
strictly positive entries, except from entry $i$. Finally, since also $v_i>0$%
, we have the proof.
\end{enumerate}

Notice that, if $\mathbf{Z}$ satisfies the conditions of Proposition \ref%
{prop:our_exist}, then it must also hold that $| \lambda_{max} (\mathbf{Z})
| <1$, because of Gershgorin circle theorem. However, the condition that $|
\lambda_{max} (\mathbf{Z}) | <1$ is in general not sufficient to guarantee
that $\left( \mathbf{I} - \mathbf{Z} \right)^{-1} \vec{1} \gg \vec{0}$.
\hfill $\blacksquare$

\section{Proofs of Propositions}

\label{app:proofs}

\subsection*{Proposition \protect \ref{Prop: SCE if LQ-OiffA text} (page 
\protect \pageref{Prop: SCE if LQ-OiffA text})}

\textbf{Proof. } By Remark \ref{Rem: LQ JOP imply OAP}, $NG$ satisfies
observability by active players. Hence, Lemma \ref{Lemma:SCE if OiffA} in
Appendix \ref{app:SCE} and the best-reply equation yield the result. \hfill $%
\blacksquare $

\subsection*{Proposition \protect \ref{prop:interiorSCE} (page \protect
\pageref{prop:interiorSCE})}

\textbf{Proof. } Conditions $1$, $2$, and $3$ correspond, respectively, to
the conditions in Propositions 
\ref{prop:our_exist}, \ref{prop:BCZ_exist}, and \ref{prop:BKD_exist} from
Appendix \ref{app:interior}. \hfill $\blacksquare$

\subsection*{Proposition \protect \ref{remark:radius} (page \protect \pageref%
{remark:radius})}

\textbf{Proof. } Let us consider separately the two sets $I\backslash I_{%
\mathbf{a}^{\ast }}$ and $I_{\mathbf{a}^{\ast }}$ of inactive and active
agents.

For every $i\in I\backslash I_{\mathbf{a}^{\ast }}$, $\alpha _{i}+\underline{%
x}_{i}<0$; thus, $a_{i}^{\ast }=0$ is a best reply to every conjecture $\hat{%
x}_{i}\in \left( \underline{x}_{i},-\alpha _{i}\right) $ and a sufficiently
small perturbation of $\hat{x}_{i}$ does not make $i$ become active.

Now, let us focus on the subset $I_{\mathbf{a}^{\ast }}$ of active agents.
For each $i\in I_{\mathbf{a}^{\ast }}$, a perturbation in $\hat{x}_{i}$
induces a change in the corresponding best reply. Let us focus on
perturbations that are small enough so that all actions of agents in $I_{%
\mathbf{a}^{\ast }}$ remain strictly positive. Since $\rho(\mathbf{Z})<1$ is
a strict inequality, Assumption \ref{ass:limited} guarantees that the
limiting points of the discrete path system defined for actions by %
\eqref{eq:network_payoffs} and \eqref{eq:learning} are locally stable,
because the non--null eigenvalues and eigenvectors of the Jacobian of this
system are the same eigenvalues and eigenvectors of $\mathbf{Z}_{I_{\mathbf{a%
}^{\ast }}}$.

Thus, there is $\epsilon >0$ such that the perturbation of beliefs given by
any $\mathbf{x}_{0}$ with $\left \Vert \mathbf{x}_{0}-\mathbf{\hat{x}}%
\right
\Vert <\epsilon $ is small enough so that inactive agents keep being
inactive and all actions of active agents in $I_{\mathbf{a}^{\ast }}$ remain
strictly positive.

In this way, the discrete system defined for actions by %
\eqref{eq:network_payoffs} and \eqref{eq:learning} converges back to $%
\mathbf{a}^{\ast }$. \hfill $\blacksquare $

\subsection*{Proposition \protect \ref{prop:stableSCE} (page \protect \pageref%
{prop:stableSCE})}

\textbf{Proof. } 
For all the action profiles considered in the proposition the inactive
players are choosing a best response for an open set of conjectures; thus,
being inactive is robust to small perturbations of justifying non-falsified
conjectures. With this, we can focus on the active agents. Note that if we
take an active agent $i$ from $I_{\mathbf{a}^{\ast }}$ and we make him
inactive, then the new matrix $\mathbf{Z}_{I_{\mathbf{a}^{\ast }}\backslash
\{i\}}$ for active players is a sub--matrix of $\mathbf{Z}_{I_{\mathbf{a}%
^{\ast }}}$ obtained deleting the row and the column corresponding to agent $%
i$. This process can be repeated removing more active agents, which means
that if we remove a subset $J\subset I_{\mathbf{a}^{\ast }}$ of the active
agents, then the new matrix $\mathbf{Z}_{I_{\mathbf{a}^{\ast }}\backslash J}$
is a sub--matrix of $\mathbf{Z}_{I_{\mathbf{a}^{\ast }}}$ obtained deleting
all the rows and the columns corresponding to every agent $j\in J$.

So, given the results from Propositions \ref{prop:interiorSCE} and \ref%
{remark:radius}, to prove the statement, we need to prove that if an
adjacency matrix satisfies one of the three conditions, then also every
sub--matrix of that matrix, which is obtained deleting one row and one
column with the same index, satisfies that condition. By induction this will
be true for every sub--matrix of that matrix, which is obtained deleting any
subset of rows and columns with the same indices.

For Point $1$ the result is clear, because a property that holds for all the
elements of a matrix will hold also for all the elements of a sub--matrix of
that matrix.

Point $2$ is based on two assumptions. Assumption \ref{ass:negative} is
still valid if we remove one column and one row of a matrix because it is a
property of all the elements of that matrix. To check for Assumption \ref%
{ass:limited}, let us consider the following implications of the
Perron--Frobenius theorem (see, e.g., \citealp{savchenko2003perron}): {%
(i) for a matrix with all positive entries, there exists a real eigenvalue
(often called the \emph{Perron root}) which is equal to its spectral radius;
(ii) the Perron root of any principal submatrix of such a matrix does not
exceed that of the original matrix. } In our case, Assumption \ref%
{ass:negative} implies that our matrix can be seen as a matrix with all
positive elements with a minus sign in front, and this proves the statement.

Point 3 holds because of a generalization of the Cauchy interlace theorem
applied to symmetrizable matrices (see \citealp{kouachi2016cauchy} and %
\citealp{mckee2020symmetrizable}). We know that the magnitude of the
eigenvalues of the sub--matrix of a symmetrizable matrix, obtained deleting
one row and one column with the same index, are between the magnitudes of
the minimal and the maximal eigenvalues of the old matrix. So, the
sub--matrix of a limited matrix, which is obtained deleting one row and one
column with the same index, is limited. The resulting sub--matrix is also
symmetrizable. That is because the original matrix was obtained as the
product of a diagonal and a symmetric matrix, and to obtain the sub--matrix
we can delete the corresponding rows and columns in those diagonal and
symmetric matrices: the two matrices will maintain their properties and the
result will be our sub--matrix. \hfill $\blacksquare $

\subsection*{Proposition \protect \ref{Prop: SCEglob} (page \protect \pageref%
{Prop: SCEglob})}

\textbf{Proof. } \noindent A selfconfirming equilibrium is such that, for
all $i\in I$, rationality implies%
\begin{equation*}
a_{i}^{\ast }=\min \{ \max \{0,\alpha _{i}+\hat{x}_{i}\},\bar{a}_i\} \text{,}
\end{equation*}%
where $\hat{x}_{i}$ is the conjecture of $i$ about the payoff state. Each
agent then thinks that%
\begin{equation*}
m^{\ast }=\alpha _{i}a_{i}^{\ast }-\frac{1}{2}\left( a_{i}^{\ast }\right)
^{2}+a_{i}^{\ast }\hat{x}_{i}+\hat{y}_{i}\ ,
\end{equation*}%
so that 
\begin{equation}  \label{yhat}
\hat{y}_{i}=m^{\ast }-\alpha _{i}a_{i}^{\ast }+\frac{1}{2}\left( a_{i}^{\ast
}\right) ^{2}-a_{i}^{\ast }\hat{x}_{i}\ .
\end{equation}%
Substituting the expression of the true actual payoff 
\begin{equation*}
m^{\ast }=\alpha _{i}a_{i}^{\ast }-\frac{1}{2}\left( a_{i}^{\ast }\right)
^{2}+a_{i}^{\ast }x_{i}+y_{i}
\end{equation*}%
into \eqref{yhat}, we get the dependence between $\hat{y}_{i}$ and $\hat{x}%
_{i}$: 
\begin{equation*}
\hat{y}_{i}=y_{i}+a_{i}^{\ast }\left( x_{i}-\hat{x}_{i}\right) \ .
\end{equation*}%
The first and second items in the proposition are derived, respectively, if $%
a_{i}^{\ast }=0$ or if $a_{i}^{\ast }>0$. \hfill $\blacksquare $

\subsection*{Proposition \protect \ref{prop:inequality_SCE} (page \protect
\pageref{prop:inequality_SCE})}

By substituting, for each $i\in I$, the subjectively rational choice into
the confirmed conjecture condition, we get the following: 
\begin{equation}
\left(\alpha+\hat{x}_i\right)\left(\hat{x}_i-\displaystyle \sum_{j\in
I\backslash \left \{ i\right \} }z_{ij} \left(\alpha+\hat{x}%
_j\right)\right)=\left(\gamma \displaystyle \sum_{k\in I\backslash \left \{
i\right \} }\left(\alpha+\hat{x}_k\right)-\hat{y}_i\right).
\label{eq:utglobal_xy}
\end{equation}
This condition holds for each $i\in I$, so that we have a non-linear system
of $n$ equations and $2n$ unknowns. Still, from \eqref{eq:utglobal_xy} we
can provide useful insights to understand how conjectures are shaped in a
SCE.\newline
First of all, note that \eqref{eq:utglobal_xy} is linear in $\hat{y}_i$.
Thus, given any profile $(\hat{x}_i)_{i\in I}$, there exists a unique
profile $(\hat{y}_i)_{i\in I}$ consistent with the confirmed conjectures
condition. Moreover, we can also compute a bound for each $\hat{y}_i$.
Indeed, for each $i\in I$, $\hat{x}_i>0$. Then, since $a_i=\alpha+\hat{x}%
_i\leq \bar{a}$, for each $i\in I$, and given other agents' conjectures, it
must be that $y_i\leq \alpha \displaystyle \sum_{j\in I\backslash \left \{
i\right \} }z_{ij}a_{j} + \gamma \displaystyle \sum_{k\in I\backslash
\left
\{ i\right \} }a_{k} \leq \bar{a}\bigl(\alpha \displaystyle %
\sum_{k\in I\backslash \left \{ i\right \} } z_{ij} +\gamma n \bigr)
$. \newline
Given a profile $\bigl(\hat{y}_i\bigr)_{i\in I}$, condition %
\eqref{eq:utglobal_xy}, also allows us to characterize the corresponding SCE
profile $\bigl(\hat{x}_i\bigr)_{i\in I}$. Solving the second-order
polynomial, we get that the only positive solution for each $\hat{x}_i$ is
given by 
\begin{equation}  \label{eq:conjeq}
\hat{x}_i = \frac{1}{2} \left( \displaystyle \sum_{j\in I\backslash \left \{
i\right \} }z_{ij} \left(\alpha+\hat{x}_j\right) - \alpha + \sqrt{ \left( %
\displaystyle \sum_{j\in I\backslash \left \{ i\right \} }z_{ij}
\left(\alpha+\hat{x}_j\right) + \alpha \right)^2 + 4 \gamma \displaystyle %
\sum_{k\in I\backslash \left \{ i\right \} } \left(\alpha+\hat{x}_j\right)-4%
\hat{y}_i } \right)
\end{equation}
Note that, at an SCE, each $\hat{x}_i$ is increasing in others' beliefs
about local externality, and decreasing in own $\hat{y}_i$. Indeed, given $%
\hat{y}_i$, an increase in any $\hat{x}_j$ increases $j$'s action and thus
it increases the global externality. Given that only positive externalities
are considered, if $\hat{y}_i$ is kept fixed, at SCE $i$ has no other option
than having a higher $\hat{x}_i$. On the contrary, if $\hat{y}_i$ increases
keeping fixed $\bigl(\hat{x}_j\bigr)_{j\in I\backslash \{i\}}$, then actual
local and global externalities for $i$ are unchanged. However, if $i$ thinks 
$y_i$ to be higher, she necessarily needs to decrease $\hat{x}_i$. Given
that equilibrium $\hat{x}_i$ is monotonically decreasing in $\hat{y}_i$, we
can also easily compute an upper bound for $\hat{x}_i$ by simply letting $%
\hat{y}_i=0$ in \eqref{eq:conjeq}.\newline
By taking the second derivative of the right hand side of \eqref{eq:conjeq},
with respect to $\hat{x}_j$, we obtain 
\begin{equation*}
\frac{\partial^2 \hat{x}_i}{\partial \hat{x}_j^2} = - \frac{ 2 \gamma }{ 
\sqrt{ \Gamma ( \hat{x}_j ) }^{3/2} } \left( z_{ij} \left( \displaystyle%
\sum_{k\in I\backslash \left \{ i,j \right \} }z_{ik} \left(\alpha+\hat{x}_k
\right) - z_{ij} \displaystyle \sum_{h\in I\backslash \left \{ i,j\right \}
} \left(\alpha+\hat{x}_h \right) + \alpha \right) + \gamma \right)
\end{equation*}
where $\Gamma ( \hat{x}_j ) $ is an always positive quadratic expression of $%
\hat{x}_j$. If, for every couple of agents $i$ and $j$ in $I$, the
inequality 
\begin{equation}
\displaystyle \sum_{k\in I\backslash \left \{ i,j \right \} }z_{ik}
\left(\alpha+\hat{x}_k \right) - z_{ij} \displaystyle \sum_{h\in I\backslash
\left \{ i,j\right \} } \left(\alpha+\hat{x}_h \right) + \alpha \geq 0 \  \ ,
\end{equation}
is satisfied, then $\hat{x}_i$ is concave in each $\hat{x}_j$. So, there is
always a unique finite solution to the system where each player has the
higher possible belief about $\hat{x}_j$. In this solution, as we assume
that either condition 1.~or 3.~of Proposition \ref{prop:interiorSCE} is
satisfied, we derive a unique $\bigl(a^*_i\bigr)_{i\in I}$ with $a_i^*<\bar{a%
}$ for each $i$. If, $\hat{x}_i$ is convex in some $\hat{x}_j$, then the
process may self-reinforce and it is possible that a corner solution is
reached. \hfill $\blacksquare$

\subsection*{Proposition \protect \ref{prop:beta} (page \protect \pageref%
{prop:beta})}

\textbf{Proof. } Before proving the result we need to consider a slight
modification of aggregator and conjectures.

Let 
\begin{equation}
\begin{tabular}{llll}
$\tilde{\ell} _{i,\mathbf{Z_0}}:$ & $\mathbf{A}_{-i}$ & $\rightarrow $ & $%
\tilde{X}_{i}$, \\ 
& $\  \mathbf{a}_{-i}$ & $\mapsto $ & $\sum_{j\neq i}z_{0,ij}a_{j}$%
\end{tabular}
\label{eq:L aggr tilde}
\end{equation}
and 
\begin{equation}
\begin{tabular}{llll}
$\tilde{g}_{i }:$ & $\mathbf{A}_{-i}$ & $\rightarrow $ & $\tilde{Y}_{i}$ \\ 
& $\  \  \  \mathbf{a}_{-i}$ & $\mapsto $ & $\displaystyle \sum_{j\neq i}a_{j} $%
\end{tabular}%
\end{equation}
be the equivalent of $\ell_{i,\mathbf{Z}}$ and $g_{i }$, when we do not
incorporate the parameters on which there is mutual knowledge. Similarly,
let $\hat{\tilde{x}}_i$ and $\hat{\tilde{y}}_i$ be the shallow conjectures
about $\tilde{x}_i$ and $\tilde{y}$, respectively. @[Skip def and just brief
discussion how to adapt standard definition.]@ Then, we need to provide a
definition of selfconfirming equilibrium coherent with the hypotheses about
the knowledge of the agents.

\begin{definition}
\label{Def: SCEglobtilde}A profile $\left( a_{i}^{\ast },\hat{\tilde{x}}_{i},%
\hat{\tilde{y}}_{i}\right) _{i\in I}\in \times _{i\in I}\left( A_{i}\times 
\tilde{X}_{i}\times \tilde{Y}_{i}\right) $ of actions and (shallow)
deterministic conjectures is a \textbf{selfconfirming equilibrium} at $%
\left( \mathbf{Z_0},\omega,\gamma \right) $ of a network game with global
externalities with mutual knowledge of $(\omega,\gamma)$ if, for each $i\in
I $,

\begin{enumerate}
\item \emph{(subjective rationality)} $a_{i}^{\ast }=r_{i}\left( \hat{\tilde{%
x}}_{i}\right) $;

\item \emph{(confirmed conjecture)} $f_{i}\left( a_{i}^{\ast },\hat{\tilde{x}%
}_{i},\hat{\tilde{y}}_{i};\omega, \gamma \right) =f_{i}\left( a_{i}^{\ast },%
\tilde{\ell}_{i}\left( \mathbf{a}_{-i}^{\ast}, \mathbf{Z_0} \right),\tilde{g}%
_{i}\left( \mathbf{a}_{-i}^{\ast} \right); w, \gamma \right) $.
\end{enumerate}
\end{definition}

We are now ready to prove the result.\newline
Consider first the Nash equilibrium of the game with payoff function %
\eqref{eq:global_game_beta} played on a complete network. For each $i\in I$, 
$a^{NE}_{\mathbf{Z}_c,i}=r_{i}(w\sum_{k\in I\backslash \left \{ i\right \}
}a^{NE}_{\mathbf{Z}_c,k})$. Because of symmetry, for each $i\in I$, $a^{NE}_{%
\mathbf{Z}_c,i}=\frac{\alpha _{i}}{1-(n-1)w}$. \newline
Given a selfconfirming equilibrium action profile $\mathbf{a}^c$, each
player $i$, by perfect recall of her own action, can correctly infer that 
\begin{equation}
a_i^c w \tilde{x}_i+\gamma \tilde{y}_i=a_{i}w\sum_{j\in I\backslash \left \{
i\right \} }z_{0,ij}a_{j}+\gamma \displaystyle \sum_{k\in I\backslash \left
\{ i\right \} }a_{k}\  \ ,  \label{proof7_1}
\end{equation}
so that, her shallow conjectures must be such that 
\begin{equation}
a_i^c w \hat{\tilde{x}}_i+\gamma \hat{\tilde{y}}_i=a_{i}w\sum_{j\in
I\backslash \left \{ i\right \} }z_{0,ij}a_{j}+\gamma \displaystyle %
\sum_{k\in I\backslash \left \{ i\right \} }a_{k}\  \ .  \label{proof7_1a}
\end{equation}

At the same time, since by deep conjecture $\bar{\mu}_i^c$ each player $i$
thinks to be linked with all the other players, then it must be $\hat{\tilde{%
x}}_i=\hat{\tilde{y}}=\hat{\tilde{x}}_i^c$, and her shallow conjectures are
such that 
\begin{equation}
a_i^c w \hat{\tilde{x}}_i+\gamma \hat{\tilde{y}}_i=(a_{i}w+\gamma )\hat{%
\tilde{x}}_i^c.  \label{proof7_2}
\end{equation}%
So, by \eqref{proof7_1a}-\eqref{proof7_2} we have that 
\begin{equation*}
\hat{\tilde{x}}^c_i=\frac{a_{i}w\displaystyle \sum_{j\in I\backslash \left
\{ i\right \} }z_{0,ij}a_{j}+\gamma \displaystyle \sum_{k\in I\backslash
\left \{ i\right \} }a_{k}}{a_{i}w+\gamma }\  \ .
\end{equation*}%
As externalities are positive and $a_i>0$, $\gamma$ and $a_i w$ are just
weights in a weighted average. If $\frac{\gamma}{w}=0$, then $\hat{\tilde{x}}%
^c_i=\sum_{j\in I\backslash \left \{ i\right \}}z_{0,ij}a_{j}^c$, i.e.,
conjecture $\hat{\tilde{x}}^c_i$ is correct, so that $\mathbf{a}^c=\mathbf{a}%
^{NE}_{\mathbf{Z}_0}$. Finally, $\lim_{\frac{\gamma}{w} \rightarrow \infty}%
\hat{\tilde{x}}^c_i=\sum_{k\in I\backslash \left \{ i\right \} }a_{k}^c$ so
that at this limit we have $\mathbf{a}^c=\mathbf{a}^{NE}_{\mathbf{Z}_c}$.
\hfill $\blacksquare $

\subsection*{Proposition \protect \ref{prop:homeo1} (page \protect \pageref%
{prop:homeo1})}

\textbf{Proof. } \textbf{First, we derive some properties.} Recall that we
assumed a common bliss point in isolation: $\alpha _{i}=\alpha $ for each $%
i\in I$, and that $c_{i}$ is the conjectured ratio of $i$. Each equation in
the system given by \eqref{system1} can be written as an upward parabola $%
b_{1}a_{i}^{2}+b_{2}a_{i}+b_{3}=0$, in the following way 
\begin{eqnarray}
H_{i}(\vec{a},\vec{c},\gamma ,\mathbf{Z}) &=&\underbrace{c_{i}}%
_{:=b_{1}}a_{i}^{2}+\underbrace{\left( 1-\alpha c_{i}-c_{i}\left( \sum_{j\in
I}z_{ij}a_{j,t}\right) \right) }_{:=b_{2}}a_{i}  \notag \\
&&-\underbrace{\left( 1+c_{i}\left( \gamma \sum_{j\neq i}a_{j,t}\right)
\right) }_{:=b_{3}}=0\ .  \label{system2}
\end{eqnarray}%
%
%
%
%
%
%
%
%
%
%
%
%
%
%
%
%
%
So, for each $i\in I$, the solution $a_{i}^{\ast }$ is such that $H_{i}(\vec{%
a},\vec{c},\gamma ,\mathbf{Z})=0$ lays in the right--arm of an upward
parabola, where $\left. \frac{dH_{i}}{da_{i}}\right \vert
_{a_{i}=a_{i}^{\ast }}>0$. Each $H_{i}(\vec{a},\vec{c},\gamma ,\mathbf{Z})$
is linear in $c_{i}$. \newline

\noindent Equation \eqref{system2} holds in the unique positive solution
(because $b_3>0$): 
\begin{equation}  \label{parabola_solution}
a^*_i = \frac{- b_2 + \sqrt{b_2^2 + 4 b_1 b_3}}{2 b_1} \  \ ,
\end{equation}
so that $a^*_i$ can be seen as a continuous function of $b_1$, $b_2$ and $%
b_3 $. Considering that $a^*_i$ is increasing in $b_1$ (which is bounded by $%
1$), decreasing in $b_2$ and increasing in $b_3$, it follows that each $%
a^*_i $ increases in each $a_j$, with $j \ne i$. Moreover, each $a^*_i$
increases in $c_i$, so that 
\begin{equation*}
\left. \frac{d a_i}{d c_i} \right|_{a_i=a_i^*} > 0 \  \ .
\end{equation*}
If $b_2$ is bounded (from below), then $a_i^*$ is bounded above by 
\begin{equation*}
\lim_{b_1 \rightarrow 1} \frac{- b_2 + \sqrt{b_2^2 + 4 b_1 b_3}}{2 b_1} = 
\frac{- b_2 + \sqrt{b_2^2 + 4b_3}}{2 } \  \ ,
\end{equation*}
which is in turn bounded above by $\sqrt{b_3}$ (because if $a$ and $b$ are
positive, $\sqrt{a+b} \leq \sqrt{a} + \sqrt{b}$). 

\bigskip

\textbf{Second, we show that there is a homeomorphism.} There is a
continuous function that assigns to each $\vec{c} \in [0,1]^n$ an element $%
\vec{a}^* \in \mathcal{A}$, that is because

\begin{itemize}
\item either $c_i=0$ and then $a^*_i=\alpha $, with (from %
\eqref{parabola_solution}): 
\begin{equation*}
\lim_{c_i \rightarrow 0} a_i^* = \alpha \  \ ;
\end{equation*}

\item or $c_i>0$ and then each $a_i^*$ is continuously increasing in each $%
a_j $ with $j \ne i$. 
$b_2$ is bounded (from below), because the system defined by \eqref{systemNE}
admits a solution, and then also any linear transformation of this system
will admit a finite solution, which means that $b_2$ is limited.\newline
Since $b_2$ is bounded (from below), then $a_i^*$ is bounded above by 
\begin{equation*}
\sqrt{ 1 + c_i \left( \gamma \sum_{j \neq i } a_{j,t} \right) } \ .
\end{equation*}
\end{itemize}

But this upper limit is sub--linear, and then the system defined by %
\eqref{system1} admits a finite solution.


\bigskip

So, applying system \eqref{system1}, for each $\vec{c} \in [0,1]^n$, we
obtain a unique profile $\vec{a}^* \in \mathcal{A}$, and this function is
continuous because \eqref{parabola_solution} is continuous. \newline
To analyze the relation between $\vec{a}^*$ and $\vec{c}$, we already know
that each $a_i^*$ is increasing in $c_i$ and in all the other $a_j^*$, with $%
j \ne i$, which in turn are increasing in $c_j$. This shows that $a_i^*$ is
strictly monotone with respect to the lattice order of the domain of all
profiles $\vec{c} \in [0,1]^n$.

\bigskip

Strict monotonicity and continuity imply that the function from $\vec{a} \in 
\mathcal{A}$ to $\vec{c} \in [0,1]^n$ is invertible. {%
\flushright{\hfill
$\blacksquare$}}

{
}

\subsection*{Proposition \protect \ref{prop:homeo2} (page \protect \pageref%
{prop:homeo2})}

\textbf{Proof. } As resting points of the paths defined by \eqref{Hdynamics}%
, we consider the system derived from \eqref{system1} for each $i$: 
\begin{equation*}
H_{i}(\mathbf{a},\mathbf{c},\gamma ,\mathbf{Z})=\alpha +c_{i}\left( \gamma
\sum_{j\neq i}a_{j,t}\right) \frac{a_{i,t}c_{i,t}^{\prime }+1}{a_{i}c_{i}+1}%
-a_{i}=0\  \ ,
\end{equation*}%
with $c_{i,t}^{\prime }=\frac{\sum_{j\in I}z_{ij}a_{j,t}}{\gamma \sum_{j\neq
i}a_{j,t}}$. We can compute its Jacobian, with respect to $\mathbf{a}$. We
know from the proof of Proposition \ref{prop:homeo1} that each entry of this
Jacobian is strictly positive. If we prove that each row of this Jacobian
sums to less than $1$, by the Gershgorin circle theorem we will have that
the Jacobian is limited (as defined in Assumption \ref{ass:limited}), so
that the process is always a contraction and the resting points are stable
(see, e.g.,~\citealp{galor2007discrete}). The Jacobian $J$ is such that, for
each $i,j\in I$: 
\begin{equation*}
\left \{ 
\begin{array}{rcll}
J_{ij} & = & \frac{c_{i}}{a_{i}c_{i}+1}\left( \gamma +a_{i}z_{ij}\right) & %
\mbox{, for } j \ne i \\ 
J_{ii} & = & c_{i}\left( \gamma \sum_{j\neq i}a_{j}\right) \left( \frac{%
c_{i}^{\prime }}{a_{i}c_{i}+1}-c_{i}\frac{a_{i}c_{i}^{\prime }+1}{%
(a_{i}c_{i}+1)^{2}}\right) -1 & \mbox{, otherwise}.%
\end{array}%
\right.
\end{equation*}

%

The sum of each row of the Jacobian is 
\begin{equation}
\sum_{j\in I} J_{ij} = \frac{ c_i }{ a_i c_i + 1} \left( \gamma \left(
\sum_{j\neq i}a_{j}\right) \left( c_i^{\prime }- c_i \frac{a_i c^{\prime }_i
+1}{a_i c_i +1} \right) + a_i \left( \sum_{j\neq i} z_{i,j} \right) +
\gamma( n-1 ) \right) - 1 \  \ .  \label{rowJ}
\end{equation}


%


\bigskip

Let us analyze expression \eqref{rowJ} with respect to $a_{i}$, for any $%
a_{i}\geq 0$. \newline
First note that 
\begin{equation}
\lim_{a_{i}\rightarrow \infty }\sum_{j\in I}J_{ij}=\sum_{j\neq i}z_{ij}-1\  \
,  \label{eq:opt1}
\end{equation}%
whose absolute value is less than one by assumption. \newline
Moreover, 
\begin{equation}
\lim_{a_{i}\rightarrow 0}\sum_{j\in I}J_{ij}=c_{i}\gamma \left( \left(
\sum_{j\neq i}a_{j}\right) \left( c_{i}^{\prime }-c_{i}\right) +(n-1)\right)
-1\  \ .  \label{eq:opt2}
\end{equation}%
An interior maximum or minimum of the numerical expression \eqref{rowJ},
with respect to $a_{i}$, must satisfy first order condition 
\begin{eqnarray}
-\left( \frac{c_{i}}{a_{i}c_{i}+1}\right) ^{2}\left( \gamma \left(
\sum_{j\neq i}a_{j}\right) \left( c_{i}^{\prime }-c_{i}\frac{%
a_{i}c_{i}^{\prime }+1}{a_{i}c_{i}+1}\right) +a_{i}\left( \sum_{j\neq
i}z_{ij}\right) +\gamma (n-1)\right) &&  \notag \\
+\frac{c_{i}}{a_{i}c_{i}+1}\left( \gamma \left( \sum_{j\neq i}a_{j}\right)
\left( \frac{c_{i}}{a_{i}c_{i}+1}\right) \left( c_{i}^{\prime }-c_{i}\frac{%
a_{i}c_{i}^{\prime }+1}{a_{i}c_{i}+1}\right) +\left( \sum_{j\neq
i}z_{ij}\right) \right) &=&0 \  \ .  \notag
\end{eqnarray}%
The last expression can be simplified and results in 
\begin{equation*}
c_{i}\gamma (n-1)=\sum_{j\neq i}z_{ij}\  \ ,
\end{equation*}%
which is independent of $a_{i}$. So, the only candidates for being minima or
maxima for expression \eqref{rowJ} are its values in the extrema, namely %
\eqref{eq:opt1} and \eqref{eq:opt2}.

\bigskip

Also, the sign of the first derivative of \eqref{rowJ} with respect to $a_i$
is equal to the sign of $\sum_{j\neq i} z_{ij} - c_i \gamma (n-1) $. So, if $%
c_i \gamma (n-1) < \sum_{j\neq i} z_{ij}$ we have that \eqref{rowJ} is
strictly increasing in $a_i$, and then \eqref{eq:opt1} is strictly greater
than \eqref{eq:opt2}.

\bigskip

The value of \eqref{eq:opt1} is between $-1 $ and $1$, by assumption,
because $0 < \sum_{j\neq i} z_{ij} < 2$.

\bigskip

The quantity in \eqref{eq:opt2} is minimized by $c_i \rightarrow 0 $%
; and $c^{\prime }_i \rightarrow 0$. In this case \eqref{eq:opt2} goes to $%
-1 $ from the right, and for any $c_i>0$ it will be greater than $-1$. 
This completes the proof, because we have shown that any row of the Jacobian 
$J$ sums to a number between $-1$ and $1$. \hfill $\blacksquare$

\bibliographystyle{ecta}
\bibliography{SCE}

%
%
%
%

\end{document}